\documentclass[12pt]{article}
\usepackage{amsmath,amsfonts}
\usepackage{hyperref}
\usepackage{graphicx}

\usepackage[dvips]{color}
\usepackage{amsmath}
\usepackage{amssymb}
\usepackage{amsxtra}
\newcommand{\vev}[1]{\left\langle #1 \right\rangle}

\newcommand {\CalE} {\mathcal E}

\newcommand {\CalO} {\mathcal O}

\newcommand {\CalN} {\mathcal N}

\newcommand {\CalL} {\mathcal L}

\newcommand {\CalM} {\mathcal M}

\newcommand {\BR}   {\mathbb R}
\newcommand {\BZ}   {\mathbb Z}
\newcommand {\BC}   {\mathbb C}

\newcommand {\ve}  {\varepsilon}
\newcommand {\ep}  {\epsilon}

\newcommand{\g}{\mathfrak{g}}

\newcommand {\p} {\partial}

\DeclareMathOperator{\coker}{coker}

\DeclareMathOperator{\tr} {tr}
\DeclareMathOperator{\Tr} {Tr}

\DeclareMathOperator{\ad} {ad}
\DeclareMathOperator{\ind} {ind}

\DeclareMathOperator{\diag}{diag}

\newcommand{\SU}{SU}
\newcommand{\U}{U}
\newcommand{\SO}{SO}

\numberwithin{equation}{section}

\usepackage[numbers,sort&compress]{natbib}
\usepackage{hypernat}

\usepackage{amsthm}
\usepackage{amssymb}
\usepackage{epic}
\usepackage{eepic}
\usepackage[matrix,arrow]{xy}
\usepackage{epsfig}

\makeatletter\@addtoreset{equation}{section}\makeatother

\def\bC {\mathbb{C}}

\def\bR {\mathbb{R}}

\newcommand{\beq}{\begin{equation}}
\newcommand{\eeq}{\end{equation}}
\newcommand{\bea}{\begin{eqnarray}}
\newcommand{\eea}{\end{eqnarray}}

\renewcommand{\bar}{\overline}

\newcommand{\cG}{{\mathcal G}}

\newcommand{\cL}{{\mathcal L}}

\newcommand{\cN}{{\mathcal N}}

\newcommand{\cR}{{\mathcal R}}

 \newcommand{\rf}[1]{(\ref{#1})}
\renewcommand{\title}[1]{\vbox{\center\LARGE{#1}}\vspace{5mm}}
\renewcommand{\author}[1]{\vbox{\center\large#1}\vspace{5mm}}
\newcommand{\address}[1]{\vbox{\center\em#1}}

\begin{document}
\bibliographystyle{utphys}
 
\begin{titlepage}
\begin{center}
%\vspace{3mm}
\hfill {\tt CERN-PH-TH/2011-100}\\
\hfill {\tt ITEP-TH-xx/xx}\\
\hfill {\tt UT-Komaba/11-2}\\
\vspace{8mm}

\title{
{\LARGE
\hskip-8pt Exact Results for 't Hooft Loops  in Gauge    Theories on $S^4$}}
\vspace{6mm}
  
Jaume Gomis${}^{a}$\footnote{\href{mailto:jgomis@perimeterinstitute.ca}
{\tt jgomis@perimeterinstitute.ca}},
Takuya Okuda${}^{b}$\footnote{\href{mailto:takuya@hep1.c.u-tokyo.ac.jp}
{\tt takuya@hep1.c.u-tokyo.ac.jp}},
and
Vasily Pestun${}^{c }$%
\footnote{%
\hspace{0.2mm}\href{mailto: pestun@fas.harvard.edu}{\tt pestun@fas.harvard.edu}
\hspace{2mm}
On leave of absence from ITEP, Moscow, 117259, Russia.}

\vskip 10mm
\address{
${}^a$Perimeter Institute for Theoretical Physics,\\
Waterloo, Ontario, N2L 2Y5, Canada}
\vskip 2mm
\address{${}^a$Theory Group, CERN,\\
  CH-1211, Geneva 23, Switzerland}
 \vskip 2mm
\address{
${}^b$Institute of Physics, University of Tokyo,\\
Komaba, Meguro-ku, Tokyo 153-8902, Japan}
\vskip 2mm
\address{
${}^c$Center for the Fundamental Laws of Nature,\\
Jefferson Physical Laboratory, Harvard University,\\
Cambridge, MA 02138, USA}
 
\end{center}

\vspace{5mm}
\abstract{\smallskip
\normalsize{
\noindent
The path integral of a general $\CalN=2$ supersymmetric gauge theory on $S^4$ is
exactly evaluated in the presence of a  supersymmetric 't Hooft loop operator. The
result we find 
-- obtained using localization techniques  --
captures all perturbative quantum corrections as well as non-perturbative 
effects due to instantons and monopoles, which are supported at the 
north pole, south pole and equator of $S^4$. 
As a by-product, our gauge theory calculations   successfully confirm the predictions
made  for
't Hooft loops   obtained from the calculation of  topological defect correlators  in Liouville/Toda conformal field theory.}
}
\vfill
\end{titlepage}

\vbox{ \vspace{-20mm}
\tableofcontents
} 
%%%%%%%%%%%%%%%%%%%%%%%%%%%%%%

\newcommand{\vm}{\text{vm}}
\newcommand{\hm}{\text{hm}}

\section{Introduction}
\label{sec:intro}

Supersymmetry -- apart from being phenomenologically appealing for physics beyond the standard model -- is a powerful symmetry which constraints the dynamics of  gauge theories. 
Investigations of supersymmetric gauge theories have yielded important  physical (and mathematical) insights and serve as  calculable models for the rich dynamics of four dimensional gauge theories. For instance, the  exact low energy effective action of $\cN=2$ super Yang-Mills constructed by Seiberg and Witten \cite{Seiberg:1994rs} provides an elegant physical realization of quark confinement in terms of the dual Meissner effect, via the condensation of magnetic monopoles. 

The correlation functions of gauge invariant operators in supersymmetric gauge theories -- despite enjoying more controlled dynamics in comparison to QCD --  are highly non-trivial to calculate. 
Even for supersymmetric observables, which preserve some of the symmetries of the theory, 
generic  correlation functions have perturbative corrections to arbitrary loop order as well as non-perturbative instanton corrections.  
Only in the past few years, exact calculations for the correlation functions of 
some supersymmetric operators started to be available.
An important early step in this recent development
was the calculation of the exact partition function of physical $\cN=2$ gauge theories on $S^4$ and  of the  expectation value of supersymmetric Wilson loop operators in these theories \cite{Pestun:2007rz}. Likewise, the computation of certain supersymmetric domain walls in $\cN=2$ gauge theories on $S^4$ -- such as Janus and duality walls -- were presented in \cite{Drukker:2010jp} (see also \cite{Hosomichi:2010vh}).

Some of the most basic observables of four dimensional gauge theories are loop operators. These operators can be classified according to  whether the loop operator is electric or magnetic, giving rise to Wilson and 't Hooft operators respectively. 
 Gauge theory loop operators -- which are supported on curves in spacetime -- are   order parameters for the phases that a gauge theory can exhibit, and serve  as probes of  the quantum dynamics of gauge theories.  Loop operators are also the most basic observables on which   S-duality is 
conjectured to act in supersymmetric gauge theories  (or certain nonsupersymmetric lattice models), and therefore are ideal probes of this remarkable symmetry exhibited by some supersymmetric gauge theories and M-theory. Calculating these observables exactly allows for a quantitative study of S-duality and serves as a theoretical playground for gaining a deeper understanding of the inner workings of dualities.

In this paper we evaluate  the exact path integral which computes  the expectation value of  supersymmetric 't Hooft loop operators in an arbitrary  $\cN=2$ supersymmetric gauge theory on $S^4$ admitting a Lagrangian description. The expectation value of 't Hooft loop operators  -- originally introduced \cite{'tHooft:1977hy} to probe the 
phase structure of gauge theories --  are calculated by explicit evaluation of the path integral using localization \cite{Witten:1988ze}. In the localization framework,  the path integral is one-loop exact with respect to an effective $\hbar$-parameter, but nevertheless  the computation yields the exact result   with respect to the   gauge theory coupling constant of the theory.  Our analysis of 't Hooft loops together with the results of  \cite{Pestun:2007rz} for Wilson loops, provide a suite of complete, exact  calculations of the most elementary  loop operators in supersymmetric gauge theories.

We find that for an $\cN=2$ gauge theory in $S^4$, the expectation value of  a supersymmetric 't Hooft operator carrying magnetic charge labeled by a coweight%
\footnote{
We recall that a coweight, denoted as $B$ here, is an element of the
Cartan subalgebra $\mathfrak t$ of $G$ such that the product $\alpha \cdot B$
is an integer for all roots $\alpha\in \mathfrak t^*$ of $G$.
} $B$ of the gauge group $G$
 takes the form
 %\begin{eqnarray}
\beq
\begin{aligned}
\vev{T(B)}_{\cN=2}=&\int\, da \,\sum_v Z_\text{north}(v)  Z_\text{south}(v)  Z_\text{equator}(B,v)\\
%\nonumber\\
=&\int\, da \,\sum_v \left|Z_\text{north}(v)\right|^2   Z_\text{equator}(B,v)\,.
%\nonumber
\label{answerintro}
%\end{eqnarray}
\end{aligned}
\eeq
The integral is over the Cartan subalgebra of the gauge group.
The coweight $B$ of $G$ can be identified with the highest weight for
a representation of the Langlands (or GNO \cite{Goddard:1976qe}) dual group ${}^L G$.%
\footnote{%
The  Cartan subalgebra ${}^L\mathfrak t$ of ${}^L G$
can be identified with the dual $\mathfrak t^*$ of
the Cartan subalgebra of $G$ and vice versa: 
${}^L\mathfrak t \simeq \mathfrak t^*$,
${}^L\mathfrak t^* \simeq \mathfrak t$.
}
The sum is then over the coweights $v$ of $G$ such that
their corresponding weights of ${}^L G$ appear in the representation 
specified by $B$.

\begin{figure}[t]
\begin{center}
 \includegraphics[scale=0.5]{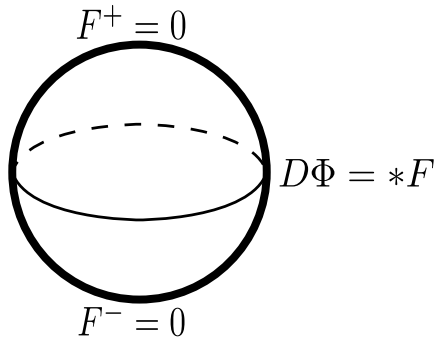}
 \end{center}
 \caption{Instanton, monopole and anti-instanton field configurations}
\end{figure}

The path integral  in the localization computation receives contributions which localize to the north and south poles of $S^4$ as well as to the equator, where the 't Hooft operator is supported.
Each factor has an elegant interpretation as arising from  specific field configurations in the effective path integral arising in the localization computation. The magic of localization is that it restricts the integral over the space of all field configurations    to the  submanifold   of field configurations 
invariant under a fermionic symmetry $Q$, which also preserves the supersymmetric 't Hooft operator. These field configurations are solutions to the localization  saddle point equations. Integrating out the fluctuations around each of the saddle points and summing over them in the path integral   yield  the exact result for the expectation value of the 't Hooft loop operator.

 The north pole factor captures the effects of point-like instantons while the south pole one incorporates the contributions of point-like anti-instantons. These configurations are the  solutions to the localization saddle point equations at the north and south poles  of $S^4$, given by $F^+=0$ and $F^-=0$ respectively.
 The result of summing over these saddle points  can be written in terms of Nekrasov's instanton partition function  \cite{Nekrasov:2002qd} of the corresponding $\cN=2$ theory  in $\bR^4$ (more precisely in the $\Omega$-background), with arguments depending on the effective magnetic charge $v$ 
 \beq
\begin{aligned}
{Z}_{\text{north}}(v)=&  Z_{\text{cl}}\left(ia-{v\over 2},q\right){Z}_{\text{1-loop},\text{pole}}\left(ia-{v\over 2}, im_f\right)Z_{\text{inst}}\left(ia-{v\over 2}, {1\over r}+ i m_f, {1\over r},{1\over r},q\right)\\  
\\
{Z}_{\text{south}}(v)=&  Z_{\text{cl}}\left(ia+{v\over 2},\bar q\right){Z}_{\text{1-loop},\text{pole}}\left(ia+{v\over 2}, im_f\right)Z_{\text{inst}}\left(ia+{v\over 2}, {1\over r}+ i m_f, {1\over r},{1\over r},\bar q\right)\,,
\end{aligned}
\eeq
with $Z_{\text{cl}}, {Z}_{\text{1-loop},\text{pole}}$ and $Z_{\text{inst}}$ given in (\ref{eq:class-contr},\ref{loopcombination},\ref{nekrinst}).
The parameters $m_f$ are the masses of the hypermultiplets in the $\cN=2$ gauge theory, $r$ is the radius of $S^4$ and $q=\exp(2\pi i \tau)$, where $\tau$ is the gauge theory coupling constant\footnote{For a semi-simple gauge group there is a coupling constant for each simple factor.}
\beq
\tau={\theta\over 2\pi}+{4\pi i \over g^2}\,.
\nonumber
\eeq

A  crucial new contribution to the 't Hooft loop expectation value arises from the equator of $S^4$, where the localization saddle point equations   
are the  Bogomolny equations $D\Phi=*F$. $Z_\text{equator}(B,v)$ captures the contribution to the path integral of field configurations which are  solutions  to the Bogomolny equations in the presence of a singular monopole  background  labeled by the magnetic charge $B$, created by the 't Hooft loop operator insertion. The sum over $v$ in \rf{answerintro} appears due to the physics of monopole screening, whereby   smooth non-abelian monopole field configurations 
screen the charge $B$ of the singular mononopole down to an effective magnetic charge $v$. 
In the path integral we must sum over all possible effective magnetic charges labeled by coweights $v$, which are attainable given a singular monopole of magnetic charge $B$.%
\footnote{The necessity to sum over such configurations was conjectured in \cite{Gomis:2009ir}, where the perturbative analysis of the expectation value of 't Hooft operators in $\cN=4$ super Yang-Mills was performed.}
The ${}^L G$-weights corresponding to $v$
precisely span the weights of the representation of ${}^L G$ for which 
$B$ corresponds to the highest weight.\footnote{We recall that regular monopoles are labeled by coroots, which when acting on the singular monopole, labeled by a coweight $B$, generate all coweights associated to $B$.}
The equatorial contribution is 
\beq
Z_\text{equator}(B,v)=Z_\text{1-loop,eq}(ia,im_f,B)\, Z_\text{mono}(ia,im_f;B,v)\,,
 \eeq
 where $Z_\text{1-loop,eq}$ is given in \rf{1-loop-equator} and $Z_\text{mono}$ in section \ref{sec:screening}. Combining all the various contributions 
 produces the exact expectation value for the supersymmetric 't Hooft loop operator in $\cN=2$ gauge theories on $S^4$.

Our gauge theory computations are in elegant agreement with the
conjectures and calculations in \cite{Alday:2009fs,Drukker:2009id,Gomis:2010kv} for 't Hooft operators  in certain $\cN=2$ gauge theories 
 using topological defect operators in two dimensional nonrational  conformal field theory. In these
papers, gauge theory loop operators in $\cN=2$ gauge theories were
identified with loop operators (topological webs more generically) in
two dimensional Liouville/Toda conformal field theory, and some correlation
functions were explicitly calculated. The Liouville/Toda conformal field theory computations are shown
to capture in detail
all the features of our gauge  theory computation, thereby 
establishing the proposal put forward in  \cite{Alday:2009fs,Drukker:2009id,Gomis:2010kv}.

The localization calculation   performed in this paper is the first
example of an exact computation of a path integral in the presence
of a genuine singularity due to a disorder operator --
an operator characterized by the singularities induced on the fields -- and of which a 
  't Hooft operator  is a prime example.%
\footnote{%
A monopole operator in three dimensions is a closely related disorder operator.
The work \cite{Kim:2009wb} performed localization computations
for monopole operators in three dimensions to compute the supersymmetry index
via radial quantization, 
thus removing the singularity by a coordinate change.
In this paper we deal with the  monopole singularity more directly.
} 
In order to treat precisely the fluctuations around the singular field
configuration, we employ the mathematical correspondence between
singular monopoles in three dimensions and $U(1)$-invariant
instantons in four dimensions \cite{Kronheimer:MTh}.
This turns out to be a particularly clean way to carry out the
relevant index calculations.

 The plan of the rest of the paper is as follows. Section 2   briefly introduces the
 key ingredients that will be needed to perform the localization computation of 't Hooft
 operators in $\cN=2$ gauge theories on $S^4$. In section 3 we derive the localization
 saddle point equations relevant for the localization computation, demonstrate that these
 equations   interpolate between the anti-self-duality, self-duality and Bogomolny equations 
 at the north pole, south pole and equator respectively, and find the most general
 non-singular solution to these equations. This section also 
describes the singular
 field configuration produced by the supersymmetric 't Hooft operator as well as the
symmetries of the theory used to carry out the localization computation. Section 4 contains the calculation
of the classical contribution of the 't Hooft loop path integral, which includes a discussion of the relevant boundary terms. In this 
section we demonstrate that the classical result can be factored into a contribution arising from the north pole
and one from the south pole. Section 5 computes the contribution 
due to the singular solutions to the saddle point equations arising at the north and south poles, described
by pointlike instantons and anti-instantons. In section 6 we calculate the localization one-loop determinants
arising from the north and south poles of $S^4$ as well as from the equator. Section $7$ describes the effect
of monopole screening in the study of the equatorial Bogomolny equations and explains how to calculate the
contribution to the 't Hooft loop expectation value due to screening. In section $8$ we compare our gauge theory
results with the Liouville/Toda computations conjectured to capture 't Hooft operators in certain $\cN=2$ gauge theores.
We finish with conclusions in section 9. The Appendices contain some technical details and computations

\section{${\cal N}=2$ Gauge Theories in $S^4$ and Localization}
\label{sec:localize}

 In this section we introduce the main ingredients of the localization analysis in  \cite{Pestun:2007rz} that we require to calculate the exact expectation value of supersymmetric 't Hooft operators in an arbitrary four dimensional ${\cal N}=2$ gauge theory on $S^4$ admitting a Lagrangian description.\footnote{Localization of
 some $\cN=2$ gauge theories was also considered in \cite{Rey:2010ry}.}
Such a  theory is completely characterized by the   choice of a gauge group $G$ and of a representation $R$ of $G$ under which the ${\cal N}=2$  hypermultiplet transforms, the  ${\cal N}=2$ vectormultiplet transforming in the adjoint representation of $G$.  This  includes gauge theories with several gauge group factors and multiple matter representations by letting $G$   be the product of several gauge groups   and by taking $R$ to be a reducible representation of $G$. It therefore applies to any gauge theory with a Lagrangian description.

The on-shell field content of the $\cN=2$  multiplets is given by
 \begin{eqnarray}
 \hbox{vectormultiplet}&:& (A_\mu, \Phi_0,\Phi_9, \Psi)\nonumber\\
  \hbox{hypermultiplet}&:& (q, \tilde q^\dagger, \chi)\nonumber\,.
 \end{eqnarray}
In this notation, the  usual  complex scalar field of the ${\cal N}=2$ vectormultiplet is
constructed out of the real fields 
 $\Phi_0$ and $\Phi_9$.
 One complication in the construction of the $\cN=2$ Lagrangian in $S^4$ overcome  in \cite{Pestun:2007rz} was  to turn on in a supersymmetric way  mass parameters for the flavour symmetries associated to the hypermultiplet. These  ${\cal N}=2$ gauge  theories  on $S^4$ are invariant under  the superalgebra $OSp(2|4)$, where $Sp(4)\simeq SO(5)$ is the isometry group on $S^4$  and $SO(2)_R$  is a subgroup of the $SU(2)$ R-symmetry of the corresponding ${\cal N}=2$ gauge theory in flat spacetime.
 
 The key idea behind localization  \cite{Witten:1988ze} exploits that the path integral -- possibly enriched with  any observables invariant under the action of a supercharge $Q$ --  is unchanged upon deforming the supersymmetric Lagrangian  of the theory by a $Q$-exact term
 \begin{equation}
 \cL\rightarrow \cL +t\, Q\cdot V\,.
 \label{deformedaction}
 \end{equation}
 The restriction on the choice of $V$ is such that if $Q^2$ generates a 
 symmetry and a gauge transformation, as will be the case in our analysis,
 then $V$ must be gauge invariant and also invariant under the action of the symmetry.
 Also we require
the path integral to be still convergent after the deformation, and
 that the contribution  from the boundary in the space of fields vanishes. In order to localize the gauge fixed path integral, the supersymmetry generated by $Q$ must be realized off-shell, and a gauge fixing procedure must be implemented. This was accomplished in  \cite{Pestun:2007rz} by introducing suitable auxiliary fields and a ghost multiplet, which plays a key role in precisely determining the measure of integration of the fluctuations.

 Since the path integral is independent of $t$, we can study it in the $t\rightarrow \infty$ limit. In this limit the saddle points of the path integral are the solutions to the localization equations, which are the saddle points of the deformed action
$ Q\cdot V$.
 In this limit, the path integral  becomes one-loop exact  with respect to the effective $\hbar=1/t$ parameter and can be evaluated by summing over all saddle points.
  Therefore, it can be calculated by evaluating  the original Lagrangian $\cL$  on the saddle points   and by  integrating out the quadratic fluctuations of all the fields in   the Lagrangian  deformation $Q\cdot V$ expanded around the solutions to the saddle  point equations.\footnote{The original Lagrangian $\cL$ is irrelevant for the localization one-loop analysis.}
   Of course, even though the path integral is one-loop exact with respect to $t$, it yields results to all orders in perturbation theory with respect to the original gauge coupling constant $\tau$ of the theory. This underlies the power of localization. In favorable situations, for a judicious choice of $V$, the deformation freezes out most of the fields that must be integrated over in the path integral, thus yielding   a path integral for a reduced model, with fewer degrees of freedom.

In the analysis in \cite{Pestun:2007rz},  as well as in our analysis, it suffices to   single out a single supersymmetry generator $Q$ of the  $OSp(2|4)$ symmetry algebra present in any ${\cal N}=2$ gauge theory on $S^4$. This supercharge generates an $SU(1|1)$ subalgebra of  $OSp(2|4)$, given explicitly by
\begin{equation}
Q^2=J+R\,,\qquad [J+R,Q]=0\,.
\end{equation}
$J$ is the generator of a $U(1)_J$ subgroup of the $SO(5)$ isometry group of the $S^4$ while $R$ is the $SO(2)_R\simeq U(1)_R$ symmetry generator in $OSp(2|4)$. If we represent the $S^4$ of radius $r$ by the embedding equation
\begin{equation}
  X_1^2+\ldots+X_5^2=r^2\,,
  \label{embedding}
\end{equation} 
then $J$ acts as follows
\beq
\begin{aligned}
X_1+iX_2&\rightarrow e^{i\ve}(X_1+iX_2)\,  \\
X_3+iX_4&\rightarrow e^{i\ve}(X_3+iX_4)\,.
\label{Jdeff}
\end{aligned}
\eeq
We note that the action of $J$ has two antipodal fixed points on $S^4$, which can be  used to define the north and south pole of $S^4$. The $U(1)$ symmetry 
associated to $J+R$ will be denoted by $U(1)_{J+R}\equiv (U(1)_J\times U(1)_R)_{\text{diag}}$.

 We conclude this section by mentioning a property of the localization   equations  that we will  exploit in the following section when studying the ${\cal N}=2$ gauge theory path integral on $S^4$ in the presence of a supersymmetric 't Hooft loop operator.
  The deformation term $Q\cdot V$ that we add to the
action naturally splits into two pieces, one giving rise to localization
equations for the vectormultiplet and one for the  hypermultiplet. In formulas 
\begin{equation}
V=V_{\vm}+V_{\hm}=\Tr(\overline{ Q\cdot \Psi}\,  \Psi)+\Tr(\overline{Q\cdot \chi}\,\chi)\,,
\end{equation}
where $\Psi$ and $\chi$ are the fermions in the vectormultiplet and hypermultiplet respectively. We represent the fermion fields in the $\cN=2$ gauge theory by sixteen component, ten dimensional Weyl spinors of $Spin(10)$ subject to the projection conditions (see appendix \ref{kill} for spinor notations and conventions)
\beq
\begin{aligned}
\Gamma^{5678}\Psi&=-\Psi\\
\Gamma^{5678}\chi&=+\chi\,.
\label{transformRR}
\end{aligned}
\eeq
Since the bosonic part of deformed action $Q\cdot V$ --    given by $\Tr(|Q\cdot\Psi|^2)+\Tr(|Q\cdot\chi|^2)$ --  is positive definite, the saddle point equations are
\beq
\begin{aligned}
Q\cdot\Psi&=0\\
Q\cdot\chi&=0\,.
\end{aligned}
\eeq

As shown in  \cite{Pestun:2007rz}, the only solution of the saddle point equations \begin{equation}
Q\cdot \chi=0
\end{equation}
  forces all the fields in the hypermultiplet to vanish.\footnote{This can be shown by writing $V_\text{hm}$ as a sum of squares. One of the terms that is generated is a mass term for the scalars in the hypermultiplet -- that is $qq^\dagger+\tilde q\tilde q^\dagger$ -- which implies that  on the saddle point  $q=\tilde q=0$.}
    Therefore, we are left to analyze the non-trivial saddle point equations for the vectormultiplet fields\footnote{\label{dimred}This formula should be dimensionally reduced to four dimensions using that $F_{mn}=[D_m,D_n]$ and that $D_{A}\cdot= [\Phi_A,\cdot]$  for $A=9,0$. See appendix \ref{sec:lie-alg-conv} for gauge theory conventions.}
      \begin{equation} 
Q\cdot \Psi={1\over 2}F_{mn}\Gamma^{mn} \epsilon_Q -{1\over 2}\Phi_A\Gamma^{A\mu}  \nabla_\mu\epsilon_Q +i K_j\Gamma^{8\,j+4}\epsilon_Q=0\,,
\label{saddle}
  \end{equation}
  where $A_m\equiv(A_\mu,\Phi_A)=(A_\mu,\Phi_9,\Phi_0)$  and $K_j\equiv(K_1,K_2,K_3)$ are the propagating bosonic fields and three auxiliary fields of the ${\cal N}=2$ vectormultiplet respectively. Therefore in  our conventions $m=1,2,3,4,9,0$, while $\mu=1,2,3,4$ and $A=9,0$. $\epsilon_Q$ is the conformal Killing spinor that parametrizes the supersymmetry transformation generated by the supercharge $Q$.

  The  equations \rf{saddle} are   Weyl invariant.  That is $Q\cdot \Psi=0$ is invariant under the Weyl  transformation 
  \begin{equation}
g_{\mu\nu}\rightarrow \Omega^2g_{\mu\nu},~~~A_\mu\rightarrow A_\mu,~~~\Phi_A\rightarrow \Omega^{-1} \Phi_A,~~~ K_j\rightarrow\Omega^{-2} K_j ,~~~ \epsilon_Q\rightarrow \Omega^{1/2} \epsilon_Q\,.
\label{transform}
\end{equation}
We will use this symmetry to study \rf{saddle} in a Weyl frame where the 
  localization equations take a simpler form.
  
\section{'t Hooft Loop in $S^4$ and Localization Equations}
\label{sec:Hooft}

 In this section we initiate our study of the expectation value of a supersymmetric 't Hooft loop operator \cite{'tHooft:1977hy} in an arbitrary $\cN=2$ gauge theory in $S^4$.   We start by constructing a   supersymmetric 't Hooft loop operator which is annihilated by $Q$ (and therefore by $J+R$). This implies that we can localize the 't Hooft loop path integral using the supercharge $Q$.  The derivation and interpretation of the localization saddle point equations $Q\cdot \Psi=0$ in \rf{saddle} follow. We will then find the most general non-singular solution to the localization equations in the presence of a supersymmetric 't Hooft loop operator.
 
A  't Hooft loop operator inserts a Dirac monopole into (an arbitrary) spacetime. The operator  has support  on the loop/curve spanned by the wordline of the monopole. In an arbitrary gauge theory, the operator is characterized by a boundary condition near the support of the loop operator that specifies the magnetic flux created  by the monopole. Since the choice of 't Hooft operator depends on the embedding of the $U(1)$ gauge group of a Dirac monopole into the gauge group $G$, these operators are labeled by a    coweight or magnetic weight vector $B$, which takes values  in the coweight lattice $\Lambda_{cw}$ of the gauge group $G$ \cite{Goddard:1976qe}.

Locally, near the location of any point on the loop -- where the loop is
locally a straight line --  the 't Hooft operator creates quantized
magnetic flux \cite{Kapustin:2005py} 
%\JG{Changed sign to be consistent with choice of supercharge}
\begin{equation}
F={B\over 4}\epsilon_{ijk} {x^i\over |\vec x|^3} dx^k\wedge dx^j\,, 
\label{magnetic}
\end{equation}
where $x^i$ for $i=1,2,3$ denote the three local transverse coordinates to any point in the loop. Since $B\equiv B^aH_a\in \mathfrak{t}$ takes values in the Cartan subalgebra  
$\mathfrak{t}$ of the Lie algebra $\mathfrak{g}$ of the gauge group $G$, the magnetic flux \rf{magnetic} is abelian.
 Locally, this operator inserts quantized flux through the $S^2$ that
 surrounds any point in the loop
%\JG{Changed sign to be consistent with choice of supercharge}
\begin{equation}
\int_{S^2}{F\over 2\pi}=-B\,.
\end{equation}

 In  order to be able to apply localization  we must consider   supersymmetric 't Hooft loop operators invariant under the action of $Q$. These operators create  a local singularity on the scalar fields of the $\cN=2$ vectormultiplet. The   singularity which will be locally compatible with our choice of $Q$ is\footnote{This follows by noting that a 't Hooft loop sourcing the scalar field $\Phi_9$ shares common supersymmetries with the Wilson loop considered in \cite{Pestun:2007rz} -- which couples to the scalar field $\Phi_0$ -- and which  by construction is annihilated by the supercharge $Q$. We will soon explicitly show that the exact 't Hooft loop singularity is invariant under the action of $Q$.}\begin{equation}
\Phi_9={B\over 2|\vec x|}\,.
\label{pole}
\end{equation}

A 't Hooft loop operator which is globally annihilated  by   $Q$ and $J+R$, can be constructed by choosing -- without loss of generality -- the support of the 't Hooft operator  to be the maximal circle on $S^4$
 \begin{equation}
 X_1^2+X_2^2=r^2, \qquad X_3=X_4=X_5=0\,,
 \label{supportcircle}
 \end{equation}
 which is located at the equator of $S^4$ and left invariant by the action of $J$ (see \rf{Jdeff}).
 We find it convenient to study the localization equations $Q\cdot \Psi=0$ in \rf{saddle}  in the presence of the circular 't Hooft loop by choosing the following coordinates on $S^4$ (see appendix \ref{coords} for various useful coordinate systems) 
 \begin{equation}
ds^2={\sum_{i=1}^3dx_i^2\over \left(1+{|\vec x|^2\over 4r^2}\right)^2} + 
r^2 {\left(1-{|\vec x|^2\over 4r^2}\right)^2\over \left(1+{|\vec x|^2\over 4r^2}\right)^2}d\tau^2\,.
\end{equation}
The coordinates $x_i$, where $|\vec x|^2\leq 4r^2$, define a three-ball $B_3$.
In these coordinates, the support of the circular 't Hooft loop  \rf{supportcircle}
is the maximal circle parametrized by the coordinate $\tau$ located at $x_i=0$.  In these coordinates the action of $J$, defined in   \rf{Jdeff},  is (see equation \rf{bs1cooord})
\beq
\begin{aligned}
x_1+ix_2&\rightarrow e^{i\ve}(x_1+ix_2)\\
\tau&\rightarrow\tau+ \ve\,.
\label{Jaction}
\end{aligned}
\eeq
%  \TO{North Pole $\rightarrow$ north pole?
% Both seem possible (google ``north pole n-sphere'')
% but see
% \href{http://answers.yahoo.com/question/index?qid=20080108162403AAqTXcv}{this
% link}.
%}
 Therefore, the north and south poles of the $S^4$  -- the fixed points of the action of $J$ -- are located at $\vec x=(0,0,2r)$ and $\vec x=(0,0,-2r)$  respectively.

Now by using the invariance of the saddle point equations  \rf{saddle}  under  the Weyl transformation  \rf{transform}, 
   the solutions to the saddle point equations on $S^4$   can be obtained from the solutions of the saddle point equations in $B_3\times S^1$
 \begin{equation}
 ds^2_{B_3\times S^1}={\sum_{i=1}^3dx_i^2} + 
r^2 \left(1-{|\vec x|^2\over 4r^2}\right)^2d\tau^2\,.
\label{ballmetric}
 \end{equation}
  They are related by the transformation  \rf{transform} with
 $\Omega= \left(1+{|\vec x|^2\over 4r^2}\right)$.

    One advantage of this choice of Weyl frame is that the exact 
 singularity produced by the  circular 't Hooft loop operator in $B_3\times S^1$  is identical to the one  produced by inserting  a static point-like monopole   in flat spacetime.  The exact circular 't Hooft loop background on $B_3\times S^1$ annihilated by $Q$ when the topological angle vanishes -- that is  when $\theta=0$ --  is given by\footnote{\label{susyproof}In appendix \ref{sec: soluns} we show that this background solves the localization saddle point equations $Q\cdot \Psi=0$ derived in the next subsection.}
 \begin{equation}
  \label{scalarpole} 
  \begin{aligned}
  F&={B\over 4}\epsilon_{ijk} {x^i\over |\vec x|^3} dx^k\wedge dx^j\\
 \Phi_9&={B\over 2|\vec x|}\,.
  \end{aligned}
\end{equation}
 When the topological angle is non-trivial -- that is when $\theta\neq 0$ --  then the   particle  inserted by the 't Hooft operator is a dyon, which acquires   electric charge through the Witten effect \cite{Witten:1979ey}. If  the 't Hooft operator is labeled by a magnetic weight $B$, the induced electric weight is ${g^2\theta B/ 4\pi}$. Moreover, the scalar field $\Phi_0$ also acquires a singularity near the loop. The exact background    created by a supersymmetric   't Hooft loop    on $B_3\times S^1$   is given by$^{\text{\scriptsize\ref{susyproof}}}$
\beq
\begin{aligned}
  F_{jk} &= -\frac B 2 \ep_{ijk} \frac{x_i}{|\vec x|^3}\,, \qquad F_{i\hat 4}=-ig^2\theta {B\over 16\pi^2} \frac{x_i}{|\vec x|^3}\,,\\
  \Phi_9 &= \frac{B}{2 |\vec x|}\,,\qquad\qquad\quad \Phi_0 =- g^2\theta {B\over 16\pi^2} {1\over  |\vec x|}\,.
  \label{dyonic}
\end{aligned}
\eeq
The corresponding singularity created by the insertion of  the circular 't Hooft loop in $S^4$ can   then be simply obtained by performing the Weyl transformation \rf{transform} with $\Omega= \left(1+{|\vec x|^2\over 4r^2}\right)$.

 \subsection{Symmetries and Fields}
% \subsection{Symmetries and Localization Equations in $B_3\times S^1$}
 \label{sec:ball}
 
 We now proceed to determining the partial differential equations for the bosonic fields in the $\cN=2$ vectormultiplet  on $B_3\times S^1$ 
  whose solutions yield the saddle points  of the localization path integral  upon Weyl transforming  them back to $S^4$.\footnote{As we mentioned earlier, the saddle point equations for the hypermultiplets force the fields in the multiplet to vanish.}  We first have to choose the supercharge $Q$ with which to localize the 't Hooft loop path integral.

  The supersymmetry transformations of an ${\cal N}=2$ gauge theory on a four manifold with metric $h_{\mu\nu}$ is  parametrized by a  sixteen component Weyl spinor of $Spin(10)$ which solves the conformal Killing spinor equation\footnote{The theory has maximal number of supersymmetries when the metric is conformally flat.}
  \begin{eqnarray}
  \nabla_\mu \epsilon  = \tilde{\Gamma}_\mu \tilde\epsilon\, 
  \label{killspin}
  \end{eqnarray}	
  subject to the projection
  \begin{equation}
  \Gamma^{5678}\epsilon=-\epsilon\,.
  \end{equation} 
 $\tilde\epsilon$ is determined in terms of $\epsilon$ by $\tilde\epsilon={1
 \over 4} \Gamma^\mu \nabla_\mu \epsilon$.\footnote{The spinor $\epsilon$ is referred to as  a conformal Killing spinor because its defining equation $\nabla_\mu \epsilon ={1\over 4} \tilde{\Gamma}_\mu\Gamma^\nu\nabla_\nu \epsilon$ is invariant under the Weyl transformation \rf{transform}.}
 It  satisfies
  \beq
  \tilde{\Gamma}^\mu \nabla_\mu \tilde\epsilon  =  -{R\over 12} \epsilon\,,
 \eeq
where    $R$ is the scalar curvature derived from $h_{\mu\nu}$.

  The equations on $B_3\times S^1$  that we need to analyze are\footnote{In our conventions   $Q$ acts on a field as a fermionic operator and therefore $\epsilon$ is a commuting spinor.}
\begin{equation}
 {1\over 2}F_{mn}\Gamma^{mn} \epsilon -{1\over 2}\Phi_A\Gamma^{A\mu}  \nabla_\mu\epsilon +i K_j\Gamma^{8\,j+4}\epsilon=0\,,
 \label{saddleball}
 \end{equation}
 for an specific choice of $\epsilon=\epsilon_Q$.
Here 
 $\epsilon$ is the (commuting) conformal Killing spinor of the $\cN=2$ gauge theory on $B_3\times S^1$ which parametrizes the supersymmetry transformation generated by the supercharges  of the  $OSp(2|4)$ symmetry of the $\cN=2$ gauge theory.  
 The   general conformal Killing spinor on $B_3\times S^1$ --  with metric \rf{ballmetric} --  is given by (see appendix \ref{kill} for details and conventions)
 \begin{equation}
 \epsilon=\cos(\tau/2)\left(\hat \ve_s+x^i\tilde\Gamma_i\,\hat \ve_c\right)+\sin(\tau/2)\,\tilde\Gamma^4\left(2r\, \hat \ve_c+{x^i\over 2r}\Gamma_i\,\hat \ve_s\right)\,,
 \label{killsol}
 \end{equation}
 where $\hat \ve_s$ and $\hat \ve_c$ are two constant ten dimensional Weyl spinors of opposite ten dimensional  chirality obeying $\Gamma^{5678}\hat \ve_s=-\ve_s$
 and $\Gamma^{5678}\hat \ve_c=-\ve_c$.

We now identify the  spinor   $\epsilon_Q$ which parametrizes the supersymmetry transformations of the supercharge $Q$ generating the $SU(1|1)$ subgroup of the $OSp(2|4)$ symmetry of an $\cN=2$ theory in $S^4$. This is the supercharge used in our localization analysis. We  take the   spinor  $\hat \ve_{s}$ to be 
 \bea
   \hat \ve_{s} = \tfrac 1 2 (1,0,0,0,0^4,1,0,0,0,0^4) 
    \eea
   and  the   spinor  $\hat \ve_{c}$
   \begin{equation}
  \hat\ve_{c} = -{i\over 2r} \Gamma^{120} \hat\ve_{s} = {1\over 4r} (0,0,0,-1,0^4,0,0,0,-1,0^4)\,.
  \label{wilson}
\end{equation}
Therefore, the conformal  Killing spinor associated to $Q$ is given by
\beq
\epsilon_Q={1\over 4r}
\begin{pmatrix}
   2 r \cos \left(\frac{\tau }{2}\right) -x_3 \cos \left(\frac{\tau }{2}\right) \\
    x_1 \sin \left(\frac{\tau }{2}\right)+x_2 \cos \left(\frac{\tau
   }{2}\right)  \\
   x_2 \sin \left(\frac{\tau }{2}\right)-x_1 \cos \left(\frac{\tau }{2}\right) \\
    x_3  \sin \left(\frac{\tau
   }{2}\right) -2 r \sin \left(\frac{\tau
   }{2}\right)  \\
   0^4 \\
    2 r  \cos \left(\frac{\tau }{2}\right) +x_3 \cos \left(\frac{\tau }{2}\right)\\
    -x_1 \sin \left(\frac{\tau }{2}\right)
    -x_2 \cos
   \left(\frac{\tau }{2}\right)\\
 x_1 \cos \left(\frac{\tau }{2}\right)-x_2 \sin \left(\frac{\tau }{2}\right)\\
    -2 r \sin
   \left(\frac{\tau }{2}\right)-x_3 \sin
   \left(\frac{\tau }{2}\right)\\
   0^4
  \end{pmatrix}\,,
  \label{killloc}
    \eeq
and has norm 
$\epsilon_Q\epsilon_Q={1\over 2}\left(1+{|\vec x|^2\over 4r^2}\right)
$.\footnote{Using \rf{transform} the norm of the spinor on $S^4$ is therefore $1/2$.}

We note that the  spinor $\epsilon_Q$
% \beq
  %\hat\ve_{c} = -{i\over 2r} \Gamma^{120} \hat\ve_{s}
 %\eeq
 generates one of the unbroken supersymmetries\footnote{Which obey $(1+i\tilde\Gamma^4\Gamma^0) \hat \ve_{s}=0$ and $(1+i\Gamma^0\tilde\Gamma^4) \hat \ve_{c}=0$.} preserved by the circular Wilson loop coupled to the scalar field $\Phi_0$   in the $\cN=2$ vectormultiplet 
  \beq
\Tr\exp\left( \oint_{S^1} \left[A_\mu {dx^\mu\over ds}+i |\dot x|\Phi_0\right] ds\right)\, 
\eeq
supported on the maximal circle at $\vec x=0$ in $B_3\times S^1$.  Therefore $Q$ can be used to localize the path integral in the presence of this Wilson loop operator, as in \cite{Pestun:2007rz}.

Given our choice of supercharge $Q$, we can now calculate $Q^2$, that is the symmetries and gauge transformation that $Q^2$ generates when acting on the  fields of the $\cN=2$ gauge theory. Due to the addition of suitable auxiliary fields,  this symmetry is  realized off-shell, as required for localization. 

The spacetime symmetry transformation induced by $Q^2$ is generated by the   Killing vector
\beq
  v^{\mu}(x,\tau) \equiv {e^{\mu}}_{\hat\mu} v^{\hat \mu}= 2 \epsilon_Q \Gamma^{\mu} \epsilon_Q = \left(- {x_2\over r},  {x_1\over r}, 0, {1\over r}\right)\,,
\eeq
where $e^{\hat i}=e^i=dx^i$ for $i=1,2,3$ and $e^{\hat 4}=r\left(1-{|\vec x|^2\over 4r^2}\right)d\tau$ is a vielbein basis for the metric on $B_3\times S^1$ given in \rf{ballmetric}.
Therefore,   $Q^2$ yields    the infinitesimal  $U(1)_J$  spacetime   transformation  \rf{Jaction} generated by $J$.

The operator  $Q^2$ also generates a $U(1)_R$  $R$-symmetry transformation. It acts on the fields of the theory as a  $U(1)_R\subset SU(2)_R$ subgroup of the $SU(2)_R$ symmetry present when the $\cN=2$ theory is in flat spacetime. Therefore, it acts on the gauginos $\Psi$ in the vectormultiplet and the scalars  $(q,\tilde q^\dagger)$ in the hypermultiplet.  The infinitesimal $R$-symmetry transformation generated by $Q^2$ is parametrized by the rotation parameter\footnote{Where 
$\tilde \epsilon_Q={1
 \over 4} \Gamma^\mu \nabla_\mu \epsilon={1\over 4r}\left(-\sin({\tau\over 2}),0^2,-\cos({\tau\over 2}),0^4,-\sin({\tau\over 2}),0^2,-\cos({\tau\over 2}),0^4\right)$.}
 \beq
v_R\equiv -4 \tilde \epsilon_Q\,   \Gamma^{56}\epsilon_Q=-4 \tilde \epsilon_Q  \,\Gamma^{78}\epsilon_Q={1\over r}\,.
\label{Rtransform}
\eeq
% \TO{I get the opposite sign
% if I use $\tilde\epsilon=\epsilon_c(\tau)$ read from (\ref{eq:epsilon_sc}).
% I can send a mathematical file.
% }
% \JG{I got the same. In the susy algebra, however,  the vector field
% associated to $R$-symmetry has a relative - sign compared to the
% vector field associated with spacetime gauge transformation (see eqn
% 2.7 in Vasily's paper). I have put back the - sign in the definition
% of the $R$-symmetry vector field, but I am not sure if it is more
% conventional to do it the other way. Please choose whatever you
% prefer.}
 As  advertised, our choice of supercharge $Q$ corresponding to the Killing spinor \rf{killloc} generates an $SU(1|1)$ subalgebra of $OSp(2|4)$
\beq
 Q^2=J+R\qquad [J+R,Q]=0\,,
\eeq
where $J+R$ generates $U(1)_{J+R}\equiv (U(1)_J\times U(1)_R)_{\text{diag}}$.

In the presence of  $N_{\text F}$ hypermultiplets transforming in a representation $R$ of $G$, the $\cN=2$ gauge theory has a  flavour symmetry group $G_{\text F}$, and the masses  $m_f$  with $f=1,\ldots N_{\text F}$ of the hypermultiplets take values in the
Cartan subalgebra of the flavour symmetry algebra, which has rank $N_{\text F}$.   The  action  of $Q^2$  on the hypermultiplets fields generates an infinitessimal  flavour symmetry transformation with parameters $m_f$, while the flavour symmetry action on vectomultiplet fields  is trivial.

Finally, the operator $Q^2$ further generates a gauge transformation with gauge group $G$ on all the fields in the theory. The gauge transformation is a function of  the scalar fields $\Phi_A=(\Phi_9,\Phi_0)$  of the $\cN=2$ vectormultiplet. The associated gauge parameter is given by
\beq
\Lambda \equiv  \Phi_A v^A\,,
\eeq
where 
\beq
v^A\equiv 2 \epsilon_Q \Gamma^{A} \epsilon_Q \qquad A=9,0\,.
\eeq
Explicit calculation using \rf{killloc} and Weyl transforming to $S^4$ using \rf{transform} gives\footnote{See below for more details.}
\beq
\Lambda= i  \Phi_0-{{x_3/r}\over 1+{|\vec x|^2\over 4r^2}} \Phi_9 \,.
\eeq
This implies that the gauge transformation parameter  at the north and south poles  of $S^4$ \newline --    which  are located  at $\vec x=(0,0,\pm 2r)$ -- are
\beq
\begin{aligned}
 \Lambda(N)&=i\Phi_0(N)-\Phi_9(N)\\
  \Lambda(S)&=i\Phi_0(S)+\Phi_9(S)\,.
  \label{actionpoles}
 \end{aligned}
\eeq
Therefore the
 gauge transformation acts differently at the north and south
 poles of the $S^4$,
%\VP{Need to swap north and south pole}
 which are the fixed points of the action of the $U(1)$ generator $J$. This observation will have far reaching consequences in our computation of the expectation value of 't Hooft operators in these theories. At the equator of $S^4$, on the other hand, we have that 
 \beq
 \Lambda(E)=i\Phi_0\,.
 \label{equagauge}
 \eeq

In summary, $Q^2$ acting on the bosonic  fields of the $\cN=2$ vectormultiplet -- whose localization equations we are after -- generates a $J+R$ and a $G$-gauge transformation     that  can be encoded in terms of the  vector field 
\beq
v^m=2\epsilon_Q\Gamma^m\epsilon_Q=\left(- {x_2\over r},  {x_1\over r}, 0, {1\over r}, -{x_3\over r} ,{i}  \left(1+{|\vec x|^2\over 4r^2}\right)\right) ~~~~~~m=1,2,3,4,9,0\,.
\label{vecfield}
\eeq
More explicitly, the action of $Q^2$ on these fields   is$^{\text{\scriptsize\ref{dimred}}}$
\beq
\begin{aligned}
Q^2\cdot A_\mu &=- [v^m D_m, D_\mu] \label{Q21}\\
Q^2\cdot\Phi_A &=- [v^m D_m,\Phi_A]\,.
\end{aligned}
\eeq
 Including the action on hypermultiplets, we conclude that  the action of $Q^2$ on all fields in the $\cN=2$ theory defined on $B_3\times S^1$   generates an $U(1)_{J+R}\times G\times G_{\text F}$ transformation.
 
 \subsection{Localization Equations in $B_3\times S^1$}
% \subsection{Symmetries and Localization Equations in $B_3\times S^1$}
 \label{sec:balla}

Given our choice of supercharge $Q$, we can now proceed to finding the saddle point equations  \rf{saddle} of the localization path integral
\begin{equation}
  Q \cdot \Psi = \frac 1 2 F_{mn} \Gamma^{mn} \epsilon_Q - 2 \Phi_{A} \tilde \Gamma^{A}
  \tilde \epsilon_Q + i K_{j} \Gamma^{8\,{j+4}} \epsilon_Q=0\,,
  \label{localll}
\end{equation}
where we have used that
\beq
\nabla_\mu \epsilon_Q=\tilde\Gamma_\mu \tilde \epsilon_Q\,,
\label{tildepinor}
\eeq
and $\epsilon_Q$ is given in \rf{killloc}. The equations can be found by projecting \rf{localll} 
on a basis of spinors generated by\footnote{Since the $\cN=2$ theory has eight supercharges, there are eight independent equations.}
\bea
&&\Gamma_{m} \epsilon_Q\qquad\quad m=1,2,3,4,9 \label{project1}\\
  &&  \overline{ \Gamma^{8\, j+4}\epsilon_Q}\qquad j=1,2,3 \label{project2}\,.
\eea
We note that the projection equations along $\Gamma_0 \epsilon_Q$ can be obtained   from a linear combination of  the projection  equations  along \rf{project1} since the conformal Killing spinor $\epsilon_Q$ satisfies the linear constraint
\beq 
v^m\Gamma_m \epsilon_Q=0\,,
\label{projectionvect}
\eeq
with $v_m$ given in \rf{vecfield}.

In order to develop intuition for the saddle point equations, we first study them in 
the point  $\vec x = 0, \tau = 0$ in $B_3\times S^1$.
  Projecting   \rf{localll}  along $\Gamma_{\hat m} \epsilon_Q$ yields
%\JG{Fixed equation}\VP{fine}
  \begin{equation}
  \begin{aligned}
 2 \epsilon_Q \Gamma_{\hat m} Q\cdot  \Psi =  ( F_{\hat m4} + i D_{\hat m} \Phi_0 )   
-   {1\over r}\Phi_9\, \delta_{\hat m3}=0
\qquad \hat m = 1 \dots 4, 9.
  \end{aligned}
  \label{invariance}
\end{equation}
These equations have a simple interpretation. They describe the $Q^2$-invariance equations of the bosonic fields in the $\cN=2$ vectormultiplet (obtained by setting equations  \rf{Q21}   to zero), which at $\vec x = \tau = 0$ are generated by the   vector field $v^m=(0,0,0,{1\over r},0,{i})$  (see equation \rf{vecfield}). 
This captures the combined action of  a $J$ and a $G$-gauge transformation  
 with vector field $v^\mu=(0,0,0,{1\over r})$   and     gauge parameter $\Lambda=i \Phi_0$ respectively. The invariance equation for the scalar field $\Phi_0$ is a linear combination of \rf{invariance}, a fact which follows from \rf{projectionvect}.

Projection of \rf{localll}  along $ \overline{ \Gamma^{8\,
    j+4}\epsilon_Q}$ gives     three dimensional  equations. They are
the Bogomolny equations
%\JG{Fixed equation}\VP{fine}
\begin{equation}
  \begin{aligned}
\label{eq:space}
 2 \overline {\epsilon_Q \Gamma_{{j+4}, 8}} Q \cdot \Psi =
  -D_{j} \Phi_9 + (*_3 F)_{j} + i K_{j} +   {i\over r} \delta_{j3} \Phi_0=0\qquad j =1,2,3\,.
  \end{aligned}
 \end{equation}

We can move to an arbitrary point $\tau\neq 0$ at $\vec x=0$ by acting on the equations  \rf{invariance} \rf{eq:space}  by the $U(1)_J$ transformation generated by $J$.  The invariance equations \rf{invariance} remain the same while the  three dimensional equations take the same form  \rf{eq:space} upon replacing the auxiliary scalar fields $K_i$ by rotated ones
\begin{equation}
  \begin{pmatrix}
  K_1 \\
  K_2 
  \end{pmatrix} 
\rightarrow 
\begin{pmatrix}
  \cos\tau & \sin\tau \\
  -\sin\tau & \cos\tau
\end{pmatrix}
\begin{pmatrix}
  K_1 \\
  K_2
\end{pmatrix}\,.
\end{equation}

We can now consider the general equations with $\vec x\neq 0$ and $\tau \neq 0$. 
 The $Q^2$-invariance equations, obtained by projecting \rf{localll} along $\Gamma_{m} \epsilon_Q$,
  are given by\footnote{As already mentioned, the invariance equation for $\Phi_0$ is a linear combinations of these equations.}
\begin{equation}
\label{eq:invariance}
  \begin{aligned}
   {1\over 2r}F_{14} + \left[D_1, {i\over 2}\left(1+{|\vec x|^2\over 4r^2}\right)\Phi_0- {x_3\over 2r} \Phi_9\right]+ {x_1\over 2r} F_{12} =0 \quad   [D_1, v^{m}D_{m}] =0\\
    {1\over 2r}F_{24} + \left[D_2, {i\over 2}\left(1+{|\vec x|^2\over 4r^2}\right)\Phi_0- {x_3\over 2r} \Phi_9\right]- {x_2\over 2r} F_{21} =0 \quad   [D_2, v^{m}D_{m}] =0\\
      {1\over 2r}F_{34} + \left[D_3, {i\over 2}\left(1+{|\vec x|^2\over 4r^2}\right)\Phi_0- {x_3\over 2r} \Phi_9\right]+ {x_1\over 2r} F_{32}- {x_2\over 2r} F_{31} =0 \quad   [D_3, v^{m}D_{m}] =0\\
       \left[D_4, {i\over 2}\left(1+{|\vec x|^2\over 4r^2}\right)\Phi_0- {x_3\over 2r} \Phi_9\right]+ {x_1\over 2r} F_{42}- {x_2\over 2r} F_{41} =0 \quad   [D_4, v^{m}D_{m}] =0\\
         {1\over 2r} [\Phi_9,D_{\tau}]+ \left[\Phi_9, {i\over 2}\left(1+{|\vec x|^2\over 4r^2}\right)\Phi_0\right]+ {x_1\over 2r} [\Phi_9,D_{2}]- {x_2\over 2r} [\Phi_9,D_{1}]  =0 \quad   [\Phi_9, v^{m}D_{m}] =0\,.
    %    2F_{1,\tau} + [D_1, i(1+x^2)\Phi_0] - 2[D_1, x_3 \Phi_9] + 2x_1
 %   F_{12} =0\quad\text{i.e.}\quad   [D_1, v^{m}D_{m}] =0 \\
% 2F_{2,\tau} + [D_2, i(1+x^2) \Phi_0] - 2 [D_2,x_3 \Phi_9] - 2 x_2
% F_{21} =0 \quad\text{i.e.}\quad [D_2, v^{m} D_{m}] =0 \\
%2F_{3,\tau} + [D_3, i(1+x^2) \Phi_0] - 2 [D_3, x_3 \Phi_9] + 2 F_{32} 
%x_1 - 2 F_{31} x_2 = 0\quad\text{i.e.}\quad  [D_3, v^m D_m]=0 \\
%[D_\tau, i(1+x)^2 \Phi_0] - 2 [D_{\tau}, x_3 \Phi_9] + 2 F_{\tau 2} x_1 
%- 2 F_{\tau 1} x_2 =0 \quad\text{i.e.}\quad [D_{\tau}, v^m D_m] = 0 \\
%2 [\Phi_9,D_{\tau}] +  [\Phi_9, i(1+x^2) \Phi_{0}] + 2 [\Phi_9,D_2] x_1 -
%2[\Phi_9,D_1] x_2  = 0\quad\text{i.e.}\quad [\Phi_9, v^m D_m] =0  
 \end{aligned}
\end{equation}

The three dimensional equations, which we call     deformed monopole equations,  are   obtained by projecting \rf{localll} along $ \overline{ \Gamma^{8\, j+4}\epsilon_Q}$. They are given by
\begin{multline}
\label{eq:sp1}
 -(4r^2+x_1^2-x_2^2 - x_3^2) [D_1 \Phi_9]  - 2 x_1 x_2 [D_2 \Phi_9]
- 2 x_1 x_3 [D_3 \Phi_9]
-4r x_2 [D_{\hat 4} \Phi_9]
 - 2 x_1 \Phi_9
 - 2 x_1 x_3 F_{12}+ \\
+ 2 x_1 x_2 F_{13}+ (4r^2- x_1^2 + x_2^2 + x_3^2) F_{2 3} - 4r x_3 F_{1\hat 4}
+ i (4r^2 + |\vec x|^2) K_1 + 4r x_1 F_{3\hat 4} =0\,.\\
%\\
%  -(1+x_1^2-x_2^2 - x_3^2) [D_1 \Phi_9] - 2 x_1 \Phi_9
% - 2 x_1 x_2 [D_2 \Phi_9]
%- 2 x_1 x_3 [D_3 \Phi_9]
%-2 x_2 [D_{\hat 4} \Phi_9]
%+ (1- x_1^2 + x_2^2 + x_3^2) F_{2 3} \\
%- 2 x_1 x_3 F_{12} + 2 x_1 x_2 F_{13} - 2 x_3 F_{1\hat 4}
%+ i (1 + x^2) K_1 + 2 x_1 F_{3\hat 4} =0,
\end{multline}

\begin{multline}
\label{eq:sp2}
 -(4r^2-x_1^2+x_2^2 - x_3^2) [D_2 \Phi_9]  - 2 x_1 x_2 [D_1 \Phi_9]
- 2 x_2 x_3 [D_3 \Phi_9]
+4r x_1 [D_{\hat 4} \Phi_9]
 - 2 x_2 \Phi_9
 - 2 x_2 x_3 F_{12}+ \\
- 2 x_1 x_2 F_{23}-(4r^2+ x_1^2 - x_2^2 + x_3^2) F_{1 3}- 4r x_3 F_{2\hat 4}  + i (4r^2 + |\vec x|^2) K_2 + 4r x_2 F_{3\hat 4} =0\,,
%\\
%   -2 x_1 x_2 [D_1 \Phi_9] + (-1 + x_1^2 - x_2^2 +x_3^2) [D_2 \Phi_9]
%- 2 x_2 x_3 [D_3,\Phi_9] + 2 x_1 [D_{\hat 4} \Phi_9] 
%-2 x_2 \Phi_9
%-2 x_1 x_2 F_{23} \\
%-2 x_2 x_3 F_{12}
%-(1+x_1^2-x_2^2 +x_3^2) F_{13} 
%-2 x_3 F_{24}
%+2 x_2 F_{34}
%+i (1+ x^2) K_2 =0
\end{multline}

\begin{multline}
\label{eq:sp3}
\hspace{-3mm}
 2 x_1 x_3  [D_1\Phi_9]
+ 2 x_2 x_3 [D_2 \Phi_9]
- (4r^2 +x_1^2+x_2^2-x_3^2) [D_3 \Phi_9] 
+ 2 x_3  \Phi_9 
+ 4 i r \Phi_0 
+(4r^2 -x_1^2-x_2^2+x_3^2)F_{12}+\\
+ 2 x_1 x_3  F_{23} 
 -2 x_2 x_3  F_{13} 
 -4r x_2 F_{2\hat 4} 
 -4r x_3 F_{3\hat4} 
 -4r x_1 F_{1\hat4} 
+ i \left(4r^2+ |\vec x|^2\right) K_3 =0\,.
\end{multline}
We note that the equations near the location of the 't Hooft loop -- at $\vec x=0$ -- reduce to the familiar Bogomolny equations in $\bR^3$, thus justifying their name. These equations are a supersymmetric extension of well known equations, which interpolate  between $F^+=0$ at the north pole, the Bogomolny equations at the equator and $F^-=0$ at the south pole.  This concludes our derivation of the saddle point equations of the 't Hooft loop path integral.

In appendix \ref{sec: soluns} we explicitly show that the background created by the insertion of a circular 't Hooft loop -- given in equations \rf{scalarpole} (and  \rf{dyonic} when $\theta\neq 0$) --  is a solution of  the localization equations derived in this section. This confirms that we can study the expectation value of    a supersymmetric circular 't Hooft loop operator in any $\cN=2$ gauge theory on $S^4$ by localizing the path integral with our choice of supercharge $Q$.

We can now anticipate   some key features in the evaluation of  the 't Hooft loop path integral of the  $\cN=2$ theory defined on $S^4$. As explained earlier, the fields and conformal Killing spinor  in $B_3\times S^1$ and $S^4$ are related by the Weyl transformation \rf{transform} with  $\Omega= \left(1+{|\vec x|^2\over 4r^2}\right)$.
We   note that the conformal Killing spinor  in $S^4$ -- which we denote by $\epsilon_Q^{\text{sphere}}$ -- has negative/positive four dimensional chirality at the fixed points of the $U(1)$ action of $J$, denoted as  north/south poles of $S^4$ respectively. In formulas\footnote{The volume form is given by $\epsilon^{\hat4\hat1\hat2\hat3}=1$.}
\beq
\begin{aligned}
\Gamma^{\hat 4\hat1\hat2\hat3}\epsilon^{\text{sphere}}_{Q}(N)&=-\epsilon^{\text{sphere}}_{Q}(N)\\
\Gamma^{\hat4\hat1\hat2\hat3}\epsilon^{\text{sphere}}_{Q}(S)&=+\epsilon^{\text{sphere}}_{Q}(S)\,,
\end{aligned}
\eeq
% \JG{I changed implies as it seems too strong. The fact that one gets to
%   leading order   Nekrasov's action in the omega background near N and S
%   pole seems more non-trivial than just saying that the spinor is
%   chiral. Is it not?}
%\VP{Fine}
and therefore instantons and anti-instantons are supersymmetric at the north and south poles of $S^4$ respectively.

Moreover, in the neighborhood of the north pole the $Q$-complex of the $\cN=2$  theory on $S^4$   generated by $\epsilon^{\text{sphere}}_Q$  reduces to the complex of the equivariant  Donaldson-Witten twist\footnote{Also known as  $\cN=2$ gauge theory in the $\Omega$-background \cite{Nekrasov:2002qd}.}   in $\bR^4$  \cite{Nekrasov:2002qd}, described by the instanton equations $F^{+} = 0$.  Likewise, in the neighborhood of the south pole the $Q$-complex of the $\cN=2$  theory on $S^4$  reduces to that of the equivariant  conjugate Donaldson-Witten twist in $\bR^4$, described by the anti-instanton equations $F^{-} = 0$. 
%This will be explicitly seen by analyzing the localization saddle point equations at the north and south poles (see equations \rf{eq:sp1}-\rf{eq:sp3}), where they  reduce to $F^+=0$ and $F^-=0$ respectively. 

 This implies that the path integral  for a  't Hooft loop  receives contributions from equivariant instantons at the north pole and equivariant anti-instantons at the south pole.
 These are singular solutions to the localization equations which must be included in the computation of the 't Hooft loop expectation value.   
 The  equivariant instanton/anti-instanton partition function in $\bR^4$ 
is captured by the so-called Nekrasov partition function \cite{Nekrasov:2002qd}, which will play a prominent role in our analysis. 

 From our expression for the action of $Q^2$ on the fields  (see \rf{vecfield}), we find that the $U(1)_{\ve_1}\times U(1)_{\ve_2}$ equivariant rotation parameters $(\ve_1,\ve_2)$  in Nekrasov's   partition function \cite{Nekrasov:2002qd}  at the north and south poles in $S^4$ are fixed  to
 \beq
 \ve_1=\ve_2=\ve={1\over r}\,,
 \eeq
since  $U(1)_{\ve_1}\times U(1)_{\ve_2}$ acts on $\bR^4$ as 
 \beq
 X_1+iX_2\rightarrow e^{i\ve_1}(X_1+iX_2)\qquad\qquad
X_3+iX_4\rightarrow e^{i\ve_2}(X_3+iX_4)\,.
 \eeq
Here $(X_1,\ldots,X_4)$ are the $S^4$ embedding coordinates   \rf{embedding} which parametrize the   local $\bR^4$ near the north and south poles.
As Nekrasov's partition function is for the $\cN=2$ topologically twisted theory in $\bR^4$ -- which mixes the $SU(2)$ Lorentz  with  $SU(2)$ R-symmetry  generators --
  the $U(1)_{J+R}$   symmetry   generated by $Q^2$ in the physical theory on $S^4$  gets identified at the north and south poles with  the 
  $(U(1)_{\ve_1}\times U(1)_{\ve_2})_{{\rm diag}}$
  symmetry   in Nekrasov's   partition function.

%We will be using notations
%\begin{equation}
%\ve_{+} =  \frac { \ve_1 + \ve_2}{2} \quad \ve_{-} = \frac{ \ve_1 - \ve_2}{2}.
%\end{equation}
%When dealing with $\ve_{-} = 0$ we also denote  $\ve = \ve_{+} = \ve_1 =
%\ve_2 = r^{-1}$.%
%\footnote{%
%\label{foot:ve+}%
%We warn the reader that in some literature the symbol $\ve$ refers to $2\ve_+$.
%} 

Moreover, it follows from equation \rf{vecfield} that the equivariant parameter $\hat a\in \mathfrak{t}$ for the action of constant $G$-gauge transformations   in $\bR^4$ in Nekrasov's partition function is fixed at the north and and south poles of the $S^4$ to\footnote{Here we note that the value of scalar fields at the north and south poles of $B_3\times S^1$ and $S^4$ are related by $\Phi_{S^4}= 2\Phi_{B_3\times S^1}$ through   Weyl rescaling, while at the equator $\Phi_{S^4}= \Phi_{B_3\times S^1}$.} 
\beq
\hat a(N)=i\Phi_0(N)-\Phi_9(N) \qquad  \hat a(S)=i\Phi_0(S)+\Phi_9(S)
\eeq
 % $\hat a(N)=i\Phi_0(N)-\Phi_9(N)$ and $\hat a(S)=i\Phi_0(S)+\Phi_9(S)$ 
 respectively.  Since the 't Hooft loop  induces a non-trivial  background for the scalar field $\Phi_9$ \rf{dyonic}, which  is non-vanishing at the north and south poles, the instanton/anti-instanton partition function contributions arising from the fixed points of $J$   explicitly depend on the magnetic weight $B$ labeling the 't Hooft operator.
We will return to the instanton and anti-instanton contributions to the 't Hooft loop path integral in section \ref{sec:inst}.

Likewise, there are singular solutions to the localization equations arising from the equator in $S^4$, where the 't Hooft loop is inserted. As we have shown, near  the equator we must consider solutions to the
Bogomolny equations in the presence of the singular monopole configuration created by the 't Hooft loop operator. We will consider the contribution of these singular
solutions to the saddle point equations in sections \ref{sec:one-loop-eq} and \ref{sec:screening}.

Our next task is to study the non-singular    solutions   of the localization equations.
 
\subsection{Completeness of Solutions
\label{sec:vanishing-theorem}}

In the evaluation of the 't Hooft loop path integral using localization we must sum over all the saddle points of the localization action $Q\cdot V$  which have a prescribed singularity,  induced by insertion of the 't Hooft operator.
 Therefore, we  wish to obtain the most general
solution  of the  localization equations (\ref{eq:invariance}-\ref{eq:sp3})
satisfying the appropriate boundary conditions imposed by the presence of the circular 't Hooft loop operator. The boundary condition  requires   that the solutions
to the localization equations approach the background
 (\ref{dyonic}) near the location of the 't Hooft loop, supported  at the equator of $S^4$.

 In this section we obtain the most general non-singular solution to these equations (besides the singularity due to the ' t Hooft operator). Singular solutions to the localization equations, however, will play a central role in our computations. We will discuss singular solutions supported at the north and south poles of the $S^4$  and their contribution to the expectation value of the 't Hooft loop operator in section \ref{sec:inst}, while the contribution of the singular solutions supported at  the equator will be analyzed in section
\ref{sec:screening}.

In appendix \ref{sec: soluns} we show that the field 
configuration
 \beq
\begin{aligned}
  F_{jk} &= -\frac B 2 \ep_{ijk} \frac{x_i}{|\vec x|^3}\,, \qquad F_{i\hat 4}=-ig^2\theta {B\over 16\pi^2} \frac{x_i}{|\vec x|^3}\,,\qquad   \Phi_9 = \frac{B}{2 |\vec x|}\,,\\
\qquad\qquad\quad \Phi_0 &= -g^2\theta {B\over 16\pi^2}{1\over  |\vec x|}+\frac{a}{1 + {|\vec x|^2\over 4r^2}}\,,\qquad K_3= -{a/r\over \left(1+ {|\vec x|^2\over 4r^2}\right)^2}\,,
\end{aligned}
  \label{dyonic-zero-mode}
\eeq
solves
the saddle point equations $Q\cdot \Psi=0$. This field configuration 
  is the   't Hooft loop background
(\ref{dyonic})
 deformed by a ``zeromode"\footnote{The corresponding field configuration  is annihilated by $D_\mu\left[\left(1+{|\vec x|^2\over 4r^2}\right)\Phi_0\right]=0$ since the background gauge field is abelian.} of $\Phi_0$, which is  labeled by $a$. The    auxiliary field $K_3$ in the $\cN=2$ vectormultiplet is also  turned on. Therefore, evaluation of the path integral requires integrating over the ``zeromode" $a\in \mathfrak{t}$, which takes values in the Cartan subalgebra $ \mathfrak{t}$
 of the gauge group $G$.

We will  now show that  the only solutions to $Q\cdot\Psi=0$ which are
smooth away from the loop 
are given by (\ref{dyonic-zero-mode}).
% smooth away from the location of
% the loop operator, is the background configuration (\ref{scalarpole}).%
% \footnote{%
% When the theta angle is non-zero, the background is
% given by
% (\ref{backgroundtotal}).
% }
% First we assume that the theta-angle is zero.
% \TO{If this is not the case, the reality condition of fields
% is modified.}
For this it   suffices to consider the deformed monopole equations, 
the differential equations
(\ref{eq:sp1}-\ref{eq:sp3}).
We find it more transparent, however, to take instead
a projection of the localization equations 
$Q\cdot \Psi=0$ along $\overline{\Gamma_{9\mu}\epsilon_Q}$.
This gives
\begin{eqnarray}
\hskip-20 pt 0 =
\overline{ \epsilon_Q \Gamma_{\mu 9}} Q\cdot \Psi
=-(*F)_{\mu\nu} v^\nu+\frac{i}{2r} D_\mu (x_3 \Phi_0)
- D_\mu
\left[\frac 1 2\left(1+\frac{|\vec x|^2}{4r^2}\right)
 \Phi_9
\right]
+i \sum_{j=1}^3 w_\mu^{(j)} K_j\,,
\label{eq:loc-eq-2}
\end{eqnarray}
% \TO{I'm taking into account the convention $\epsilon_{4123}>0$
% in taking the Hodge dual.}\JG{That's good, this is the one used throughout the text.}
where we have used $[\Phi_9,\Phi_0]=0$,
which follows from the imaginary part of the last equation
 in (\ref{eq:invariance}). 
We have also defined three real one-forms
$  w^{(j)}_\mu =\overline{\epsilon_Q \Gamma_{\mu 9}} \Gamma^{8\,j+4}\epsilon_Q$.

The field strength  $F=F^{(r)}+i F^{(i)}$ has real and imaginary parts.
The imaginary part  is due to the presence of the 't Hooft operator background
with $\theta\neq 0$, while the fluctuating part of the field that we integrate over in the path integral must be real. The imaginary part of equations \rf{eq:sp1} and \rf{eq:sp2} imply that
\beq
K_1=K_2=0\,,
\eeq
while the imaginary part of (\ref{eq:invariance}) requires that
\beq
\begin{aligned}
{1\over 2r} F^{(i)}_{j4}+\left[D_j, {i\over 2}\left(1+{|\vec x|^2\over 4r^2}\right)\Phi_0\right]&=0\qquad j=1,2,3\\
x_1F^{(i)}_{42}-x_2F^{(i)}_{41}&=0\,.
\end{aligned}
\eeq
Therefore, these equations completely determine $\Phi_0$ in terms of the electric field 
produced by the 't Hooft operator when $\theta\neq 0$ up to a zeromode, which we parametrize by $a$ in \rf{dyonic-zero-mode}. Moreover, the imaginary part of  \rf{eq:sp3} locks in the value of the auxiliary field $K_3$ in terms of the zeromode part of $\Phi_0$. Therefore, the most general solution to the localization equations for   the electric field $F_{j4}$  and the scalar fields $\Phi_0,K_1,K_2$ and $K_3$ is given in   \rf{dyonic-zero-mode}. Now it remains to show that the most general solution to the localization equations for the magnetic field $F_{ij}$ and the scalar field $\Phi_9$ is also given by \rf{dyonic-zero-mode}.

From the real part of (\ref{eq:loc-eq-2}) 
%into the real($=$anti-hermitian)
%and imaginary($=$hermitian) parts, 
we obtain
\begin{eqnarray}
i_v  *F^{(r)}
-D\left[
%\frac{1}{2}
\left(1+\frac{|\vec x|^2}{4r^2}\right)
 \Phi_9\right]
&=&0\,.
\label{eq:QPsi-real}
% \\
% -i_v *F^{(i)}+  D\left[\frac{x_3}{2r} \Phi_0\right]
% + \sum_{j=1}^3 w^{(j)} K_j\,
% &=&0\,,
\end{eqnarray}
%where $F=F^{(r)}+i F^{(i)}$.
% \TO{Check that the imaginary part leads to the known
% solutions for $K_j$ and $\Phi_0$.}
% \VP{yes, it is good}
% We have used $[\Phi_9,\Phi_0]=0$,
% which follows from the imaginary part of the last line
%  in (\ref{eq:invariance}). 
%\JG{Sorry,where is this being used?}\VP{In
%  \rf{eq:loc-eq-2}  Takuya dropped commutator terms $[\Phi_9, \Phi_0]$ because they are
%   set to zero by 3.39. But if you just explicitly expand $\overline{
%     \epsilon_Q \Gamma_{\mu 9}} Q\cdot \Psi$ you get $[\Phi_9,\Phi_0]$ terms in
%   RHS of\rf{eq:loc-eq-2}. I suggest to insert explanation right after (3.44) saying
%   $[\Phi_9,\Phi_0]$ is there but was dropped } 
We also note that the real part
of the $Q^2$-invariance equations (\ref{eq:invariance}) implies that
\beq
\label{eq:u1-invar}
 - i_v F+D(v^9\Phi_9)=0\,,\qquad \qquad
~~~~~~~~
i_v D \Phi_9=0\,.
\eeq
Let us define
a $1$-form
 $\tilde v= dx^\mu v_\mu/(v_\nu v^\nu)$ dual to the four-vector $v$, so that
 $i_v \tilde v = 1$. Now, in terms of the redefined   gauge field
\beq
\hat
 A=A+v^9\Phi_9\,\tilde v\,,
 \label{redefined}
\eeq
the  $Q^2$-invariance equations \rf{eq:u1-invar} imply that
\begin{eqnarray}
\label{eq:U1-hat-inv}
%  v^\nu \hat F_{\mu\nu}=0\,,
i_v \hat F=0\,,
\end{eqnarray}
where $\hat F=d\hat A+\hat A\wedge\hat A$.
Indeed, 
\begin{multline}
  i_v \hat F = i_v F + i_v D( v^9 \Phi_9\, \tilde v) = 
i_v F + i_v ( D(v^9 \Phi_9) \wedge \tilde v) + 
i_v (v^9 \Phi_9 D \tilde v) =\\
 = i_v F + (i_v D(v^{9} \Phi_9))\wedge \tilde v - D(v^9 \Phi_9) \wedge (i_v \tilde v) + v^9 \Phi_9\, i_v (D \tilde
v)\,,
\end{multline}
which using    $i_v \tilde v = 1$ and  the   equations in
 (\ref{eq:u1-invar}) makes the  first three terms vanish.
The last term vanishes for any
Riemannian metric invariant under the
action generated by the vector field $v$. In this situation $\CalL_{v} \tilde v = 0$ and since 
$\CalL_{v}  = D i_v + i_v D $ we have that indeed
$ i_v D \tilde v = - D(i_v \tilde v) = - D(1) = 0$. Therefore, the $Q^2$-invariance equations  (\ref{eq:U1-hat-inv})  
 reduce  the whole system of equations in $S^4$
to equations in the three-dimensional space $M_3=S^4/U(1)$, since $v$ generates the $U(1)_J$ spacetime transformation corresponding to $J$.

The scalar field in the $S^4$ conformal frame is
$\Phi=\Phi_9\left(1+\frac{|\vec x|^2}{4r^2}\right)$.
In  the $S^4$ metric\footnote{%
The orientation 
is such that the volume form is proportional to
$d\tau dx^1 dx^2 dx^3\propto d\vartheta  d\psi {\rm vol}(S^2) $.
}  (see appendix \ref{coords})
\begin{eqnarray}
 ds^2_{S^4}=
r^2d\vartheta^2+ \frac {r^2} 4\sin^2\vartheta d\Omega_2
+r^2\sin^2\vartheta (d\psi+\omega)^2
\end{eqnarray}
% rather than
%in the $B_3\times S^1$ metric.%
%In this frame 
equation (\ref{eq:QPsi-real})   reads 
\begin{eqnarray}
  i_v * F^{(r)}=
%\frac 1 2
 D \Phi\,.
  \label{eq:QPsi-realn}
\end{eqnarray}
In this metric, the  $1$-form $\tilde v$ is given by $\tilde v=r(d\psi+\omega)$, and the  
 redefined gauge field  \rf{redefined} is    
\begin{eqnarray}
  \hat A=A^{(r)}-\Phi r\cos\vartheta (d\psi+\omega)\,,
\end{eqnarray}
in terms of which equation (\ref{eq:QPsi-realn})  is
\begin{eqnarray}
  i_v *\hat F-
\frac{1}{\sin^2\vartheta} D(\Phi\sin^2\vartheta)=0\,.
\label{iv*Fhat}
\end{eqnarray}

Let us introduce a 1-form $\lambda$ and a function $h$
as quantities that appear in the background values
of $\hat A$ and $\Phi$ specified by (\ref{dyonic}):
\begin{eqnarray}
\hat A= -B \lambda\,,
\hspace{5mm}
\Phi
=\frac{B}{2 \sin^2\vartheta} h
\hspace{5mm}
 \text{(in the background)}\,.
\label{hatted-background}
\end{eqnarray}
Since the background solves the equation (\ref{iv*Fhat}),
$\lambda$ and $h$ satisfy the relation
\bea
0=i_v * d\lambda+\frac{dh}{2\sin^2\vartheta}\,.
\eea

In order to derive useful identities, we square the
left-hand side of the equation  (\ref{iv*Fhat})
and integrate it with an appropriate measure:
\bea
\begin{aligned}
0=&\int_{S^4}
\frac{1}{2h}
\left |\left|  
i_v * \hat F-
\frac{1}{\sin^2\vartheta} D ( \Phi\sin^2\vartheta)
\right |\right|^2
\\
=&
\int_{S^4}
\frac{1}{2h}
\left(
\left |\left|  
i_v *\hat F 
-\frac{\Phi}{ h} dh
\right |\right|^2
+
\left |\left|  
\frac{h}{\sin^2\vartheta} 
D\left(\frac{ \Phi \sin^2\vartheta}h\right)
\right |\right|^2\right)
\\
&
\hspace{10mm}
-\int_{S^4}
\frac{1}{\sin^2\vartheta}
\Tr\left[
 D\left(
\frac{\Phi\sin^2\vartheta}h\right)\wedge
*
\left( i_v *\hat F-\frac{\Phi}h dh\right)
\right]\,.
\end{aligned}
\eea
The measure is chosen so that the cross-term becomes
a total derivative:
\begin{eqnarray}
&&
-\frac{1 }{\sin^2\vartheta}
 D\left(\frac{\Phi\sin^2\vartheta}h\right)
\wedge
*
\left( i_v *\hat F-\frac{\Phi}h dh\right)
\nonumber\\
% &=&
%   -\frac{2 }{\sin^2\vartheta}
%  D(\Phi\sin^2\vartheta/h)\wedge
% \left(
% -v_\mu dx^\mu\wedge \hat F- \frac{\Phi}{h} 2\sin^2\vartheta v_\mu dx^\mu \wedge
% d\lambda
% \right)
% \nonumber\\
% &=&
%  D
% \left(
% \frac{\Phi\sin^2\vartheta}{h}
% \right)\wedge
% \tilde v\wedge
% \left(
%  \hat F
% + \frac{2 \Phi}h
% \sin^2\vartheta
% \,
% d\lambda
% \right)
% \nonumber\\
% &=&
% \frac{32}{r}
%  D\left[
% \frac{\Phi\sin^2\vartheta}{h}
%  \hat F
% +
% \left(\frac{\Phi\sin^2\vartheta}{h}\right)^2d\lambda
% \right]
% \wedge
% \tilde v
%\nonumber\\
&=&
 D\left[
\frac{\Phi\sin^2\vartheta}{h}
\left(
 \hat F
+
\left(\frac{\Phi\sin^2\vartheta}{h}\right)
d\lambda
\right)
\right]
\wedge
\tilde v
\nonumber\\
&=&
 D\left[
\frac{\Phi\sin^2\vartheta}{h}
\left(
 \hat F
+
\left(\frac{\Phi\sin^2\vartheta}{h}\right)
d\lambda
\right)
\wedge
\tilde v
\right]
-
\left[
\frac{\Phi\sin^2\vartheta}{h}
\left(
 \hat F
+
\left(\frac{\Phi\sin^2\vartheta}{h}\right)
d\lambda
\right)
\right]
\wedge
 d\tilde v\,.
\nonumber
\end{eqnarray}
The second term in the last line
 has to vanish because it is a $4$-form annihilated by $i_v$.
Thus the cross term is a total derivative, and the only potential
contribution to its integral
is from the equatorial $S^1$ where the 't Hooft loop
is inserted.
Let us consider a tubular neighborhood $S^1\times B_3$ of thickness $\delta>0$
and denote its boundary by $\Sigma_3$.
Then
%\begin{eqnarray}
\beq
\begin{aligned}
&  \int_{S^4}
\frac{1}{2h}
\left(
\left |\left|  
i_v *\hat F 
-\frac{\Phi}{ h} dh
\right |\right|^2
+
\left |\left|  
\frac{h}{\sin^2\vartheta} 
D\left( 
\frac{\Phi \sin^2\vartheta}h
\right)
\right |\right|^2
\right)
\\
% &=&-
% \int_{S^4}
% d\Tr\left[
% \frac{\Phi\sin^2\vartheta}{h}
% \left(
%  \hat F
% +
% \left(\frac{\Phi\sin^2\vartheta}{h}\right)
% d\lambda
% \right)
% \wedge
% \tilde v
% \right]
% \nonumber\\
&
\hspace{40mm}
=
\lim_{\delta\rightarrow 0}\int_{\Sigma_3}
\Tr\left[
\frac{\Phi\sin^2\vartheta}{h}
\left(
 \hat F
+
\left(\frac{\Phi\sin^2\vartheta}{h}\right)
d\lambda
\right)
\wedge
\tilde v
\right]\,.
\end{aligned}
\label{squares-comp}
\eeq
%\end{eqnarray}
Because the fields must obey the  boundary conditions associated 
to a 't Hooft operator at the equator of $S^4$,
their values on $\Sigma_3$ must approach the background values
(\ref{hatted-background}), for which the
integrand vanishes.
Thus the squares in the first line of (\ref{squares-comp})
must vanish separately, and 
 we have in particular
\begin{eqnarray}
% i_v *\hat F 
% -\frac{\Phi}{ h} dh
% =0\,,
% \quad
D\left(\frac{ \Phi \sin^2\vartheta}h\right)
=0\,.
\end{eqnarray}
The boundary condition near the operator
 then requires that $\Phi=B h/(2\sin^2\vartheta)$ up to a gauge transformation,
corresponding to the original 't Hooft operator background we started with.

In summary, the most general non-singular solution to the localization equations is the field configuration
\rf{dyonic-zero-mode}.

\section{Classical Contribution}
\label{sec:classical}

 In this section we calculate the classical contribution to the
 localization path integral computing the expectation value of a
 supersymmetric   't Hooft loop in an arbitrary $\cN=2$ gauge theory on
 $S^4$. 
  The classical contribution  to the   path integral is obtained by evaluating the $\cN=2$ gauge theory action on $S^4$ -- including suitable boundary terms -- on the saddle point solutions of the localization equations.

Using that the localization equations set to zero the scalar fields in the $\cN=2$ hypermultiplet, the classical contribution to the path integral arises from evaluating the bosonic action of the $\cN=2$ vectormultiplet on $S^4$ on the Weyl transformed \rf{transform} saddle point solution \rf{dyonic-zero-mode}.  The relevant part of the $\cN=2$ gauge theory action on $S^4$  of radius $r$  is given by
\begin{equation}
  \label{eq:YM-action}
  S_{\cN=2} =-\frac {1}{g^2} \int_{S^4} \sqrt{h}\Tr \left  ( \frac 1 2 F_{\mu\nu} F^{\mu\nu} + D_\mu \Phi_A D^\mu \Phi_A +
     \frac{R}{6} \Phi_A\Phi_A +  K_3^2\right)-{i\theta\over 8\pi^2} \int_{S^4}\Tr \left( F\wedge F\right)\,,
\end{equation}
where we have denoted by $h_{\mu\nu}$ the $S^4$ metric and $R=12/r^2$ is the  scalar curvature. The classical action   \rf{eq:YM-action} is invariant under the Weyl transformation  \rf{transform}. Therefore, we can calculate the classical contribution to the expectation value of the 't Hooft loop by computing the $\cN=2$ gauge theory action on $B_3\times S^1$ \rf{ballmetric} evaluated on the  background \rf{dyonic-zero-mode}.  The non-topological part of the action is thus\footnote{Where we have used  the $SO(3)\times SO(2)$ symmetry of the background  \rf{dyonic-zero-mode} and that the  $B_3\times S^1$  metric  \rf{ballmetric} has $R/6={1/\left(2r^2\left(1-{x^2\over 4r^2}\right)\right)}$. $x$ is the radial coordinate in $B_3$}
 \beq
 -\frac {2\pi\cdot 4\pi}{g^2} \int_{0}^{2r} dx\, r\left(1-{x^2\over 4r^2}\right)x^2\Tr\left[{1\over 2}F_{ij}F_{ij}+F_{i\hat 4}F_{i\hat 4}+D_i\Phi_AD_i\Phi_A+{\Phi_A\Phi_A\over   2r^2\left(1-{x^2\over 4r^2}\right)}  +K_3^2 \right]\,,
 \eeq
 while the topological term is\footnote{The volume form is given by $\epsilon^{\hat 4123}=1$.} 
 \beq
 {i\theta\over 8\pi^2}\cdot 2\pi\cdot 4\pi \int_{0}^{2r} dx\, r\left(1-{x^2\over 4r^2}\right)x^2\epsilon^{ijk}\Tr\left[F_{i\hat4}F_{jk}\right]\,.
 \eeq
 Explicit computation using the saddle point configuration \rf{dyonic-zero-mode} gives 
\beq
 S^{(0)}_{\cN=2}=-{8\pi^2\over g^2} r^2\Tr a^2 + \theta\, r  \Tr(a B) -\Tr B^2\left(\frac {4\pi^2 }{g^2}+{g^2\theta^2\over 16\pi^2}\right)r \int_{\delta}^{2r}dx\, {1\over x^2}\,.
 %\left[{1\over x^4}+{1\over x^4}+{1\over   2r^2 x^2\left(1-{x^2\over 4r^2}\right)}\right]
 \label{onshelly}
\eeq
The unregulated on-shell action is clearly divergent, as it measures the infinite self-energy of a point-like monopole. This  divergence -- which is proportional to the length of the curve on which the 't Hooft loop is supported -- can be regulated by introducing a cutoff $\delta$ in the integration over $x$, and subtracting terms in the action proportional to $1/\delta$. This subtraction can be implemented by adding to the action  
\rf{eq:YM-action}  covariant boundary terms supported on the $x=\delta$ hypersurface $\Sigma_3$. 

The relevant boundary terms are  
\beq
-{2\over g^2}\int_{\Sigma_3} \Tr\left(\Phi_9\, F\right)\wedge d\tau+i{2\over g^2}\int_{\Sigma_3} \Tr\left(\Phi_0 *_4\hskip-2pt F\right)\wedge d\tau\,.
\eeq
Evaluating them on the saddle point solution \rf{dyonic-zero-mode} we get
\beq
-{2\over g^2}\int_{\Sigma_3}  \Tr\left(\Phi_9\, F\right)\wedge d\tau=-{2\over g^2}  \Tr\left({B\over 2\delta} \cdot(-2\pi B)\right) 2\pi r={1\over \delta}\frac {4\pi^2 r }{g^2}\Tr B^2
\eeq
and
\beq
\begin{aligned}
i{2\over g^2}\int_{\Sigma_3}  &\Tr\left(\Phi_0 *_4\hskip-2pt F\right)\wedge d\tau=i{2\over g^2}  \Tr\left(\left(-g^2\theta {B\over 16\pi^2}{1\over  \delta}+a\right) \cdot\left(ig^2\theta {B\over 4\pi}\right)\right) 2\pi r=\\
 &={1\over \delta}{g^2\theta^2 r\over 16\pi^2}\Tr B^2-\theta\, r \Tr(aB)\,.
\end{aligned}
\label{finiteb}
\eeq
The terms proportional to $1/\delta$ in the boundary terms cancel the self-energy divergences in the bulk on-shell action in \rf{onshelly}. Moreover, the on-shell boundary term \rf{finiteb} generates a finite contribution,   which precisely cancels the corresponding one appearing in the bulk on-shell action \rf{onshelly}.
Therefore, the leading classical action for the  circular 't Hooft loop in the $\cN=2$ gauge theory is given by
\beq
 S^{(0)  \rm{total}}_{\cN=2}=-{8\pi^2\over g^2} r^2\Tr a^2   +\Tr B^2\left(\frac {2\pi^2 }{g^2}+{g^2\theta^2\over 32\pi^2}\right)\,.
 \label{clasicalon}
\eeq

The classical action \rf{clasicalon} can be split into the sum  of two terms, which are the complex conjugate of each other
\begin{eqnarray}
&&S^{(0) \rm{total}}_{\cN=2}= -\frac{1}{2} r^2\left[ \left(-\frac{8
   \pi ^2}{g^2}+i \theta
   \right) \Tr\left(ia-ig^2\theta {B\over 16\pi^2r}-\frac{B}{2
   r}\right)^2
\right.
\nonumber\\
&&
\hspace{28mm}
\left.+\left(-\frac{8 \pi
   ^2}{g^2}-i \theta \right)
    \Tr\left(ia-ig^2\theta {B\over 16\pi^2r}+\frac{B}{2 r}\right)^2\right].
    \label{calculetclasic}
\end{eqnarray}
This observation leads to an illuminating interpretation.  The classical result for the 't Hooft loop path integral on $S^4$ is captured by the classical contribution  to Nekrasov's equivariant instanton and  anti-instanton partition functions on $\bR^4$ \cite{Nekrasov:2002qd}
  localized at the north and south poles of the $S^4$ respectively. As we shall see, the classical, one-loop and instanton factors in Nekrasov's  equivariant instanton/anti-instanton partition function in $\bR^4$ \cite{Nekrasov:2002qd} will enter  in the computation of the  't Hooft loop on $S^4$.

We first recall that the classical contribution to the $\cN=2$ equivariant instanton partition function
 in $\bR^4$    -- or the partition function of the $\cN=2$ theory in the $\Omega$-background -- is given by \cite{Nekrasov:2002qd}
 \begin{equation}
\label{eq:class-contr}
 Z_{\text{cl}}(\hat a, q)=\exp\left[{1\over 2\ve_1\ve_2}  \,2\pi i   \tau\Tr  \hat a^2\right]\,.
 \end{equation}
The constant field  $\hat a \in \mathfrak{t}$ is  the equivariant parameter for the action of  $G$-gauge transformations on the moduli space of instantons in $\bR^4$, 
while  $\ve_1$ and $\ve_2$ are the equivariant parameters of the $U(1)_{\ve_1}\times U(1)_{\ve_2}$ action on $\bR^4=\bC\oplus \bC$
\beq
\begin{aligned}
z_1\rightarrow &\,e^{i\ve_1}z_1\\
z_2\rightarrow &\,e^{i\ve_2}z_2\,.
\end{aligned}
\eeq
 The parameter $q= \exp\left(2\pi i \tau\right)$ is the instanton fugacity   while $\bar q$ is the fugacity   for anti-instantons,  where $\tau$ 
is  the complexified coupling constant of the $\cN=2$ gauge theory 
\beq
\tau={\theta\over 2\pi}+{4\pi i \over g^2}\,.
\eeq
In section \ref{sec:balla} we have already mentioned that the supercharge $Q$ with which we localize the 't Hooft loop path integral becomes near the north and south poles of the $S^4$ the supercharge which localizes the equivariant instanton and anti-instanton partition function  in $\bR^4$  \cite{Nekrasov:2002qd} respectively, with the following equivariant parameters 
\beq
\begin{aligned}
 \ve_1&=\ve_2=\ve={1\over r}\\
\hat a(N) &=  i\Phi_0(N)-\Phi_9(N)\\
 \hat a(S)&=   i\Phi_0(S)+\Phi_9(S)\,.
%\qquad \ve_1&=\ve_2={1\over r}\,.
\label{equivvv}
\end {aligned} 
\eeq 
Therefore, inspection of the solution  of  the localization saddle point equations at the north and south poles\footnote{We have evaluated the scalar field $\Phi_0$ and $\Phi_9$   at the north and south poles of $S^4$. From equation \rf{dyonic-zero-mode} we find that the value of the field $\Phi_9$ ($\Phi_0$) at the north and south poles of $B_3\times S^1$, which are located at $\vec x=(0,0,\pm2r)$, is $\Phi_9={B\over 4r}$  ($\Phi_0=i{a\over 2}-ig^2\theta {B\over 32\pi^2r}$). Weyl transforming to $S^4$ using $\Phi_{S^4}= 2\Phi_{B_3\times S^1}$, we get the formula \rf{scalaratpoles}.} in \rf{dyonic-zero-mode} yields\footnote{In section \ref{sec:one-loop-eq}
we will need the value of $\hat a$ at the equator of $S^4$. It is given by $\hat a(E)=i\Phi_0(E)=ia-ig^2\theta {B\over 16\pi^2r}$.}
\beq
\hat a(N) =ia-ig^2\theta {B\over 16\pi^2r}-  {B\over 2r} \qquad \hat a(S) = ia-ig^2\theta {B\over 16\pi^2r}+{B\over 2r}\,.
\label{scalaratpoles}
\eeq
This implies that the classical  equivariant instanton/anti-instanton partition functions arising from the north and south poles are given by
\beq
Z_{\text{north},\text{cl}}=Z_{\text{cl}}(\hat a(N), q)\qquad\qquad Z_{\text{south},\text{cl}}=Z_{\text{cl}}(\hat a(S), \bar q)\,.
\label{classsic}
\eeq
Therefore,  the classical   expectation value \rf{calculetclasic} for the 't Hooft loop operator with   magnetic weight $B$ in any $\cN=2$ gauge theory on $S^4$  factorizes into a classical contribution associated to the north and south poles respectively
\beq
\exp\left(-S^{(0) \rm{total}}_{\cN=2}\right)= Z_{\text{north},\text{cl}}\cdot Z_{\text{south},\text{cl}}=
\left|Z_{\text{cl}}\left(ia-ig^2\theta {B\over 16\pi^2r}- {B\over 2r}, 
%\tau
q
\right)\right|^2\,,
\eeq
the south pole contribution being
 the complex conjugate of the north pole one
 \beq
  Z_{\text{south},\text{cl}}=\overline Z_{\text{north},\text{cl}}\,.
 \eeq
 
   The identification of the integrand of the 't Hooft loop path integral with contributions arising from  the north and south poles of $S^4$ will be a recurrent theme in our computation of the 't Hooft loop expectation value. As we shall see, however, an important contribution also arises from the equator of $S^4$.

\section{Instanton Contribution}
\label{sec:inst}

In the previous section we have calculated the classical contribution to the expectation value of a 't Hooft loop with magnetic weight $B$ on $S^4$ due to the non-singular solutions of 
the localization equations (besides the obvious singularity created by the insertion of the 't Hooft operator),  which are labeled by $a\in\mathfrak{t}$  \rf{dyonic-zero-mode}.
As discussed earlier, however, there exist  singular solutions to the localization equations supported at the north and south poles. In this section we determine their contribution to the 't Hooft loop expectation value.

The 
localization equations \rf{eq:sp1}-\rf{eq:sp3} at the north and south pole of the $S^4$ become, respectively,   the instanton and anti-instanton equations
\beq
\hbox{north}:\ \  F^+=0\qquad \qquad \hbox{south}:\ \  F^-=0\,.
\eeq
These equations describe singular field configurations, corresponding to point-like instantons, which are localized at the poles of $S^4$. The inclusion of these singular field configurations  in the localization computation implies that 
we must enrich the result in section \ref{sec:classical} with the  contribution of point-like instantons and anti-instantons arising at the north and south pole respectively.   We now  identify these contributions and include their effect in the computation of the 't Hooft loop path integral.

Nekrasov's equivariant instanton (anti-instanton) partition function in $\bR^4$ \cite{Nekrasov:2002qd} computes the contribution of instantons (anti-instantons) to the path integral of an  $\cN=2$ gauge theory in the so-called $\Omega$-background. We denote it by  \cite{Nekrasov:2002qd}
\beq
Z_{\text{inst}}(\hat a , \tilde m_f, \ve_1, \ve_2, q)\,,
\label{nekrinst}
\eeq
where $(\ve_1, \ve_2,\hat a,\tilde m)$ are the equivariant parameters for the $U(1)_{\ve_1}\times U(1)_{\ve_2}\times G\times G_{\text F}$ symmetries of the $\cN=2$ gauge theory.
$\tilde m_f$ with $f=1,\ldots, N_{\text{F}}$ denote the equivariant parameters for the flavour symmetry group $G_{\text F}$ associated to the hypermultiplet and $q$ is the instanton fugacity. 

Since the $\cN=2$ gauge theory action on $S^4$ and  $Q$-complex near the poles reduces to those of the  $\cN=2$ gauge theory in the $\Omega$-background, 
%and equivariant instanton (anti-instanton) complex respectively, 
the contribution of the singular field configurations in our localization computation due to point-like instantons  and anti-instantons at the north and south poles respectively, are precisely captured by Nekrasov's instanton and anti-instanton   partition function.

As we have already mentioned, the $Q$-complex of the $\cN=2$ theory near the north (south)  pole of $S^4$ reduces to that describing Nekrasov's equivariant instanton (anti-instanton)  partition function  on $\bR^4$ with $U(1)_{\ve_1}\times U(1)_{\ve_2}$ equivariant parameters
%However, the equivariant parameters $(\ve_1,\ve_2)$ for the $U(1)\times U(1)$ rotation action 
%\beq
%\begin{aligned}
%z_1&\rightarrow e^{i\ve_1}z_1\\
%z_2&\rightarrow e^{i\ve_2}z_2
%\end{aligned}
%\eeq
%in Nekrasov's instanton partition function are fixed to 
$\ve_1=\ve_2=1/r$. Furthermore, the equivariant parameter $\hat a\in {\mathfrak t}$ for the action of the gauge group $G$ on the instanton moduli space is given respectively by equations (\ref{equivvv}, \ref{scalaratpoles})
\beq
\begin{aligned}
  \hat a(N)&=i\Phi_0(N)-\Phi_9(N)=ia-ig^2\theta {B\over 16\pi^2r}-  {B\over 2r}\\
  \\
  \hat a(S)&=i\Phi_0(S)+\Phi_9(S)=ia-ig^2\theta {B\over 16\pi^2r}+ {B\over 2r} \,.
  \end{aligned}
\label{gauge-param-poles}
\eeq
%Now let 
%\beq
%Z_{\text{Nekr}}(\hat a , \hat m, \ve_1, \ve_2, \tau)\,,
%\eeq
%be Nekrasov's instanton partition function
%\cite{Nekrasov:2002qd}. Comparing the equivariant parameters for the
Therefore, the contribution to the 't Hooft loop expectation arising from   the solutions to the $F^+=0$ equations at the north pole is
given by
\begin{equation}
\label{eq:Nekr-inst}
  Z_{\text{north},\text{inst}} = Z_{\text{inst}}\left( ia - ig^2 \theta \frac{B}{16
    \pi^2 r} - \frac{B}{2r}, {1 \over r}+ i m_f, {1\over r}, {1\over r},q\right),
\end{equation}
while that due to the solutions of the $F^-=0$ equations at the south  pole is 
\begin{equation}
  Z_{\text{south},\text{inst}} = Z_{\text{inst}}
\left(
ia - i g^2 \theta \frac{B}{16
    \pi^2 r} +   \frac{B}{2r}, {1 \over r}+im_f, {1\over r}, {1\over r},\bar q
\right)\,.
\end{equation}
%where $m$ denote the    mass parameters for the flavour symmetries
%associated to   the hypermultiplet in  the $\cN=2$ gauge theory. 
% In particular our $\hat m$ is the equivariant parameter $m$ in  \cite{Nekrasov:2002qd} and our $m$ is
%the physical mass parameter $m_E$ in
%\cite{Pestun:2007rz}, with the relation between the two being\footnote{See
%discussion in \cite{Okuda:2010ke} on the difference in the conventions
%of the hypermultiplet mass in the literature.}
We have used the relation 
\begin{equation}
  \tilde m_f ={\ve_1 + \ve_2\over 2}+ i m_f \qquad \quad f=1,\ldots, N_{{\text F}}\,
  \label{massmap}
\end{equation}
  derived in \cite{Okuda:2010ke} between the physical mass $m_f$ of a hypermultiplet 
and the equivariant parameter $\tilde m_f$ in Nekrasov's instanton partition function. 

Taking into account the following identity obeyed by the instanton
partition function \cite{AGT,Okuda:2010ke}
\begin{equation}
\label{eq:symmetry}
  Z_{\text{inst}}(\hat a, \tilde m_f, \ve_1, \ve_2, q) =
  Z_{\text{inst}}(- \hat a, \ve_1 + \ve_2 - \tilde m_f, \ve_1, \ve_2, q)\,,
\end{equation}
we find that the anti-instanton south pole contribution is the complex conjugate of the one in the instanton north pole one
\begin{equation}
  Z_{\text{south},\text{inst}} = \bar Z_{\text{north},\text{inst}}\,.
\end{equation}

We can now combine the  results of this section with the ones found in the previous one and write down the  ``classical" contribution to the expectation value of a 't Hooft loop with magnetic weight $B$. Summing over all saddle points of the localization equations -- including both non-singular and singular solutions at the north and south poles -- which are labeled by $a\in \mathfrak{t}$,  leads to\footnote{We have trivially shifted the integration variable $ia\rightarrow ia +ig^2 \theta \frac{B}{16\pi^2 r}$.}
\begin{equation}
\label{eq:thooft-result}
 \vev{T(B)}\simeq \int da \left | Z_{\text{cl}}\left( ia - \frac{B}{2r},q\right) 
Z_{\text{inst}}\left( ia  - \frac{B}{2r}, {1\over r}+ i m, {1\over r},{1\over r},q\right)\right|^2,
\end{equation}
with $Z_{\text{cl}}$ and $Z_{\text{inst}}$ given in \rf{eq:class-contr} and \rf{nekrinst} respectively.

\section{One-Loop Determinants}\label{sec:one-loop}

The calculation of a  path integral using localization enjoys the drastic simplification of 
reducing the computation to one-loop order with respect to the deformation parameter $t$, while
being exact with respect to the original gauge theory coupling constant. In this section we calculate the relevant  determinants required for computing
  the expectation value of  't Hooft operators on $S^4$. 
Computation of the one-loop determinants in the  $\cN=2$ gauge fixed action is performed by   expanding to quadratic order in  all field fluctuations  -- which include vectomultiplet, hypermultiplet and ghost multiplet fields --  the deformation term  $\hat Q\cdot \hat V$ around  the saddle point configuration background \rf{dyonic-zero-mode}. In the gauge fixed theory,   the supercharge $Q$  combines with the BRST operator $Q_{BRST}$  as $\hat Q =Q+Q_{BRST}$, such that  the deformed action $Q\cdot V$ \rf{deformedaction} together with gauge fixing terms can be written as  $\hat Q\cdot \hat V$, with $\hat V=V+V_{\rm{ghost}}$  \cite{Pestun:2007rz}. As shown in \cite{Pestun:2007rz},  the saddle points of $\hat Q\cdot \hat V$ coincide with those of $Q\cdot V$, and we can borrow the  saddle point configuration in  \rf{dyonic-zero-mode} for the calculation of the determinants.

  Direct evaluation of the determinants by
diagonalization of the quadratic fluctuation operator in the saddle point background  is rather
complicated. Instead, we calculate the relevant one-loop determinants
using an index theorem. More precisely we use the Atiyah-Singer index
theorem for transversally elliptic operators \cite{MR0482866}, which was also used in \cite{Pestun:2007rz} to compute the partition function of   $\cN=2$ gauge theories on $S^4$.

Even though we are considering the physical $\cN=2$ gauge theory on $S^4$ (not a topologically twisted theory), the combined supersymmetry and BRST transformations generated by $\hat Q$ can be written in cohomological form \cite{Pestun:2007rz}. Fields of opposite statistics are paired into doublets under the action of $\hat Q$. Schematically, denoting the fields of even and odd statistics with a subindex $e$ and $o$ respectively, we have that
\beq
\begin{aligned}
\hat Q\cdot \varphi_{e,o}&=\hat \varphi_{o,e}\\
\hat Q\cdot \hat\varphi_{o,e}&= \cR\cdot\varphi_{e,o}\,.
\label{pairing}
\end{aligned}
\eeq
Here $\cR$ is the generator of the  $U(1)_{J+R}\times G\times G_{\text F}$  symmetries discussed in section \ref{sec:ball}, 
corresponding to the group  $U(1)_{J+R}$ combining
the $U(1)_J$ rotation on $S^4$  \rf{Jdeff} with an $SO(2)_R$ $R$-symmetry transformation,
the $G$-gauge and the $G_{\text F}$ flavour symmetries respectively. Therefore, $\hat Q$ acts
as an equivariant cohomological  operator since 
\beq
\hat Q^2\cdot \varphi_{e,o}=\cR\cdot\varphi_{e,o}\,,
\eeq
and $\hat Q^2$ is nilpotent   on $\cR$-invariant field configurations. The invariance of the deformation term $\hat Q\cdot \hat V$  under the action of $\hat Q$ and the pairing of of the fields   as in \rf{pairing} leads to cancellations between  bosonic and fermionic fluctuations. The remainder of this cancellation is the following ratio of determinants over non-zeromodes \cite{Pestun:2007rz}
\beq
%\sqrt{
{{{\rm det}_{{\rm Coker}D^{{\rm vm}}}\cR|_{o}}\over {\rm det}_{{\rm Ker}D^{{\rm vm}}}\cR|_e}\cdot {{{\rm det}_{{\rm Coker}D^{{\rm hm}}}\cR|_{o}}\over {\rm det}_{{\rm Ker}D^{{\rm hm}}}\cR|_e}
%}
\,.
\label{oneloopdets}
\eeq
%where $R$ is the generator of the $U(1)\times U(1)\times G\times G_{\text F}$ symmetries discussed in section \ref{sec:ball}, which are  generated by the action of $\hat Q^2$ on the fields of the theory.
 The differential operators $D^{{\rm vm}}$ and $D^{{\rm hm}}$ are obtained from the expansion of the deformation term $\hat Q\cdot \hat V$ for the vectormultiplet and hypermultiplet fields respectively.

Therefore, the one-loop determinants
that appear in the localization computation of  the partition function of an  $\cN=2$ gauge theory on $S^4$ are given by the product of weights for the group action
$\cR$ generated by $\hat Q^2$ on the vectormultiplet and hypermultiplet fields. Furthermore, the weights appearing in the determinants \rf{oneloopdets} can be determined from the computation of the $\cR$-equivariant index
\beq
{\rm ind}\,D={\rm tr}_{{\rm Ker}D} e^\cR-{\rm tr}_{{\rm Coker}D} e^\cR\,,
\label{indexcharacter}
\eeq
for   $D=D^{{\rm vm}}$ and $D=D^{{\rm hm}}$. 
In order to  convert the index
(Chern character)
$\text{ind}\,D$ in \rf{indexcharacter}
into a   fluctuation determinant (Euler character),
we read off the weights $w_\alpha(\ve_1,\ve_2,\hat a,m_f)$  from the index and combine them to get the determinant 
according to the rule
\begin{eqnarray}
\label{rule-appA}
\sum_j c_j e^{w_j(\ve_1,\ve_2,\hat a,\hat m_f)}
\rightarrow \prod_j w_j(\ve_1,\ve_2,\hat a,\hat m_f)^{c_j}\,,
\end{eqnarray}
where $(\ve_1,\ve_2,\hat a,m_f)$ denote the equivariant parameters for $U(1)_{\ve_1}\times U(1)_{\ve_2}\times G\times G_{\text F}$.\footnote{We recall that
$U(1)_{J+R}= (U(1)_{\ve_1}\times U(1)_{\ve_2})_{{\rm diag}}$.}
The relevant $\cR$-equivariant indices can then be calculated from
the equivariant Atiyah-Singer index theorem for transversally elliptic operators \cite{MR0482866}, to which we now turn.

The index theorem  localizes contributions to the fixed points of the action of $\cR$, that is to the north and south poles of $S^4$.  Therefore, the relevant index 
corresponds to the equivariant index of the vectormultiplet and hypermultiplet complexes of the $\cN=2$ theory in the $\Omega$-background, to which the $\cN=2$ gauge theory on $S^4$ reduces at the poles.
The presence
of a   't Hooft loop, however,  introduces a further contribution, arising from the equator, where the operator is supported.

\subsection{Review of the Atiyah-Singer Equivariant Index Theory}
\label{sec:review-index}

Consider a pair of vector bundles  $(E_0, E_1)$  over a manifold $M$.
Let  $V_i = \Gamma(E_i)$ be the space of sections of $E_i$, $i = 1,2$.

Let $T=U(1)^n$ be the maximal torus of
a compact Lie group  $\cG$  acting on $M$ and the bundles $E_i$,
and let  $D: V_0 \to V_1$ be an elliptic differential operator commuting
with the $\cG$-action.
 In this situation we can
define the $\cG$-equivariant index of the operator $D$ as a formal character
\begin{equation}
  \ind D (t) = \tr_{H^0} t - \tr_{H^1} t \quad t=(t_1,t_2,\ldots, t_n) \in T\,,
  \label{indice}
\end{equation}
where $H^0 = \ker D, H^1 = \coker D$. If  $D$ is elliptic
 and $M$ is compact, $H^0$ and $H^1$ 
 are finite dimensional vector spaces.

 The index does not depend on small deformations of
 the operator $D$ and, therefore, is a topological invariant.
 If the action of $\cG$ on $M$ has a discrete set of fixed points,
 Atiyah and Singer 
represent the index as a sum over the set of fixed points $F$ 
\begin{equation}
  \ind D(t) = \sum_{p \in F} \frac { \tr_{E_0(p)} t - \tr_{E_{1}(p)} t} 
{ \det_{TM_p}(1 - t)} \,.
\label{ASindice}
\end{equation}

Each fixed point contribution to the Atiyah-Singer index formula \rf{ASindice} 
is a rational function in $t$.
  For  an elliptic operator $D$ on  a compact manifold $M$ 
 the  sum over all of the fixed point contributions to the index is 
a finite  Laurent polynomial in $t=(t_1,\ldots, t_n)$,
% and $t^{-1}$, 
since the spaces $H^i$ are finite dimensional.

The basic example is the equivariant index of the Dolbeault operator
$\bar \p: \Omega^{0,0}(\BC) \to \Omega^{0,1}(\BC) $ from the space of
functions to the space of $(0,1)$-forms on the complex plane $M=\BC$
 under the $T = \U(1)$ action  $ z \mapsto t z$. Computing the index of $\ind(\bar \p)(t)$
directly using \rf{indice}   we just need to evaluate the $\U(1)$ character 
on the space of holomorphic functions  
\begin{equation}
  f(z) = \sum_{k\geq 0} c_k z^k\,,
\end{equation}
since $\coker \bar\p$  is trivial in $\BC$.
Under the $U(1)$ action the functions transform as
$\tilde f(\tilde z) = f(z)$ for $\tilde z = t z$, that is $\tilde
f(z) = f( t^{-1} z)$. Hence $c_k \mapsto \tilde c_k = t^{-k} c_k$.
Therefore
\begin{equation}
  \ind (\bar \p)(t)  = \sum_{n=0}^{\infty} t^{-n} = \frac {1}{1 - t^{-1}},
\label{ind-Dolb}
\end{equation}
where the last equality should be understood formally since for $|t| =
1$ the series does not actually converge. 

On the other hand, we can evaluate $\ind (\bar \p)(t)$   using the Atiyah-Singer fixed point theorem \rf{ASindice}.
Since there is a single fixed point at $z=0$ of the $U(1)$ action, we
get\footnote{
The fiber  $(E_0)_{z=0}$  transforms trivially, the fiber
 $(E_1)_{z=0}$  transforms as 
$\tilde f_{\bar z} =
f_{\bar z} (d \bar z /d  \tilde {\bar z}) = f_{\bar z} \bar t^{-1} =
f_{\bar z}  t$ for $|t|^2 =1$, hence the numerator in the Atiyah-Singer
theorem is $(1-t)$. The denominator is  $\det_{TM_p}(1 - t) =
(1-t)(1-t^{-1})$ as $(z,\bar z) \mapsto (t z, t^{-1} z)$. 
}
\begin{equation}
  \ind (\bar \p)(t) = \frac{ 1- t}{ (1 - t)(1 - t^{-1})} =
  \frac{1}{1 - t^{-1}}\,,
\end{equation}
thus reproducing the previous computation.

The index theory for elliptic operators can be generalized to
transversally elliptic operators \cite{MR0482866}. 
Let $T$ be the maximal torus of a Lie group $\cG$  that acts on a manifold $M$.
An operator $D$ on $M$  is called transversally elliptic with
respect to the $\cG$ action on $M$
%, where $T$ is a Lie group,  
if it is elliptic in all
directions transversal to the $\cG$-orbits on $M$.
As in the elliptic case, the index of $D$ 
possesses the excision property.
Therefore the index can be computed
as a sum of local contributions, a sum over the fixed points of the $\cG$ action.
% \JG{need to supply
%   reference. Should we add a sentence defining transversally elliptic? I
%   have to confess that before Vasily's paper I had never heard of this
%   notion :(~}. 
% For the application of the index theorem to the
% localization calculation on $S^4$,  we take $T=U(1)$.
%In the transversally elliptic case t
The total index $ \ind D (t) $ is an
infinite formal  Laurent
series $\sum_{n} c_{n} t^{n}$ with $n \in \BZ$, since the cohomology spaces $H^i$
can be infinite dimensional. However, for  each $c_n$, 
the multiplicity of the representation $n$
 in  $ \oplus (-1)^{i} H^{i}$, is finite.
Atiyah-Singer theory allows us to find $c_{n}$ unambiguously 
since the theory specifies
whether each fixed point contribution is
to be expanded in   powers of $t$ 
 or $t^{-1}$,
after  choosing a deformation of the symbol for $D$.%
\footnote{%
After summing over fixed points, $c_n$ is independent
of the choice of deformation.
}

 In the paper \cite{Pestun:2007rz}, the partition function
and the Wilson loop expectation value
were computed,
with the one-loop contributions evaluated using an index theorem.
In the set-up of  \cite{Pestun:2007rz} and the current paper, the manifold is $M=S^4$ and the spacetime part of the relevant group $\cG=U(1)_{J+R}\times G\times G_{\text F}$ is generated by $J$ \rf{Jdeff}.
The differential operators that appear in the quadratic part of $\hat Q\cdot \hat V$
fail to be elliptic on the equatorial $S^3$, 
but they are still transversally elliptic and the generalized
index theorem can be applied.
In \cite{Pestun:2007rz} the index is a sum of local contributions from the north and south poles of $S^4$,
which are the fixed points of $J$.

When we turn on  the singular monopole background
(\ref{dyonic-zero-mode}), there is an extra complication since
some of the fields are singular along the equatorial $S^1$
where the loop operator is inserted. This gives rise to an
extra contribution to the one-loop determinant, associated with the equator of $S^4$.
We believe that the index theorem for transversally
elliptic operators can be generalized to the situation 
where such singular monopoles are present. 
A similar index theorem was established in \cite{MR1624279}
using a relation between singular monopoles and $U(1)$-invariant
instantons \cite{Kronheimer:MTh}.
Assuming the existence of such an index theorem,
we will compute local contributions from the equatorial $S^1$,
for which there is a natural expansion.
The specific choice of a deformation of the symbol
made in \cite{Pestun:2007rz} led to the 
expansion in positive and negative powers of $t$ at the north
and south poles, respectively.
In the presence of a 't Hooft loop we will apply the same deformation,
and therefore obtain the same rules for expansion at the north and south poles.

\subsection{North and South Pole Contributions}

%\subsubsection{Vectormultiplet Determinant}
%\label{sec:1-loop-vect}

We wish to compute the vectormultiplet and hypermultiplet one-loop determinant contributions
from the north and south poles, which are  the fixed point set of $J$.
The relevant complex for the vectormultiplet calculation is the self-dual complex while for the hypermultiplet
 it is the Dirac complex.
%We recall that  the $\hat Q$ complex near the north/south pole reduces to the
%equivariant self-dual/anti-self-dual  complex  in $\bR^4$ with equivariant parameters $\ve_1=\ve_2=1/r$.
We now consider the associated equivariant indices and one-loop determinants.

\subsubsection*{Vectormultiplet Determinant}

%Let us begin by reviewing the one-loop calculation in \cite{Pestun:2007rz}.
Near the north pole,  we consider the complex\footnote{The complex \rf{eq:self-dual} can be turned into
the two-term complex in \rf{indice} by ``folding" the complex as \newline
$ D_\text{SD}: \Omega^{1} \stackrel{d^* \oplus d_{+}}{\longrightarrow} \Omega^{0}\oplus\Omega^{2+}$,
where $d^*$ is the   conjugate of $d$.}
  of vector bundles
associated with linearization of the anti-self-dual
equation $F^{+} = 0$ on $\BR^{4}$
\begin{equation}
\label{eq:self-dual}
 D_\text{SD}:    \Omega^{0} \stackrel{d}{\to} \Omega^{1} \stackrel{d_{+}}{\to} \Omega^{2+}\,,
\end{equation}
where $d$ is the de Rham differential and $d_{+}$ is the composition of
the de Rham differential and self-dual projection operator.
We want to compute the equivariant index of $D_\text{SD}$
with respect the $T=\U(1)_{\ve_1}\times \U(1)_{\ve_2}$ action which
rotates $\BR^{4} = \BC \oplus \BC$ as $(z_1, z_2) \mapsto (t_1 z_1, t_2
z_2)$.
For the moment we take $t_1$ and $t_2$ generic though we will set
$t_1=t_2$ in the end, as  $U(1)_{J+R}$   corresponds  to $(U(1)_{\ve_1}\times U(1)_{\ve_2})_{{\rm diag}}$ in the self-dual/anti-self-dual  complex at the north/south pole.
The Atiyah-Singer formula \rf{ASindice} for the complexification
of \rf{eq:self-dual} gives\footnote{The weights of the $\U(1)^2$ action are: $\{(0,0)\}$ for $\Omega^{0}$, $\{(\pm1,0), (0,\pm1)\}$ for $\Omega^{1}$, and $\{(0,0), (1,1), 
(-1,-1)\}$ for $ \Omega^{2+}$.}
\begin{eqnarray}
  \ind (D_{\text{SD},\BC})(t_1, t_2) 
&=& \frac { (t_1 t_2 + t_1^{-1} t_2^{-1} +2 ) -
    (t_1+t_1^{-1} + t_2 + t_2^{-1})}{ (1-t_1)(1-t_1^{-1})
    (1-t_2)(1-t_2^{-1})} 
\nonumber\\
% = \\
%  = \frac { (1-t_1)(1-t_2) + (1-t_2^{-1})(1-t_1^{-1}) }{  (1-t_1)(1-t_1^{-1})
%     (1-t_2)(1-t_2^{-1})}  
&=& \frac {  1+ t_1 t_2 } { (1 -t_1)(1-t_2)}\,.
\label{indexvect}
\end{eqnarray}
% \begin{multline}
%   \ind (D_{\BC})(t_1, t_2) = \frac { (t_1 t_2 + t_1^{-1} t_2^{-1} +2 ) -
%     (t_1+t_1^{-1} + t_2 + t_2^{-1})}{ (1-t_1)(1-t_1^{-1})
%     (1-t_2)(1-t_2^{-1})} = \\
%  = \frac { (1-t_1)(1-t_2) + (1-t_2^{-1})(1-t_1^{-1}) }{  (1-t_1)(1-t_1^{-1})
%     (1-t_2)(1-t_2^{-1})}  = \frac {  1+ t_1 t_2 } { (1 -t_1)(1-t_2)}
% \end{multline}
The index for the real complex (\ref{eq:self-dual})
is the half of (\ref{indexvect}).

Unless
there is a further input from
the transversally elliptic Atiyah-Singer theory, we can expand the function \rf{indexvect} in various ways depending on whether we take $|t_i|>1$ or $|t_i|<1$.  
For example, expanding in positive powers of $t_1, t_2$ we get
\begin{equation}
\label{eq:positive}
  \frac{1 + t_1 t_2}{ (1-t_1)(1-t_2)} = \sum_{n_1,n_2 \geq 0} (1+t_1t_2)
  t_1^{n} t_{2}^n\,,
\end{equation}
while expanding in negative powers of $t_1, t_2$ we   get
\begin{equation}
  \frac{1 + t_1 t_2}{ (1-t_1)(1-t_2)} = \sum_{n_1,n_2 \geq 0}
  (1+t_1^{-1} t_2^{-1})
  t_1^{-n} t_{2}^{-n}\,,
\end{equation}
and there are several other available expansions as well.

In order to calculate the one-loop determinant for the $\cN=2$ vectormultiplet, we must consider the self-dual complex (\ref{eq:self-dual}) tensored with 
the adjoint
representation  of the gauge  group $G$, and study  the $U(1)_{\ve_1}\times U(1)_{\ve_2} \times
G\times G_{\text F}$-equivariant index for such a complex (see  \rf{pairing}). It is given by\footnote{Recall that $G_{\text F}$ acts trivially on vectormultiplet fields.}  
\begin{equation}
\label{eq:ind2}
  \ind (D^\vm) =
 \frac{ (1 + t_1 t_2)}{2 (1-t_1)(1-t_2)} \chi_{\text{adj}}(g), \quad\qquad
    g \in G \,,
\end{equation}
where $\chi_{\text{adj}}(g)$ is the character of $G$ in the adjoint representation.
More explicitly, let us denote   $t_1 = \exp ( i \ve_1)$,   $t_2 = \exp ( i \ve_2) $  and   $g = \exp( i\hat a)$, 
where $\ve_1, \ve_2$ and $\hat a$ are the elements of the Lie algebra of $\U(1)_{\ve_1} \times
\U(1)_{\ve_2}$ and of the Cartan subalgebra $\mathfrak{t}$  of $G$ respectively. Denoting by $w$ be the weights of
the adjoint representation of $G$, the index (\ref{eq:ind2}) can be written as
\begin{equation}
  \ind (D^\vm)(\ve_1, \ve_2, \hat a) =
 \frac{ (1 + e^{ i \ve_1 + i \ve_2})}{ 2(1-e^{i \ve_1}
    )(1-e^{i \ve_2})} \sum_{w\in\text{adj}} e^{ i w\cdot \hat a }\,.
    \label{newindex}
\end{equation}

% \JG{I think it is better to remove usually, which is a
%   bit wishy-washy. If there is a counterexample to this statement, we
%    can mention it in a footnote.}
% \VP{Ok, I've put ``the'' instead of ``a'' and removed
%   ``usually''. We can be concrete.}
As mentioned earlier, the one-loop determinant in the  localization computation 
of the 't Hooft loop path integral  
% \JG{ why only usually?} \VP{I don't know
%   rigorous theorem and a reference, but this was true in the localization problems I
%   have learned } 
can be computed as   the  product over all  the weights of the generator $\cR$ of the $U(1)_{J+R}\times G\times G_{\text F} $ action on the space of
fields (see \rf{pairing}). Mathematically, the product of weights computes the 
equivariant Euler
class of the normal bundle to the fixed point set.
 The corresponding index   or equivariant Chern character    determines
the one-loop determinant  or equivariant Euler character by
taking the weighted product of all weights extracted
from the exponents in the Chern character (using \rf{rule-appA}). Therefore, we will calculate the one-loop
determinant of the $\cN=2$ vectormultiplet by determining the weights under the action of $U(1)_{J+R}\times G\times G_{\text F}$ from the index \rf{newindex}.
We remove the terms with $w=0$ because they are independent of $\hat a$, so that we are only left with the sum over the roots $\alpha$  of $\mathfrak{g}$.

Let us now consider the north pole contribution 
 to the index and the associated one-loop determinant for the vectormultiplet.
%an illustrative example of how to compute the one-loop determinant from the index. 
% It sufficed to consider a weight of the adjoint representation labeled 
% by a root $\alpha$ since other weights only lead to $a$-independent factors. 
As we mentioned earlier, the deformation of the
symbol requires that the index
   in \rf{newindex}
be defined  by taking the   positive expansion for the $\U(1)_{\ve_1}\times \U(1)_{\ve_2}$ weights as in (\ref{eq:positive}).
This uniquely determines the weights under the action of $\U(1)_{\ve_1}\times \U(1)_{\ve_2}\times G\times G_{\text F}$ to be
\beq
\begin{aligned}
n_1  \ve_1 +
  n_2   \ve_2  + 
\alpha \cdot\hat a&\qquad \hbox{for}\qquad n_1,n_2\geq 0\,, \\
  (n_1 + 1)\ve_1 + (n_2 + 1) \ve_2 +
\alpha  \cdot \hat a&\qquad \hbox{for}\qquad n_1,n_2\geq 0
  \end{aligned}
\eeq
with multiplicities $1/2$.
The one-loop determinant contribution from the north pole of the $\cN=2$ vectormultiplet labeled 
by 
a root $\alpha$   of the Lie algebra $\mathfrak{g}$  is therefore
\newcommand{\Eu}{\mathrm{Eu}}
\begin{equation}
\prod_{n_1, n_2 \geq 0}  \left[ n_1  \ve_1 +
  n_2   \ve_2  +  
\alpha 
\cdot\hat a\right]^{1/2}
\left[ (n_1 + 1)\ve_1 + (n_2 + 1) \ve_2 +
  \alpha
 \cdot \hat a\right]^{1/2}\,.
\end{equation}

In our localization calculation on $S^4$, we must specialize to
the values $\ve_1=\ve_2=\ve=1/r$, which correspond to the $U(1)_{J+R}$ symmetry.
% As shown in \cite{Pestun:2007rz},
% the ghost zero-modes contribute a cancelling factor $[\alpha\cdot\hat a]^{-1/2}$.
% Thus the one-loop factor reads
% \begin{eqnarray}
% && Z_{\text{1-loop},\text{vec},N,\alpha, +} %(\ep_s, a,N) 
% \nonumber\\
% &\sim &
% [\alpha\cdot\hat a(N)]^{-1/2}
% \prod_{n_1, n_2 \geq 0}  \left[ n_1   +
%   n_2     +  \alpha\cdot\hat a(N)\right]^{1/2}
% \left[ (n_1 + 1) + (n_2 + 1) +
%   \alpha\cdot \hat a(N)\right]^{1/2}
% \nonumber\\
% &=&
% \prod_{n\geq 1}\left[n+\alpha\cdot\hat a(N)\right]^{n}\,.
% \end{eqnarray}
The expression is divergent, and
we regularize it by
identifying it with the Barnes G-function \cite{Ba} (see for example section 
5.17 in \cite{MR2723248}).
It is an analytic function that has 
a zero of
order $n$ at $x = -n$ for all integers $n > 0$, and can be defined 
by the infinite product formula  
\begin{equation}
  G(1+z) = (2\pi)^{z/2} e^{-((1+ \gamma )z^2 +z)/2} \prod_{n=1}^{\infty}  \left(1 + \frac z n \right)^n e^{-z + 
\genfrac{}{}{}{1}{ z^2}{2n}}\,.
\label{Barnesdef}
\end{equation}
Therefore,  the corresponding vectormultiplet   one-loop determinant   is given by\footnote{\label{asymptotically} For  asymptotically free gauge theories see discussion after equation \rf{loopcombination}.}

\begin{eqnarray}
Z^\vm_{\text{1-loop},\text{pole}} (\hat a)
=\prod_{\alpha}G^{1/2}\left(
\frac{\alpha \cdot\hat a }{\ve}
\right)
G^{1/2}\left(2+
\frac{
\alpha \cdot\hat a}{\ve}
\right)
\,.
\label{1-loop-pole}
\end{eqnarray}

At the other fixed point -- at the south pole -- we need to consider the anti-self-dual complex
and an expansion in negative powers of $t_1$ and $t_2$. However, the index of the anti-self-dual complex
at the south pole coincides with the index of the self-dual complex
at the north pole. Relative to the north pole, the difference amounts to the sign change
$(\ve_1, \ve_2) \rightarrow (- \ve_1,-\ve_2)$,
which can be absorbed into the redefinition of roots
 $\alpha \rightarrow -\alpha$, which just exchanges positive and negative roots, yielding once again \rf{1-loop-pole}.

Therefore, recalling that the equivariant parameters for the $G$-action at the north and south poles are fixed (\ref{equivvv}, \ref{scalaratpoles})
\beq
\hat a(N) =ia-ig^2\theta {B\over 16\pi^2r}-  {B\over 2r} \qquad \hat a(S) = ia-ig^2\theta {B\over 16\pi^2r}+{B\over 2r}\,,
\label{scalaratpolesn}
\eeq
 we obtain that the vectormultiplet one-loop determinant contributions from the north and south poles are
\beq
Z^\vm_{\text{north},\text{1-loop}}=Z^\vm_{\text{1-loop},\text{pole}} (\hat a(N))\qquad\qquad Z^\vm_{\text{south},\text{1-loop}}=Z^\vm_{\text{1-loop},\text{pole}} (\hat a(S))
\,,
\eeq
with $Z^\vm_\text{1-loop,pole} (\hat a)$ given in (\ref{1-loop-pole}).
% \begin{eqnarray}
% Z^\vm_{\text{1-loop,S}} (\hat a)
% =
% \prod_{\alpha:\text{root}}
% G\left(
% \frac{\alpha \cdot\hat a }{\ve}
% \right)^{1/2}
% G\left(2+
% \frac{
% \alpha \cdot\hat a}{\ve}
% \right)^{1/2}
% \,.
% \label{1-loop-south}
% \end{eqnarray}
Furthermore, the south pole contribution is the complex conjugate of the north pole 
\beq
Z^\vm_{\text{south},\text{1-loop}}=\overline{Z}^\vm_{\text{north},\text{1-loop}}\,,
\eeq
precisely the same relation that we found earlier for the classical and instanton contributions.

Let us now compare these results with the computation in  \cite{Pestun:2007rz}. In the absence of a 't Hooft loop we have $\hat a(N) =\hat a(S) =ia$,
% and the product of (\ref{1-loop-north})
% and 
% (\ref{1-loop-south})
and
\beq
|Z^\vm_{\text{1-loop,pole}} (\hat a)|^2
\eeq 
 is precisely
the one-loop determinant for the vectormultiplet
 obtained in \cite{Pestun:2007rz},
up to the ghosts-for-ghosts contributions.
The ghosts-for-ghosts were introduced to gauge-fix
the constant gauge transformations on $S^4$,
and they had the effect of removing the Vandermonde
 $\prod_{\alpha >0} \alpha\cdot \hat a$
from the one-loop factor, while the square of the Vandermonde
reappeared as the volume of the adjoint orbit $\{ g \hat a g^{-1}|g\in G\}$.
In the approach of this paper,
we do not introduce ghosts-for-ghosts,
and the Vandermonde is included in the one-loop factor
(\ref{1-loop-pole}).

\subsubsection*{Hypermultiplet Determinant}

% \subsubsection{Hypermultiplet Determinant}
% \label{sec:1-loop-hyper}
The index of the complex for the Dirac operator $D_{\text{Dirac}}$ 
that maps the space of positive-chirality 
spinors $S^{+}$ to the space of negative-chirality
spinors $S^{-}$  in $\bR^4$
% \JG{Should we use $\slash\hskip-8pt D$?}
% \TO{$\slash\hskip-8pt D$, $D_\text{Dirac}$, $D^\text{hm}$, whichever
% we use, we want to be consistent.
% The same remark applies to the vectormultiplet.
% Perhaps because these are auxiliary systems that do not
% appear in the theory in the forms we write,
% it may be less confusing if we are more explicit than usual.
% I prefer $D^\text{hm}$ or $D_\text{Dirac}$ to $\slash\hskip-8pt D$.
% }
\begin{equation}
  \label{eq:Scomplex}
D_{\text{Dirac}}: S^{+} \to S^{-}\,,
\end{equation}
 with a suitable   inversion of the  grading, computes the contribution of a hypermultiplet 
to the one-loop determinant  \cite{Pestun:2007rz}. 
By applying the fixed-point formula (\ref{ASindice}) 
to the Dirac complex, we obtain%
\footnote{%
The index can also be obtained by noting that the Dirac complex in 
 $\BC^2=\BR^4$ is related to the Dolbeault
  operator $\bar \p: \Omega^{0,0} \to \Omega^{0,1} \to \Omega^{0,2}$. The bundle $S^{+}$
  is given by $\Omega^{0,0} \oplus \Omega^{0,2}$ twisted by
  $K^{1/2}$ while $S^-$ is given by $\Omega^{0,1}$ twisted by $K^{1/2}$, where $K$
is canonical bundle.  We want to compute the equivariant index of $D_{\text{Dirac}}$
with respect the $T=\U(1)_{\ve_1}\times \U(1)_{\ve_2}$ action  $(z_1, z_2) \mapsto (t_1 z_1, t_2
z_2)$. Hence up to the twist by $K^{1/2}$, which
contributes a factor of $(t_1t_2)^{1/2}$ to the index, the Dirac complex 
(\ref{eq:Scomplex}) is isomorphic to standard Dolbeault
complex in $\BC^2$.
The factor $ t_1^{1/2} t_2^{1/2}$ in (\ref{eq:index-Dirac})
accounts for the twist by $K^{1/2}$.}
%Taking into account the inverted grading, we get 
%\TO{What fixes the convention for right versus left?
%This seems to be the opposite of the standard convention,
%but it may be necessary for the correct answer.
%}
%\VP{You obtain Dirac operator on $\BC^2=\BR^4$ from Dolbeault
  %operator $\bar \p: \Omega^{0,0} \to \Omega^{0,1} \to \Omega^{0,2}$,
  %with  $p = 0,1,2$,  mapping  $\Omega^{0,0} \oplus \Omega^{0,2}$ twisted by
 % $K^{1/2}$ to $S^{+}$ and $\Omega^{0,1}$ twisted by $K^{1/2}$ to $S^{-}$, where $K$
%is canonical bundle. Hence up to the twist by $K^{1/2}$, which
%contributes $(t_1t_2)^{1/2}$ to the index, the Dirac complex 
%(\ref{eq:Scomplex}) is isomorphic to standard Dolbeault complex, with
%conventions }
%\JG{Sign in second formula corrected}\TO{Looks fine to me.} \VP{OK, let
%  it be}
\begin{eqnarray}
  %\ind D^{\text{hm}}(t) = - 
  \ind D_{\text{Dirac}} & =&
 \frac{ t_1^{1/2} t_2^{1/2} + t_1^{-1/2} t_2
    ^{-1/2} 
 - ( t_1^{1/2} t_2^{-1/2} + t_1^{-1/2} t_2^{1/2})}
{ (1-t_1)(1 -t_1^{-1})(1-t_2)(1-t_2^{-1})} 
\nonumber\\
&=& \frac { t_1^{1/2} t_2^{1/2}}{ (1 -t_1)(1-t_2)}\,.
\label{eq:index-Dirac}
\end{eqnarray}

The kinetic operator for a hypermultiplet in the adjoint
representation of the gauge group and the one-loop factor were analyzed in \cite{Pestun:2007rz} in detail.
% \JG{Do you really want to keep the next sentence?}There the 1-loop factor was derived, and it has been well tested
%  by comparing with two-dimensional theories.
The corresponding index is 
given by tensoring the Dirac bundle with the adjoint bundle.
We also need to remember that
%as well as with the rank-two bundle with the action of 
the $G_{\text F}=SU(2)$
 flavour symmetry associated to an 
%$R=$
adjoint hypermultiplet acts on the
bundle. 
Therefore the 
  $\U(1)_{\ve_1}\times \U(1)_{\ve_2} \times
G\times G_{\text F}$ equivariant index for this complex, taking into account the inversion of the grading, is given by
 % \JG{Flavour symmetry may be more appropriate name?} \VP{I
%   agree with JG}.
% \TO{``Flavour symmetry'' is fine with me.
% Choose one of the British/Canadian convention ``flavour'' and the American one ``flavour'' and stick to it.}
\begin{equation}
  \ind D^\hm_\text{adj} (\ve_1,\ve_2,\hat a, \hat m) 
 =  -  \frac { 
e^{\frac 1 2 (i \ve_1 + i \ve_2) }
}
 {(1 -e^{i\ve_1})(1-e^{i\ve_2})}
\frac{e^{i \hat m} + e^{-i \hat m}}{2}
\sum_{w\in \text{adj}} e^{i w\cdot \hat a}\,.
\label{eq:index-adj}
\end{equation}
We recall that the equivariant parameter $\hat m=im$
for the $SU(2)$  flavour symmetry, which takes values in the $SU(2)$ Cartan subalgebra,  is interpreted as the mass $m$
of the adjoint hypermultiplet.

Given the formula for the equivariant index for the  hypermultiplet in the adjoint representation, group theory
completely determines the corresponding index
for an arbitrary representation $R$ of the gauge group.
To explain this claim,
let us recall that the precise flavour symmetry
depends on the type of matter representation,
and that in general we need to consider half-hypermultiplets
although in the end half-hypermultiplets
pair up into full hypermultiplets.
For a complex irreducible representation $R$,
half-hypermultiplets always appear as copies of conjugate pairs
$N_\text{F}\cdot (R\oplus \bar R)$, and the flavour symmetry is $U(N_\text{F})$.
Half-hypermultiplets in a real irreducible representation $R$
can only arise in an even number $2N_\text{F}$, in which case
the flavour symmetry is enhanced to $Sp(2N_\text{F})$.%
\footnote{%
In our convention $Sp(2N)$ has rank $N$. Also $Sp(2)=SU(2)$.}
If the irreducible representation $R$ is   pseudo-real,
classically an arbitrary number $n$ of half-hypermultiplets can appear
with $SO(n)$ as the flavour symmetry group, but
for odd $n$ an anomaly renders the theory inconsistent \cite{Witten:1982fp}.
Thus  $n=2N_\text{F}$ has to be even and the flavour symmetry is enhanced to $SO(2N_\text{F})$.
% \footnote{%
% If $n$ is odd, 
%  $\sum_f\sum_w(\ldots)$
% in  (\ref{eq:index-univ})
% is replaced by $\sum_w \left(
% e^{i w\cdot\hat a}+\sum_f\left( e^{ iw\cdot\hat a-i\hat m_f}
% + e^{ -iw\cdot\hat a+i \hat m_f}
% \right)\right)$.
% For $G=SU(2)$ and matter in the ${\bf 2}$,
% there is no known Liouville counterpart for $n$ odd.
% }
In every case, the flavour symmetry group  acts in the defining representation
% \JG{shouldn't defining equation$\rightarrow$ adjoint representation.}
% \TO{I guess I meant ``defining representation'' rather than ``defining
% equation'' if I wrote it, so I corrected it.
% Not adjoint, I believe.},
and there are $N_\text{F}$ mass parameters $\hat m_f=i m_f$ with $f=1,\ldots N_\text{F}$  parametrizing 
the Cartan subalgebra of $G_{\text F}$.
As   shown in appendix \ref{sec:matter-general},  
the following expression for the index holds 
for an hypemultiplet in an arbitrary  representation $R$ of $G$:
\begin{equation}
  \ind D^\hm_{R}(\ve_1, \ve_2,\hat a,\hat m_f) 
 =  -  \frac { 
e^{\frac 1 2 (i \ve_1 + i \ve_2) }
}
 {2(1 -e^{i\ve_1})(1-e^{i\ve_2})}
\sum_{f=1}^{N_\text{F}}
\sum_{w\in R} \left(
e^{i w \cdot\hat a-i \hat m_f}
+e^{-i w \cdot\hat a+i \hat m_f}
\right)
\,.
\label{eq:index-univ}
\end{equation}
At the north and south poles, we expand the index \rf{eq:index-univ} in positive 
and negative powers of $(t_1, t_2)$ respectively, from which we read the weights of the 
the $\U(1)_{\ve_1}\times \U(1)_{\ve_2} \times
G\times G_{\text F}$ action. Both expansions give rise to identical   one-loop determinants, given in terms
of the weights by \rf{rule-appA}.

The relevant hypermultiplet one-loop determinant of the theory on $S^4$ is obtained by setting $\ve_1=\ve_2=\ve=1/r$,   the $G$-equivariant parameters at the north and south poles to \rf{scalaratpolesn}
and $\hat m_f=i m_f$, where $ m_f$  with $f=1,\ldots,N_{\text F}$ are the masses   of the $N_{\text F}$ hypermultiplets. 
Therefore the one-loop determinants of  $N_{\text F}$ massive  hypermultiplets  in a representation $R$ of $G$   arising from the north and south poles are given  by
\beq 
Z^\hm_{\text{north},\text{1-loop}}=Z^\hm_{\text{1-loop},\text{pole}} (\hat a(N),im_f)\qquad\qquad Z^\hm_{\text{south},\text{1-loop}}=Z^\hm_{\text{1-loop},\text{pole}} (\hat a(S),im_f)
\,.
\eeq
with$^{\text{\scriptsize\ref{asymptotically}}}$ 
%\beq
%\hat a(N) =ia-ig^2\theta {B\over 16\pi^2r}-  {B\over 2r} \qquad \hat a(S) = ia-ig^2\theta {B\over 16\pi^2r}+{B\over 2r}\,,
%\label{scalaratpolesnn}
%\eeq
%$Z^\vm_\text{1-loop,pole} (\hat a)$ given in (\ref{1-loop-pole}).
 %\begin{eqnarray}
 \beq
%&&  Z^\hm_{\text{1-loop},\text{pole}}(\hat a, \hat m)
% =  Z^\hm_\text{1-loop,$R$,S}
%  (\hat a, \hat m)
%\nonumber\\
%&&
%\hspace{15mm}
Z^\hm_{\text{1-loop},\text{pole}}(\hat a, \hat m_f)=
\prod_{f=1}^{N_\text{F}}\prod_{w\in R}
G^{-1/2}
\left(1+\frac{w\cdot\hat a}{\ve}-\frac {\hat m_f}\ve
\right)
G^{-1/2}
\left(1-\frac{w\cdot\hat a}{\ve}+
\frac{\hat m_f}\ve 
\right)\,,
\label{loophyper}
%\end{eqnarray}
\eeq
 where $w$ are the weights of the representation $R$.
%, and $Z^\hm_\text{1-loop,$R$,pole} (\hat a, \hat m)$
%at the south pole
 We note that for an arbitrary representation $R$, the      hypermultiplet  one-loop determinant  at the south pole is the complex conjugate of the determinant at the  north pole\footnote{The expression $|{Z}^\hm_{\text{north},\text{1-loop}}|^2$ reproduces the one-loop determinant obtained in \cite{Pestun:2007rz}
when there is no 't Hooft loop, corresponding to $\hat a(N)=\hat a(S)=ia, \hat m_f=i m_f$.} 
\beq
Z^\hm_{\text{south},\text{1-loop}}=\overline{Z}^\hm_{\text{north},\text{1-loop}}\,.
\eeq

 We can now start gathering the results obtained until now. 
 Combining the vectormultiplet and
hypermultiplet determinants  given in \rf{1-loop-pole} and \rf{loophyper}, we conclude that the pole contribution
to the  one-loop determinant
 for an arbitrary $\cN=2$ Lagrangian theory in $S^4$ in the presence of a 't Hooft operator can be written in terms of 
\beq
{Z}_{\text{1-loop},\text{pole}}(\hat a, \hat m_f) =
\frac{
%\displaystyle
\prod_{\alpha}\left[
G\left(
%\frac{\alpha \cdot\hat a }{\ve}
 r\alpha \cdot\hat a 
\right)
G\left(2+
%\frac{\alpha \cdot\hat a}{\ve}
r \alpha \cdot\hat a
\right)\right]^{1/2}
}{
%\displaystyle
\prod_{f=1}^{N_\text{F}}\prod_{w\in R}
\left[G
%\left(1+\frac{w\cdot\hat a}{\ve}-\frac {\hat m_f}\ve
\left(1+ r w\cdot\hat a- r\hat m_f
\right)
G
\left(1-
%\frac{w\cdot\hat a}{\ve}+
r w\cdot\hat a+
%\frac{\hat m_f}\ve 
r\hat m_f
\right)\right]^{1/2}
}
\,,
\label{loopcombination}
\eeq
where we recall that $\ve=1/r$.  Formula (\ref{loopcombination}) holds for an
arbitrary $\cN=2$ gauge theory admitting a Lagrangian description, and
can be explicitly calculated given the choice of gauge group $G$ and of
a representation  $R$ of $G$ under which the hypermultiplet transforms.

For  asymptotically  free gauge theories, the localization calculation  is most accurately
performed by embedding such a theory into one that is ultraviolet finite,   which then flows 
to the asymptotically free theory upon taking the mass parameters of the finite theory to be very large. As a prototype of this construction,   $\cN=2$ pure super Yang-Mills
with arbitrary gauge group $G$ can be regulated by embedding it in the  $\cN=2^*$ theory, consisting of a vectormultiplet
and massive hypermultiplet in the adjoint representation of $G$, by then  taking  the mass of the hypermultiplet
to be very large. This construction exists for an arbitrary asymptotically free four dimensional $\cN=2$ gauge theory. Given an asymptotically free $\cN=2$ gauge theory, the end result of this procedure in the localization computation
is that 
the one-loop determinants are given by (\ref{loopcombination}) for the field content of the asymptotically free theory,  together with the replacement 
of the bare coupling constant $\tau$ of the theory with the familiar one-loop corrected running coupling constant $\tau_\text{ren}$.

The complete one-loop determinants in our localization computation arising at  the north and south poles
 are thus given by
 \beq
 {Z}_{\text{north},\text{1-loop}}={Z}_{\text{1-loop},\text{pole}}(\hat a(N), im_f)   \qquad {Z}_{\text{south},\text{1-loop}}={Z}_{\text{1-loop},\text{pole}}(\hat a(S), im_f)\,,
 \eeq
 with $\hat a(N)$ and $\hat a(S)$ in \rf{scalaratpolesn}.
Combining the one-loop result with the classical and instanton contributions computed in the previous two sections, we have that the expectation value of a 't 
Hooft loop labeled by a coweight $B$ in an $\cN=2$ gauge theory with gauge group $G$ and $N_{{\text F}}$ massive hypermultiplets
in a representation $R$ of $G$  is given by\footnote{After trivially shifting the integration variable 
$ia\rightarrow ia +ig^2 \theta \frac{B}{16\pi^2 r}$.}
% \beq
% \label{eq:thooft-result-extra}
% \hskip-16pt \vev{T(B)}\hskip-3pt \simeq\hskip-5pt 
%  \int\hskip-4pt da\hskip-1pt \left | Z_{\text{cl}}\hskip-3pt\left(\hskip-2pt ia - \frac{B}{2r},q\hskip-2pt\right)\hskip-3pt
%  {Z}_{\text{1-loop},\text{pole}}\hskip-3pt\left(\hskip-2pt ia - \frac{B}{2r}, im_f\hskip-2pt\right)\hskip-3pt Z_{\text{inst}}\hskip-3pt\left(\hskip-2pt ia  - \frac{B}{2r}, {1\over r}+ i m_f, {1\over r},{1\over r},q\hskip-2pt\right)\right|^2\hskip-5pt.
%   \eeq
\begin{eqnarray}
&&\vev{T(B)} \simeq
\nonumber\\
&&
\hspace{5mm}
\int da\left | Z_{\text{cl}}\left( ia - \frac{B}{2r},q\right)
 {Z}_{\text{1-loop},\text{pole}}\left(\hskip-2pt ia - \frac{B}{2r}, im_f
\right) Z_{\text{inst}}\left( ia  - \frac{B}{2r}, {1\over r}+ i m_f, {1\over r},{1\over r},q\right)\right|^2
\nonumber\\
\label{eq:thooft-result-extra}
\end{eqnarray}
with $Z_{\text{cl}}$, ${Z}_{\text{1-loop},\text{pole}}$ and $Z_{\text{inst}}$ given in \rf{eq:class-contr}, \rf{loopcombination} and \rf{nekrinst} respectively.

Therefore, the path integral completely factorizes into north and south pole contributions as
\begin{equation}
\label{eq:thooft-result-extra-more}
 \vev{T(B)}\simeq \int da \,  {Z}_{\text{north}}\cdot {Z}_{\text{south}} =\int da\, \left |  {Z}_{\text{north}}\right|^2\,,
\end{equation}
with 
\beq
\begin{aligned}
{Z}_{\text{north}}=& Z_{\text{cl}}\left(\hat a(N),q\right){Z}_{\text{1-loop},\text{pole}}(\hat a(N), im_f)Z_{\text{inst}}\left(\hat a(N), {1\over r}+ i m_f, {1\over r},{1\over r},q\right)\\
{Z}_{\text{south}}=& Z_{\text{cl}}
\left(\hat a(S),\bar q\right){Z}_{\text{1-loop},\text{pole}}(\hat a(S), im_f)Z_{\text{inst}}\left(\hat a(S), {1\over r}+ i m_f, {1\over r},{1\over r},\bar q\right)\,,
\end{aligned}
\eeq
which furthermore are complex conjugate to each other
\beq
Z_{\text{south}}=\overline{Z}_{\text{north}}\,.
\eeq
When the gauge theory is asymptotically free, we must replace the bare instanton fugacity $q$ by the renormalized one $q_\text{ren}$ in ${Z}_{\text{north}}$ 
and $Z_{\text{south}}$.
The $\simeq$ symbol  is used in (\ref{eq:thooft-result-extra}) and (\ref{eq:thooft-result-extra-more}) since in the presence of a 't Hooft loop operator, an extra contribution supported on the 
loop must be included, to which we now turn.\footnote{In section \ref{sec:screening} we will include yet another contribution due to monopole screening, which is non-perturbative in nature.}

\subsection{Equator Contribution}
\label{sec:one-loop-eq}
In the absence of a 't Hooft loop, 
the index is a sum of local contributions
from the north and south poles
\cite{Pestun:2007rz}, which are the fixed points of $J$.
In defining the 't Hooft loop path integral in gauge theory,
we must impose  boundary conditions 
along the loop compatible with the field  configuration of a singular monopole. In this subsection we calculate the contributions to the vectormultiplet and hypermultiplet indices 
as well as one-loop determinants  from the equatorial $S^1$ where the 't Hooft loop is located,
which are
 functions of the weights for the group action  $U(1)_{J+R}\times G\times G_{\text F}$ generated by $Q^2$ (also by $\hat Q^2$).

Let us recall from section \ref{sec:ball}
that the isometry generator $J$ in $Q^2$ acts on $B_3\times S^1$ as a spatial rotation
along the $x_3$-axis as well as a shift in the periodic coordinate $\tau$.
The conformal killing spinor   $\ep_Q$ in (\ref{killsol}) with which we localize the 't Hooft loop
path integral  
can be written as
\begin{eqnarray}
  \ep_Q&=&
e^{-\tau \frac{\Gamma^{56}+\Gamma^{78}}4}
  \left(
1-i \frac{x^i}{2r} \tilde \Gamma_i \Gamma^{120}
\right)
\hat \ve_s\,.
\end{eqnarray}
Note that $\ep_Q$ changes its sign when going around the $S^1$,  under $\tau \rightarrow \tau+2\pi$.
Therefore while all the bosons are periodic,
all the fermions in the vielbein basis are antiperiodic.
In particular, within each supermultiplet bosons and fermions
obey different boundary conditions around $S^1$.

Recall that $Q^2$ also includes the $U(1)_R$ transformation (see \rf{Rtransform}),
which is generated   by $J_{56}+J_{78}$.
When we apply the index theorem it is convenient
to redefine fields of the theory and $\ep_Q$ using $U(1)_R$ as\footnote{Here we are using ten dimensional notation for the
bosonic fields of the $\cN=2$ theory, so that \newline $A_{M}=\{A_\mu,q,\tilde q, \Phi_9,\Phi_0\}$ with $M=1,\ldots,9,0$.}
\beq
%\begin{eqnarray}
\begin{aligned}
\ep_Q&\rightarrow e^{\tau \frac{\Gamma^{56}+\Gamma^{78}}4} \ep_Q\,,
\\
  A_M&\rightarrow 
\left
(e^{\tau
 \frac{J_{56}+J_{78}}2 }
\right
){}_{\hspace{-1mm}MN} A_N \,,
\label{field-redef}\\
\Psi&\rightarrow e^{\tau \frac{\Gamma^{56}+\Gamma^{78}}4} \Psi\,,\\
\chi&\rightarrow\chi\,,
\end{aligned}
\eeq
%\end{eqnarray}
where we have normalized the ten-dimensional Lorentz
generators in the vector representation as $(J_{MN})_{PQ}=\delta_{MP}\delta_{NQ}-\delta_{MQ}\delta_{NP}$
and used that $U(1)_R$ is generated by $\Gamma^{56}+\Gamma^{78}$ when  acting on
 spinors.
After the field redefinition,
the whole vectormultiplet is periodic, and all fields in the hypermultiplet are antiperiodic.\footnote{The R-symmetry group $U(1)_R$ 
  acts non-trivially on the fermions in the vectormultiplet 
 and on the scalars in the hypermultiplet.
}
This redefinition makes the  spinor $\ep_Q$    independent of $\tau$.
The shift in $\tau$ now induces an
R-symmetry transformation in addition to the $S^1$ part of isometry $J$.

\subsubsection*{Vectormultiplet Determinant}

As we saw in section \ref{sec:balla},
the localization equations near the location of the 't Hooft operator  -- which wraps the $S^1$ at the origin in $B_3$ -- are approximately the Bogomolny equations
\begin{eqnarray}
  *_3 F=D\Phi
\end{eqnarray}
in $B_3 \times S^1$.
The differential operator that appears in the kinetic term
for the vectormultiplet is obtained by linearizing the Bogomolny equations.
Linearization of the gauge transformation and the Bogomolny equations
is described by the complex\footnote{This complex can also be turned
into a two-term complex as in \rf{indice} by folding the complex.}
\begin{eqnarray}
D_\text{Bogo}:  \Omega^0 \rightarrow \Omega^1\oplus \Omega^0
\rightarrow \Omega^2
\label{linearization}
\end{eqnarray}
in $\mathbb R^3-\{0\}$.
In appendix \ref{sec:Kronheimer}, 
we explain Kronheimer's observation that
the Bogomolny equations in $\mathbb R^3$
with a monopole singularity at the origin -- where the 't Hooft operator resides -- is equivalent to
the anti-self-duality equations for gauge fields
in $\mathbb R^4$
invariant under the action of a spacetime symmetry group $U(1)_K$.
Using Kronheimer's correspondence, we can
obtain this complex
 by projecting the self-dual complex (\ref{eq:self-dual})
to the $U(1)_K$-invariant sections.
We can compute the index of the complex
(\ref{linearization})
by averaging the index of  the self-dual complex
over the $U(1)_K$ action, picking up the contributions only
from the $U(1)_K$ invariant sections.%
\footnote{%
A similar computation was done in \cite{MR1624279}, 
where more than one singular monopole was
considered on a compact manifold.
Though our integrand to be averaged
is a rational function with poles on the integration contour,
the integrand in \cite{MR1624279}
was a polynomial due
to cancellations among singular monopoles.
}

In equation (\ref{ind-Dolb}), the index for the Dolbeault operator
$\bar\partial$ on $\mathbb C$
was obtained as the $U(1)$ character
on the space of holomorphic functions.
In this toy example the index is an infinite power series
corresponding to infinitely many monomials.
The same logic can be used to derive 
the index (\ref{indexvect}) for the complex (\ref{eq:self-dual})
through an expansion in a basis of local sections.
Among such sections, those which are invariant under $U(1)_K$
 correspond to the ordinary
spherical harmonics for the bundles in three dimensions.
We can  keep track of
the original expansion by introducing an infinitesimal
positive parameter $\delta>0$:
\begin{eqnarray}
  \ind_\delta (D_\text{SD})(t_1, t_2) 
&=& \frac {
(1+t_1^{-1} t_2^{-1})(1-t_1)(1-t_2)
}{2 (1-e^{-\delta}t_1)(1- e^{-\delta} t_1^{-1})
    (1- e^{-\delta} t_2)(1-e^{-\delta} t_2^{-1})} \,.
\end{eqnarray}

We now parametrize the $U(1)\times U(1)$ weights as
\begin{eqnarray}
  t_1=e^{-i\nu+i\frac 1 2 \ve}\,,~~~~~t_2=e^{i\nu+i \frac 1 2 \ve}\,,
\end{eqnarray}
where $\nu$ is the parameter for the group $U(1)_K$ used in Kronheimer's
construction: $(\mathbb C^2-\{0\})/U(1)_K\simeq \mathbb R^3-\{0\}$.
The parameter $\ve$ is the angle for a rotation along the $x_3$-axis
in $\mathbb R^3$, and the factors of $1/2$ ensure that
for $\ve =2\pi$ this rotation acts as $-1$ on $\mathbb C^2$
even though it acts as $+1$ on $\mathbb R^3$.

In order to  describe the singular monopole background due to the 't Hooft operator, we also need to twist by the adjoint gauge bundle
on which the gauge group $G$ and $U(1)_K$
act as $e^{\hat a+ B\nu}$, with $B$ being the magnetic weight labeling the operator.
The four-dimensional sections invariant under $U(1)_K$
can be identified with the monopole harmonics \cite{Wu:1976ge} of the corresponding
bundles over $\mathbb R^3-\{0\}$.
The index for the self-dual complex twisted by the gauge bundle is given by
\begin{eqnarray}
\text{ind}_\delta (D_\text{SD})(\nu,\ve,\hat a)
&=&\frac{
(1+e^{-i\ve})
(1-e^{-i\nu+i \ve/2})(1-e^{i\nu+i\ve/2})
}{2
(1-e^{-\delta}e^{i\nu-i\ve/2})
(1-e^{-\delta}e^{-i\nu+i\ve /2 })
(1-e^{-\delta}e^{-i\nu-i\ve/2})
(1-e^{-\delta}e^{i\nu+i\ve/2})
}
\nonumber\\
&&~~~~~\times
\sum_{w\in \text{adj}} e^{i w\cdot\hat a+ i w\cdot B \nu}\,.
\label{ind-SD-reg}
\end{eqnarray}
By averaging over $U(1)_K$, 
we get the desired index for the complex (\ref{linearization})
\begin{eqnarray}
\text{ind}(D_\text{Bogo})
&=& \lim_{\delta\rightarrow 0} \int_0^{2\pi}
 \frac{d\nu}{2\pi}
\text{ind}_\delta( D_\text{SD})(\nu,\ve,\hat a)
\nonumber\\
% &=&  \oint_{|z|=1} \frac{d z}{2\pi i z}
% \frac{
% (1+e^{-i\ve})
% (1-e^{i\ve/2}/z)(1-e^{i\ve/2}z)
% }{
% (1-e^{-\delta}e^{-i\ve/2} z)
% (1-e^{-\delta}e^{i\ve/2}/z)
% (1-e^{-\delta}e^{-i\ve/2}/z)
% (1-e^{-\delta}e^{i\ve/2}z)
% }
% \sum_{w\in \text{adj}} e^{i w\cdot a} z^{ w\cdot B}
% \nonumber\\
&=&
 \lim_{\delta\rightarrow 0}  \oint_{|z|=1} \frac{d z}{2\pi i }
\frac{
(1+e^{-i\ve})
(z-e^{i\ve /2})(1-e^{i\ve/2}z)
}{2
(1-e^{-\delta}e^{-i\ve/2}z)
(z-e^{-\delta} e^{i\ve /2})
(z-e^{-\delta}e^{-i\ve/2})
(1-e^{-\delta}e^{i\ve/2}z)
}
\nonumber\\
&&~~~~~\times
\sum_{w\in \text{adj}} e^{-i w\cdot \hat a} z^{- w\cdot B}\,,
\end{eqnarray}
where we have renamed $w$ as $w \rightarrow -w$.
We can evaluate the integral by summing over residues for the poles
inside the unit circle.
For $w\cdot B >0$ a pole at $z=0$
contributes
\begin{eqnarray}
&&\sum_{w\cdot B>0} e^{-i w\cdot\hat a}
\frac{(1+e^{-i\ve})}{2(w\cdot B-1)!}
\left.
\left(\frac{\partial}{\partial z}\right)^{w\cdot B-1}
\right|_{z=0}\frac{1}{(1-e^{-i\ve/2}z)(z-e^{-i\ve/2})}
\nonumber\\
 &=&
-\frac{e^{i\ve/2}+e^{-i\ve/2}}2
\sum_{w\cdot B>0}
e^{-i w\cdot \hat a}
\left(
e^{i\frac{w\cdot B-1}2\ve}
+
e^{i\frac{w\cdot B-3}2\ve}
+\ldots
+
e^{-i\frac{w\cdot B-1}2\ve}
\right)
 \nonumber\\
&=&
-\frac{e^{i\ve/2}+e^{-i\ve/2}}2
\sum_{w\cdot B>0} e^{-i w\cdot \hat a}
\frac{e^{i (w\cdot B)\ve/2}-e^{-i (w\cdot B)\ve/2}}{e^{i\ve/2}-e^{-i\ve/2}}\,.
\end{eqnarray}
In addition there are always two  poles at $z= e^{-\delta} e^{i\ve/2},
 e^{-\delta} e^{-i\ve/2}$.
In the limit $\delta \rightarrow 0$,
the contribution of
the pole at $z=e^{-\delta}e^{i\ve/2}$ is given by
\begin{eqnarray}
&& 
 \frac{
(1+e^{-i\ve})(e^{-\delta}e^{i\ve/2}-e^{i\ve/2})(1-e^{i\ve}e^{-\delta})
}{
2
(1-e^{-2\delta})(e^{-\delta}e^{i\ve/2}-e^{-\delta} e^{-i\ve/2 })(1-e^{-2\delta}e^{i\ve})
}
\sum_{w\in \text{adj}} e^{-i w\cdot \hat a} e^{-w\cdot B(- \delta+i\ve/2) }
\nonumber\\
&\rightarrow&
-\frac 1 4
\frac{e^{i\ve/2}+e^{-i\ve/2}}
{e^{i\ve/2}-e^{-i\ve/2}}
\sum_{w\in \text{adj}} e^{-i w\cdot\hat a} e^{-i w\cdot B \ve/2}
\,,
\end{eqnarray}
while the pole at $z=e^{-\delta}e^{-i\ve/2}$ contributes
\begin{eqnarray}
&& 
 \frac{
(1+e^{-i\ve})
(e^{-\delta}e^{-i\ve/2}-e^{i\ve/2})(1-e^{-\delta})
}{2
(1-e^{-2\delta}e^{-i\ve})(e^{-\delta}e^{-i\ve/2}-e^{-\delta}e^{i\ve/2})
(1-e^{-2\delta})
}
\sum_{w\in \text{adj}} e^{-i w\cdot \hat a} e^{-  w\cdot B (-\delta -i\ve/2)}
\nonumber\\
&\rightarrow&
\frac 1 4 \frac{
e^{i\ve/2}+e^{-i\ve/2}
}{
e^{i\ve/2}-e^{-i\ve/2}
}
\sum_{w\in \text{adj}} e^{-i w\cdot \hat a}
e^{i (  w\cdot B)\ve/2}\,.
\end{eqnarray}
Combining the residues we get
\begin{eqnarray}
  \text{ind}(D_\text{Bogo})
&=&
-\frac{e^{i\ve/2}+e^{-i\ve/2}}2
\sum_{w\cdot B>0} e^{-i w\cdot \hat a}
\frac{e^{i (w\cdot B)\ve/2}-e^{-i (w\cdot B)\ve/2}}{e^{i\ve/2}-e^{-i\ve/2}}
\nonumber\\
&&~~~~~~~
+\frac 1 4
(e^{i\ve/2}+e^{-i\ve/2})
\sum_{w\cdot B\neq 0}
e^{-i w\cdot \hat a}
\frac{e^{i (  w\cdot B)\ve/2}-e^{-i (  w\cdot B)\ve/2}}
{
e^{i\ve/2}-e^{-i\ve/2}
}
% (1+e^{i\ve}+\ldots+e^{i (w\cdot B-1)\ve})
% +\frac 1 2 
% e^{i\ve}(1+e^{i\ve})
% \sum_{w\cdot B> 0}
% e^{i w\cdot a}
% (1+e^{i\ve}+\ldots+e^{i (w\cdot B-1)\ve})
% \,,
% \nonumber\\
% &=&
% -e^{i\ve}(1+e^{-i\ve})
% \sum_{w\cdot B>0} e^{-i w\cdot a}
% \frac{1-e^{i (w\cdot B)\ve}}{1-e^{i\ve}}
% -e^{i\ve}(1+e^{i\ve})
% \sum_{w\cdot B>0} e^{-i w\cdot a}
% \left(
% 1+2 e^{i\ve}+3 e^{2i\ve}+\ldots+ (w\cdot B) e^{i (w\cdot B-1) \ve}
% \right)
% \nonumber\\
% &&~~~~~~~
% +\frac 1 2 
% e^{i\ve}(1+e^{-i\ve})
% \sum_{w\cdot B> 0}
% (e^{-i w\cdot a}-e^{iw\cdot a}e^{-i (w\cdot B)\ve})
% \frac{e^{i (  w\cdot B)\ve}-1}
% {
% e^{i\ve}-1
% }
\nonumber\\
&=&
-\frac 1 4
(e^{i\ve/2}+e^{-i\ve/2})
\sum_{\alpha >0}
(e^{i \alpha\cdot \hat a}+e^{-i\alpha\cdot\hat a})
\frac{
e^{i (  \alpha\cdot B)\ve/2}
-
e^{-i (  \alpha\cdot B)\ve/2}
}
{
e^{i\ve/2}
-
e^{-i\ve/2}
}\,.
\label{ind-Bogo}
\end{eqnarray}
In the last line we replaced the sum over the adjoint 
weights satisfying $w\cdot B>0$
by the sum over positive roots $\alpha>0$.
This is possible because
by taking  $B$  to be in the Weyl chamber all such $w$'s are 
positive roots.

For the vectormultiplet  one-loop determinant computation, 
we also need to tensor with the space of periodic functions on $S^1$.
Thus we need to compute $ \sum_{n\in \mathbb Z} e^{i n \ve} 
  \text{ind}(D_\text{Bogo})$.
A simplification arises because the parameter $n$ is summed over,
and can be shifted by an integer freely.
Finally,  the equatorial index for the vectormultiplet is
\begin{eqnarray}
&&
 \ind (D_\text{eq}^\vm)(\ve,
%_1, \ve_2, 
\hat a)=  \sum_{n\in \mathbb Z} e^{i n \ve} \,
  \text{ind}(D_\text{Bogo})
\nonumber\\
&=&
-
  \sum_{n\in \mathbb Z} e^{i n \ve} 
\frac{e^{i\ve/2}+e^{-i\ve/2}}4
\sum_{\alpha>0}
(e^{i\alpha\cdot \hat a}
+e^{-i\alpha\cdot\hat a})
% \nonumber\\
% &&
% \hspace{20mm}
% \times
\left(
e^{i\frac{\alpha\cdot B-1}2\ve}
+
e^{i\frac{\alpha\cdot B-3}2\ve}
+\ldots
+
e^{-i\frac{\alpha\cdot B-1}2\ve}
\right)
\nonumber\\
&=&
- \label{indexeqvm}
\sum_{\alpha>0}
(\alpha \cdot B)
\frac{
e^{i \alpha \cdot \hat a}+e^{-i \alpha\cdot \hat a}}2
\times
\sum_{n\in \mathbb Z} 
\left\{
\begin{array}{ll}
 e^{i n \ve} 
 &\text{ if $\alpha\cdot B$ is even,}
\\ 
  e^{i (n+1/2) \ve}
 &\text{ if $\alpha\cdot B$ is odd.}
\end{array}
\right.
\end{eqnarray}
Note that we can  write the last sum as $\sum_{n\in \mathbb Z}
e^{i (n+\alpha\cdot B/2)\ve}$ in both cases.%
\footnote{%
Physically, half-odd integer coefficients appear in the exponential for odd $\alpha\cdot B$ because the relation between the angular momentum and statistics
is reversed when the monopole charge measures in the unit amount is odd
\cite{Wu:1976ge}.
}

Setting $\epsilon=1/r$ and using that  at the equator $\hat a(E)=ia-ig^2\theta {B\over 16\pi^2r}$ \rf{equagauge}, we find that
the   equatorial one-loop determinant for the vectormultiplet is given by
\beq
Z^\text{vm}_\text{equator}= Z_\text{1-loop,eq}^\text{vm}\left(ia-ig^2\theta {B\over 16\pi^2r},B\right)\,
\eeq
where\footnote{We regulate the product 
by identifying it with the product representation of the $\sin$ function.}
\begin{eqnarray}
  Z_\text{1-loop,eq}^\text{vm}
(\hat a, B)
&=&
\prod_{n\in \mathbb Z}\prod_{\alpha> 0}
\left( n\ve
+\frac{\alpha\cdot B}2 \ve
+\alpha \cdot \hat a
\right)^{-\alpha\cdot B/2}
\left( n\ve
+\frac{\alpha\cdot B}2 \ve
-\alpha \cdot \hat a
\right)^{-\alpha \cdot B/2}
\nonumber\\
&=&
\prod_{\alpha>0}
\left[\sin
\left(\pi \alpha \cdot \left(\frac{\hat a} \ve+\frac B 2\right)
\right)\right]^{-\alpha\cdot B}
\,,
\label{vectorequator}
\end{eqnarray}
computed from the index \rf{indexeqvm} using equation \rf{rule-appA},

\subsubsection*{Hypermultiplet Determinant}

We deal with the hypermultiplet in a similar way.
The relevant differential operator is the Dirac operator
plus a coupling to the Higgs field $\Phi_9$.
In Kronheimer's correspondence,
this lifts simply to the Dirac operator on $\mathbb C^2$ given in \rf{eq:index-Dirac}.
We regularize the index (\ref{eq:index-Dirac})
by specifying the expansion in a local basis 
as
\begin{eqnarray}
\text{ind}_\delta(D_\text{Dirac})(t_1,t_2)
&=&
\frac{t_1^{-1/2}t_2^{-1/2}(1-t_1)(1-t_2)
}
{
(1- e^{-\delta} t_1^{-1})(1- e^{-\delta} t_1)
(1-e^{-\delta} t_2^{-1})(1-e^{-\delta} t_2)
}
% \nonumber\\
% &=&
% \frac{t_1^{-1/2} t_2^{-1/2}}{1+t_1^{-1} t_2^{-1}}
% \ind_\delta (D_{\BC})(t_1, t_2) 
\,.
\end{eqnarray}
We can twist the Dirac complex by a vector bundle
whose sections transform in representation $R$ of the gauge group.
Including the action of the gauge and flavour groups $G\times G_{\text F}$
as in (\ref{eq:index-univ}),
and then averaging over $U(1)_K$, we obtain
\begin{eqnarray}
  \text{ind}(D_\text{DH})&=&
\lim_{\delta\rightarrow 0}\int_0^{2\pi }
\frac{d\nu}{2\pi}
\text{ind}_\delta(D_\text{Dirac})(t_1,t_2, \hat a,\hat m_f)
\nonumber\\
&=&
- \frac 1 4
\sum_{f=1}^{N_\text{F}}
 \sum_{w\in R, w\cdot B>0}
( e^{i w\cdot \hat a-i \hat m_f}
+
 e^{-i w\cdot \hat a+i \hat m_f}
)
\frac{e^{i (w\cdot B)\ve/2}-e^{-i (w\cdot B)\ve/2}}{e^{i\ve/2}-e^{-i\ve/2}}
\nonumber\\
&&~~~~~~~
+
\frac 1 4
\sum_{f=1}^{N_\text{F}}
  \sum_{w\in R, w\cdot B<0} 
(e^{i w\cdot \hat a-i \hat m_f}
+e^{-i w\cdot \hat a+i \hat m_f}
)
\frac{e^{i (w\cdot B)\ve/2}-e^{-i (w\cdot B)\ve/2}}{e^{i\ve/2}-e^{-i\ve/2}}
\,.
\end{eqnarray}

% \TO{$m\rightarrow \hat m$, etc.}
% Including the action of the gauge and flavour groups,
% and then averaging over $U(1)_K$,
% we obtain
% \begin{eqnarray}
%   \text{ind}(D_\text{DH})&=&
% (e^{im}+e^{-i m})
% \oint \frac{d\nu}{2\pi} \text{ind}_\delta(D_\text{Dirac})
% \nonumber\\
% &=&
% -\frac 1 2 
% (e^{im}+e^{-i m})
% \sum_{\alpha >0}
% (e^{i \alpha\cdot a}+e^{-i\alpha\cdot a})
% \frac{
% e^{i (  \alpha\cdot B)\ve/2}
% -
% e^{-i (  \alpha\cdot B)\ve/2}
% }
% {
% e^{i\ve/2}
% -
% e^{-i\ve/2}
% }\,.
% \end{eqnarray}
%We need to take into account the $\tau$ dependence of the fields
%for the 1-loop computation.
We noted above that the hypermultiplet fields
are antiperiodic in $\tau$.
Thus we must tensor with the space of anti-periodic functions on $S^1$, and
change the sign for the index
because we shift the degrees for physical fields
in the complex (as we did already for the hypermultiplet contribution at the poles).
 The equatorial index for the hypermultiplet is thus
\begin{eqnarray}
& & \ind (D_{R, \text{eq}}^\hm)(\ve,
%_1, \ve_2, 
\hat a,\hat m_f)=-  \sum_{n\in \mathbb Z}e^{i (n+1/2) \ve}\, 
\text{ind}(D_\text{DH})
\nonumber\\
&=&
\frac 1 4
\sum_{f=1}^{N_\text{F}}
 \sum_{w\in R}
|w\cdot B|
( e^{i w\cdot \hat a-i \hat m_f}
+
 e^{-i w\cdot \hat a+i \hat m_f}
)
% \nonumber\\
% &&~~~~~~~
% +
% \frac 1 2
% \sum_{f=1}^{N_\text{F}}
%   \sum_{w\in R, w\cdot B<0} 
% |w\cdot B|
% (e^{i w\cdot \hat a-i \hat m_f}
% +e^{-i w\cdot \hat a+i \hat m_f}
% )
%  \\
% &=&
% \frac{e^{im}+e^{-im}}2
% %\sum_{n\in \mathbb Z} 
% %e^{i n \ve}
%  \sum_{\alpha>0} (\alpha \cdot B)
% (e^{i\alpha\cdot a}
% +e^{-i \alpha\cdot a})
\times \sum_{n\in \mathbb Z} 
\left\{
\begin{array}{ll}
 e^{i n \ve} 
 &\text{ if $w \cdot B$ is even,}
\\ 
  e^{i (n+1/2) \ve}
 &\text{ if $ w \cdot B$ is odd.}
\end{array}
\right.
\end{eqnarray}

Therefore, the  one-loop determinant contribution  from the equator of   $N_{\text F}$ hypermultiplets
in a representation $R$ of the gauge group  is
\beq
Z^\text{hm}_\text{equator}= Z_\text{1-loop,$R$,eq}^\text{hm}(ia,im_f,B)\,,
\eeq
where
\begin{eqnarray}
%&& 
 Z_\text{1-loop,$R$,eq}^\text{hm}(\hat a, \hat m_f, B)
&= &
\prod_{f=1}^{N_\text{F}}
 \prod_{w\in R}
\left[
\sin\left(
\pi w\cdot\left (\frac {\hat a} \ve + \frac B 2\right)
-\pi \frac{ \hat m_f} \ve
\right)
\right]^{|w\cdot B|/2}
\,.
\label{1-loop-equator-hyper}
\end{eqnarray}

Combining the vectormultiplet \rf{vectorequator} and hypermultiplet \rf{1-loop-equator-hyper} determinants,
 the complete  equator contribution is given by
\beq
 Z^\text{1-loop}_\text{equator}=Z_\text{1-loop,eq}\left(ia-ig^2\theta {B\over 16\pi^2r},  im_f, B\right)
\eeq
with%
\footnote{%
Up to a phase, this expression is valid even if $B$ is not
in the Weyl chamber.
}
\begin{eqnarray}
%&& 
 Z_\text{1-loop,eq}(\hat a, \hat m_f, B)
&=&
\frac{  \prod_{f=1}^{N_\text{F}}
 \prod_{w\in R}
\left[
\sin\left(
\pi w\cdot\left (\frac {\hat a} \ve + \frac B 2\right)
-\pi \frac{ \hat m_f} \ve
\right)
\right]^{|w\cdot B|/2}
}{
\prod_{\alpha>0}
\left[\sin
\left(\pi \alpha \cdot \left(\frac{\hat a} \ve+\frac B 2\right)
\right)\right]^{|\alpha\cdot B|}
}
\,.
\label{1-loop-equator}
\end{eqnarray}

We are now in the position of writing the exact expectation value of a 't Hooft loop in an $\cN=2$ gauge theory on $S^4$ with magnetic weight $B$. Multiplying the contributions associated to the poles and the equator, we have that\footnote{Shitfing variables $ia\rightarrow ia +ig^2 \theta \frac{B}{16\pi^2 r}$.}
\beq
\vev{T(B)}=\int da\, Z_\text{north}\cdot Z_\text{south}\cdot Z^\text{1-loop}_\text{equator}=\int da \left|Z_\text{north}\right|^2\cdot Z^\text{1-loop}_\text{equator}
\label{finfin}
\eeq
where
\beq
\begin{aligned}
{Z}_{\text{north}}=& Z_{\text{cl}}\left(a(N),q\right){Z}_{\text{1-loop},\text{pole}}(a(N), im_f)Z_{\text{inst}}\left(a(N), {1\over r}+ i m_f, {1\over r},{1\over r},q\right)\\
{Z}_{\text{south}}=& Z_{\text{cl}}\left(a(S),\bar q\right){Z}_{\text{1-loop},\text{pole}}(a(S), im_f)Z_{\text{inst}}\left(a(S), {1\over r}+ i m_f, {1\over r},{1\over r},\bar q\right)\\
Z^\text{1-loop}_\text{equator}=& Z_\text{1-loop,eq}(a(E), i m_f, B)\,,
\end{aligned}
\eeq
with $Z_{\text{cl}}, {Z}_{\text{1-loop},\text{pole}}, Z_{\text{inst}}$ and $Z_\text{1-loop,eq}$ given in \rf{eq:class-contr}, \rf{loopcombination},  \rf{nekrinst} and \rf{1-loop-equator}.

In section \ref{sec:screening} we will identify further  non-perturbative corrections to this result arising due to monopole screening.

\subsection{Examples
}
%\VP{Either say that in this section $\ve =r  =1$, or restore powers of
%  $\ve$}

%The total one-loop factor is the product of the pole contribution
%(\ref{1-loop-poles}) and the equator contribution (\ref{1-loop-equator}).

The formulae  we have found for the one-loop determinants in the localization computation is valid for an arbitrary $\cN=2$ gauge theory on $S^4$ admitting a Lagrangian description.
Combining the contributions from the north pole, south pole and equator we get for a 't Hooft operator of magnetic weight $B$
\begin{eqnarray}
   Z_\text{1-loop,pole}\left(
ia-\frac B 2, im
\right)
   Z_\text{1-loop,pole}\left(
ia+\frac B 2, im
\right)
  Z_\text{1-loop,eq}
(ia, im)\,.
\label{1-loop-total}
\end{eqnarray}
The choice of gauge group $G$ and representation $R$ characterizing the  $\cN=2$ theory
 is encoded in the one-loop determinant formulae \rf{loopcombination} and \rf{1-loop-equator} in the choice of the root system $\{\alpha\}$, which characterizes the gauge group,    and of the weights $\{w\}$ of $R$.
Here we write explicitly these formulae for two simple $\cN=2$ gauge theories with $G=SU(N)$: $\cN=2^*$ and $\cN=2$ conformal SQCD. We also
consider $\cN=4$ super Yang-Mills, which is  a special case of  $\cN=2^*$.
From now on we  set $\ve=r=1$.

% For $\mathcal N=2^*$ with general gauge group,
% we have
% \begin{eqnarray}
%    Z_\text{1-loop,N}\left(
% \hat a, \hat m \right)
% =
%    Z_\text{1-loop,S}\left(
% \hat a,\hat m \right)
% =
% \left(
% \prod_{\alpha:\text{root}}
% \frac{
% G(\alpha\cdot \hat a)
% G(2+\alpha\cdot \hat a)
% }{
% G(1+\alpha\cdot \hat a+\hat m)
% G(1+\alpha\cdot \hat a-\hat m)
% }
% \right)^{1/2}\,,
% \end{eqnarray}
% and
% \begin{eqnarray}
%  Z_\text{1-loop,Eq}
% (\hat a, \hat m)  
% =
% \prod_{\alpha>0}
% \left[
% \frac{
% \displaystyle
% \sin
% \left(\pi \alpha \cdot \left(\hat a +\frac B 2
% \right)
% +\pi  \hat m
% \right)
% \sin
% \left(\pi \alpha \cdot \left(\hat  a+\frac B 2\right)
% -\pi \hat m
% \right)
% }{
% \displaystyle
% \sin^2
% \left(\pi \alpha \cdot \left(i a +\frac B 2
% \right)
% \right)
% }
% \right]^{\alpha\cdot B/2}\,.
% \end{eqnarray}

\subsubsection*{The $\cN=2^*$ $\SU(N)$ theory}

For this theory the hypermultiplet is in the adjoint representation and has mass $m$. We   parametrize
\beq
a=i\,{\rm diag}(a_1,\ldots, a_N)\,,
\eeq
with  $\sum_i a_i=0$.
The magnetic weight $B$ of   an arbitrary 't Hooft loop is 
\beq
B= i\,  \text{diag}(n_1,\ldots,n_N)-i 1_{N\times N}  {1\over N} \sum_i n_i  
\qquad  n_i \in \mathbb Z\,.
\eeq
Therefore, the pole one-loop contribution  \rf{loopcombination} is given by   
\begin{eqnarray}
\hspace{-5mm}
  Z_\text{1-loop,pole}(\hat a,\hat m)
% =
%   Z_\text{1-loop,pole}(\hat a,\hat m)
=
\left(
\prod_{i\neq j} 
\frac{
G(\hat a_i-\hat a_j)
G(2+\hat a_i-\hat a_j)
}{
G(1+\hat a_i-\hat a_j-\hat m)
G(1+\hat a_i-\hat a_j +\hat m)
}
\right)^{1/2}\,.
\label{n=2test}
\end{eqnarray}
Up to a phase, we  have for the equator one-loop contribution \rf{1-loop-equator} 
\begin{eqnarray}
&&  
\hspace{-10mm}
Z_\text{1-loop,eq}(ia, im,B)
\nonumber\\
&&
\hspace{-8mm}
=
\left(
\prod_{i<j}
\frac{
\sinh\left[\pi (a_i - a_j)-\pi m-  \pi i \frac{n_i-n_j}2\right]
\sinh\left[\pi (a_i - a_j)+\pi m-  \pi i \frac{n_i-n_j}2\right]
}{
\sinh^2
\left[\pi  (a_i-a_j) - \pi i \frac{n_i-n_j} 2
\right]
}
\right)^{\frac{|n_i-n_j|}2}
\hskip-10pt \,.
\end{eqnarray}

If we further restrict to the    special case of $G=SU(2)$, so that 
\beq
a=i\, {\rm diag}(a,-a)\qquad B=i\, {\rm diag}(p/2,-p/2)\,,
\quad 
\alpha=i\, {\rm diag}(1,-1)\,,
\eeq
%and since $\alpha=i\, {\rm diag}(1,-1)$, 
we have that $\alpha\cdot B\equiv -\Tr(\alpha B)=p$.
Here the new $a$ is a real number, and $p$ is a non-negative integer (it is twice the usual $SU(2)$ spin). The pole contribution \rf{1-loop-equator} is thus
%\begin{eqnarray}
 %Z_\text{1-loop,pole}(i a+p/4, im)
 % Z_\text{1-loop,pole}(i a-p/4, im)\,,
%\end{eqnarray}
%where
\beq
%\begin{eqnarray}
%&&
\hskip-8pt Z_\text{1-loop,pole}(\hat a, \hat m)
% =
%  Z_\text{1-loop,pole}(\hat a, \hat m)
% \nonumber\\
%&& \hspace{10mm}
=
\left(
\frac{
G(2\hat a)G(2+2\hat a)
G(-2\hat a)G(2-2\hat a)
}{
G(1+2\hat a+\hat m)G(1+2 \hat a-\hat m)
G(1-2 \hat a+ \hat m)G(1-2\hat a-\hat m)
}
\right)^{1/2}\hskip-3pt,
\label{SU2-1loop-pole}
%\end{eqnarray}
\eeq
while the equator
contribution  \rf{1-loop-equator} is
\begin{eqnarray}
  Z_\text{1-loop,eq}
(ia, im, p)=
\left\{
  \begin{array}{ll}
\displaystyle
\frac{\sinh^{p/2}[\pi (2 a+m)]
\sinh^{p/2}[\pi (2 a-m)]
}{
\sinh^p(2\pi a)
}
&\text{ for $p$ even,}
\\
\displaystyle
\frac{\cosh^{p/2}[\pi (2 a+m)]
\cosh^{p/2}[\pi (2 a-m)]
}{
\cosh^p(2\pi a)
}
&
\text{ for $p$ odd.}
  \end{array}
\right.
\label{Eq-SU2}
\end{eqnarray}

\subsubsection*{The $\CalN=4$ $\SU(N)$  theory}

We note that for the  $\cN=4$ super Yang-Mills theory, obtained by setting $m=0$ in the $\cN=2^*$ expressions,  the equatorial one-loop contribution  \rf{1-loop-equator}  becomes trivial for arbitrary gauge group $G$.
Furthermore, in  $\cN=4$ super Yang-Mills, the one-loop pole contribution \rf{loopcombination}  reduces to 
%the square root of 
the Vandermonde determinant corresponding to the gauge group $G$
\beq
\prod_{\alpha>0}\alpha\cdot \hat a\,.
\eeq
For $\cN=4$ super Yang-Mills, the one-loop factors trivialize. This result was already demonstrated in the perturbative computation of the 't Hooft loop path integral   in 
\cite{Gomis:2009ir} (see also \cite{Gomis:2009xg,Giombi:2009ek}).

\subsubsection*{Conformal SQCD}

This theory has gauge group $SU(N)$ and $N_\text{F}=2 N$
massive hypermultiplets in the fundamental representation of $SU(N)$ with masses $m_f$ with $f=1,\ldots, 2N$.
We are interested in the 't Hooft loop
specified by the magnetic weight
\beq
B= i\,  \text{diag}(n_1,\ldots,n_N) \qquad \sum_i n_i=0\,.
\eeq  
Dirac quantization requires that $n_i \in \mathbb Z$.
The one-loop pole contribution \rf{1-loop-equator} is given by  
\begin{eqnarray}
  Z_\text{1-loop,pole}(\hat a,\hat m_f)
=
\left(
\frac{
\prod_{i\neq j} 
G(\hat a_i-\hat a_j)
G(2+\hat a_i-\hat a_j)
}{
\prod_{f=1}^{N_\text{F}}
\prod_{i=1}^N
G(1+\hat a_i-\hat m_f)
G(1-\hat a_i+\hat m_f)
}
\right)^{1/2}\,.
\end{eqnarray}
Up to a phase, the equatorial one-loop  contribution  \rf{1-loop-equator}  is given by
\begin{eqnarray}
 Z_\text{1-loop,eq}(ia, im_f,B)=
\frac{
\prod_{f=1}^{2N}
\prod_{j=1}^N 
\left(\sinh\left[
\pi a_j
-\pi m_f
-  \pi i \frac{n_j}2
\right]
\right)^{|n_j|/2}
}{
\prod_{i<j}
\left(
\sinh
\left[\pi  (a_i-a_j) - \pi i \frac{n_i-n_j} 2
\right]
\right)^{|n_i-n_j|}
}
\,.
\end{eqnarray}
As in (\ref{Eq-SU2}), each $\sinh$ becomes $\cosh$
when $n_j$ in the numerator or  $n_i-n_j$ in the denominator
is odd.

Specializing further to $G=SU(2)$, we have $N_\text{F}=4$ fundamental hypermultiplets.
With the same parametrization as in the $\mathcal N=2^*$ case,
$p=2n$ needs to be even for Dirac quantization.
Up to a phase, the one-loop factor   \rf{loopcombination} is
\beq
\begin{aligned}
&
Z_\text{1-loop,pole}(\hat a,\hat m_f)
\\
&
\hspace{5mm}
=
\left(
\frac{
G(2\hat a)G(2+2\hat a)
G(-2\hat a)G(2-2\hat a)
}{
\prod_{f=1}^{4}
G(1+\hat a-\hat m_f)
G(1-\hat a-\hat m_f)
G(1-\hat a+\hat m_f)
G(1+\hat a+\hat m_f)
}
\right)^{1/2}
\end{aligned}
\eeq
for the north and south poles,
and
\begin{eqnarray}
  Z_\text{1-loop,eq}
(ia, im_f, 2n)=
\left\{
  \begin{array}{ll}
\displaystyle
\frac{\prod_{f=1}^4 
\sinh^{n/2}[\pi (a+m_f)]
\sinh^{n/2}[\pi (a-m_f)]
}{
\sinh^{2n}(2\pi a)
}
&\text{ for $n$ even,}
\\
\displaystyle
\frac{\prod_{f=1}^4 
\cosh^{n/2}[\pi (a+m_f)]
\cosh^{n/2}[\pi (a-m_f)]
}{
\sinh^{2n}(2\pi a)
}
&
\text{ for $n$ odd}
  \end{array}
\right.
\label{Eq-NF=4}
\end{eqnarray}
for the equator.
%\TO{Need to make consistent the notation}

%\subsection{Comparison with Toda and Liouville Theories}
 %
  We note that for real values of $a_i$ and $m_f$, one encounters
no branch point upon integrating over $a_i$ in \rf{finfin}.
When the exponent of a $\sinh$ is a half-odd integer,
the $\sinh$ actually becomes a $\cosh$ and has no zero.

\section{Non-Perturbative Effects of Monopole Screening
}\label{sec:screening}

\subsection{Physical Picture of Monopole Screening}

In the absence of a 't Hooft loop, $Q$-invariance requires the curvature
$F$ to vanish everywhere on $S^4$, except at the north and south poles.%
\footnote{%
This is a special case of the completeness statement
in section \ref{sec:vanishing-theorem}.
}
If we allowed only smooth configurations, we would conclude
that only trivial gauge field configurations contribute.
%the theory receives non-perturbative corrections 
% from field configurations that are singular at the 
% north and south poles of $S^4$.
As shown in \cite{Pestun:2007rz}, however,
localization permits instanton corrections at the north and south poles,
which are precisely captured
by the Nekrasov partition function.
% From the vanishing theorem discussed in section \ref{sec:one-loop},
% one might conclude that there are no saddle points other than the original background configuration (?).  

One can argue in two steps  that such corrections are necessary 
\cite{Pestun:2007rz}.
First, $Q$-invariance requires that
the field strength $F$ vanish only away from the north and south poles.
If singular configurations arise as a limit of smooth configurations,
there can be contributions to the path integral 
localized at the poles. Second, the localization Lagrangian $Q\cdot V$ in the neighborhood of the poles is approximately that of the twisted $\mathcal N=2$ Lagrangian in the $\Omega$-background in $\mathbb R^4$ with the specific values of the equivariant parameters $\ve_1=\ve_2=1/r$.
The approximation becomes exact at the poles.  
Building  on the earlier work \cite{Moore:1997dj,Losev:1997tp,Lossev:1997bz}, Nekrasov showed that the path integral  of such a theory computes
the equivariant integral of certain differential forms defined on the
instanton moduli space \cite{Nekrasov:2002qd}.
  The integral can be computed by a localization formula as a sum over fixed points.  The fixed points in the moduli space of instantons   indeed correspond to gauge field configurations that are non-trivial only at the origin of $\mathbb R^4$.
We studied these instanton corrections
in the presence of a 't Hooft loop in section \ref{sec:inst} and found that the
instanton contributions are given by the Nekrasov partition function at the north and south poles with
its arguments shifted  due to the insertion of the 't Hooft operator at the equator.

In this section we study another type of non-perturbative
corrections due to the screening of magnetic charge
associated to a 't Hooft operator.
We begin by explaining how such corrections arise
in our localization framework.

% The vanishing theorem proved in () implies that 
% smooth fields satisfying the $Q$-invariance equation (?)
% must be given by the 't Hooft background ().
% This theorem, however, assumes that the fields are all smooth
% with the boundary condition ().
% Just as in the case of instanton corrections,
% we should be careful about the possible contributions
% that exhibit singular behaviors%
% in the infinitesimal
% neighborhood of the loop.

% \noindent
% \underline{Approximate solutions as fixed points on the monopole moduli space}

\subsubsection*{Monopole Screening}

As we showed in section \ref{sec:vanishing-theorem}
the only possible field configurations
that can contribute to the path integral  are
those of the form (\ref{dyonic}) in the bulk of $S^4$.
They are only allowed to deviate from (\ref{dyonic}) 
in an infinitesimal neighborhood of either the poles or 
the equator.
The deviations at the poles were considered in \cite{Pestun:2007rz} 
and have been reviewed in section \ref{sec:inst};
they are the small instanton solutions of the anti-self-dual/self-dual equations that approximate the $Q$-invariance equations
near the poles.
In the neighborhood of the loop, we saw in section \ref{sec:Hooft} that the $Q$-invariance equations
are  approximately the Bogomolny equations.
Therefore
we should study the monopole moduli space and 
look for the analog of small instanton field configurations.  

The monopole moduli space $\mathcal M_{\rm mono}$ relevant for us is the space of solutions on $\mathbb R^3$, up to gauge transformations, to the Bogomolny equations with a prescribed singularity at the origin corresponding to the insertion of a 't Hooft operator.  Since we are only interested in the behavior at the origin, the boundary condition at the infinity of $\mathbb R^3$ is irrelevant.  It is simplest to consider the solutions that have a vanishing Higgs expectation value at infinity.
The vanishing Higgs vev will allow us to use the 
ADHM construction of instantons to describe the 
monopole moduli space in section \ref{sec:ADHM}.%
\footnote{%
When the gauge group $G$ is a classical group
the moduli space can be constructed using the ADHM construction.
In this paper we focus on the case where $G$ is $U(N)$ or $SU(N)$.}
We will describe the moduli space explicitly in the case $G=SU(2)$.
For the moment we proceed assuming that $G$ is a general Lie group.

The magnetic charge of the singular monopole configuration created by the 't Hooft operator is specified by 
a coweight $ w\equiv B$.
 Generally, smooth monopoles that surround
the singular monopole  screen its  magnetic charge so that the asymptotic
behavior of the fields at infinity is that of the background
configuration  (\ref{dyonic})  with the coweight $w$ replaced by a smaller
coweight $v$. 
This is because the magnetic charge of a   smooth monopole
is labeled by a
coroot of $G$, and can screen the charge of the singular
monopole by that amount.
The coweight $v$ is such that its corresponding weight 
appears in
the irreducible ${}^L G$-representation specified by the highest weight 
corresponding to $w$.
In the terminology of \cite{Kapustin:2006pk}, 
such $v$ is said to be associated to $w$.
One can show that the magnetic charge $v$ seen at infinity must have
a smaller norm than $w$ by applying a method similar to the one we used
to prove completeness of solutions in section \ref{sec:vanishing-theorem}.%
\footnote{%
The difference $w\cdot w-v\cdot v$ can be expressed
in terms of the integral of the instanton density,
upon lifting the field configuration to instantons in $\mathbb C^2$
using Kronheimer's correspondence explained in appendix \ref{sec:Kronheimer}.
}

 Denoting by $\mathcal M(w; v)$ 
the moduli space of solutions whose asymptotic magnetic charge is given
by $v$, we have that the relevant moduli space to consider is 
\begin{eqnarray}
\mathcal M_{\rm mono}(w)=\bigcup_{v}   
\mathcal M( w; v)\,,
\label{M-direct-sum-sing}
\end{eqnarray}
where the union is over coweights $ v$ such that (if we identify
coweights with weights using a metric) $v$ is a weight that appears in the highest weight representation specified by $w$.

The spaces $\mathcal M_{\rm mono}( w)$ and $\mathcal M(w;v)$ 
are in general singular.
To understand the nature of the singularities in these spaces,
it is useful
to recall the situation with instantons.
The Uhlenbeck compactified instanton moduli space $\mathcal M_{\rm
  inst}$  \cite{Uhlenbeck:1982zm}  is singular 
due to instantons of zero size.
For $G=U(N)$ the moduli space $\mathcal M_{\rm
  inst}$ can be conveniently resolved by turning on a Fayet-Illiopolous
parameter for the real ADHM equation. 
The resolved space $\overline\CalM_\text{inst}$ is known to be isomorphic
to the moduli space of
non-commutative instantons \cite{Nekrasov:1998ss}, or 
the Gieseker resolution \cite{MR466475}
in terms of torsion free sheaves. 
% The resolved space can be described concretely in the ADHM construction
% %
% \footnote{%
% For other gauge groups a similar compactification is not known.
% %\TO{Nekrasov and Shadchin  in p.3 of their paper 
% %may be saying  that there cannot be such a resolution.
% %What is the precise statement?}
% In practice, for classical groups one can apply a pole prescription
% to the contour integrals that arise from the ADHM construction.
% This procedure circumvents the absence of appropriate regularization
% \cite{Nekrasov:2004vw}.
% }
%\VP{We don't need classical groups in this work, I removed footnote}

As explained in \cite{Kapustin:2006pk}, a natural resolution
$\overline{\mathcal M}_{\rm mono}$ 
of the moduli space of monopoles with a monopole singularity labeled by $w$ 
at the origin involves
all the coweights $w'$ associated to the coweight $w$.
%that appears in the definition (\ref{scalarpole}) of the 't Hooft operator.
The coweights $ w'$ represent the magnetic charge
at the origin
reduced by the smooth monopoles that are attracted to 
the singular monopole there.
This effect was called {\it monopole bubbling} in
\cite{Kapustin:2006pk}.
% \footnote{%
% The term  ``monopole bubbling'' in \cite{Gomis:2009ir}
% refers to the sum over $v$
% in (\ref{M-direct-sum-sing}) rather than the sum over $w'$.
% }
This means that 
a natural resolution
$\overline{\mathcal M}( w; v)$
of a component in (\ref{M-direct-sum-sing})
contains smaller moduli spaces in its boundary,
\begin{eqnarray}
\partial\overline{\mathcal M}(w; v)
\supset 
\bigcup_{w'}
\overline{\mathcal M}( w'; v)
\,,
\end{eqnarray}
with $ w'$ being the coweights such that $w'$ is associated to $w$ while $v$ is associated to $w'$.
%\TO{One could give a more precise relation that involves equality
%rather than inclusion, but to be on the safe
%side, I chose to write this less precise version of the relation.}
In the case $G = U(N)$, one can see this structure explicitly using the ADHM
construction.
Thus we have the resolution of the whole moduli space
\begin{eqnarray}
\overline{\mathcal M}_{\rm mono}( w)=\bigcup_{v}   
\overline{\mathcal M}(w;v)\,,
\label{moduli-decomp-bar}
\end{eqnarray}
where the union is over the coweights $v$ associated to $w$.

We only need to study the neighborhood of the monopole bubbling locus,
 where all the smooth monopoles are almost on top
of the singular monopole, because only these would be the approximate 
solutions to our genuine $Q$-invariance equation. 
%\footnote{%
For
example, for gauge group $\SU(2)$  and for  a 't Hooft operator with $w=(1,-1)$  (spin 1)
%($\Leftrightarrow p=2$) 
and $v=(0,0)$ (spin 0)
%($\Leftrightarrow q=0$), 
the bubbling locus is 
the zero-section $\mathbb{ P}^1$ in the resolved $A_1$ space
$\overline{\mathcal M}( w;  v)=
T^* \mathbb P^1$ (see section \ref{sec:SU2ex} for details).
%}
%\TO{We can quantify this by computing the action.}
Because $Q$-invariance implies in particular the invariance under the $U(1)_{J+R}$
generated by  $Q^2$ we are only interested in the $U(1)_{J+R}$ fixed points.
%\footnote{%
Such fixed points are necessarily in the bubbling locus
because when lifted by one dimension so that monopoles become instantons, 
the fixed points of the $U(1)_{J+R}\times U(1)_K$ action
% one $U(1)$($=U(1)_K$) 
%used for dimensional reduction, the other ($U(1)_{J+R}$)
%generated by $Q^2$  
sit in the small-instanton locus; see section \ref{sec:ADHM}.
%} 
%Since the only non-trivial part of the gauge field is at the singularity, they are solutions to the deformed Bogomolny equation that arise as the $Q$-invariance equation.  
Thus these fixed points represent 
all the subleading saddle points of the original gauge theory path integral.
Upon evaluating the path integral, 
we need to {\it sum over the fixed points}.

At each fixed point of $\overline{\mathcal{M}}(w;v)$, we need to compute the fluctuation determinants.
The common factor that appears for fixed magnetic weight $v$ was computed in
section \ref{sec:one-loop-eq},
where it was called  $Z_\text{1-loop,eq}(ia,im_f, v)$. Let us denote by 
$Z_{\text{mono}}(i a,i m_f; w; v)$
the sum of contributions from fluctuations
at the fixed points in a single component $\overline{\mathcal M}( w;
 v)$ 
divided by $Z_\text{1-loop,eq}(i a, i m_f,v)$.
The function $Z_{\text{mono}}(i a,i m_f; w; v)$ is the monopole analog of the
Nekrasov instanton partition function,
whose computation is reviewed in appendix \ref{sec:Nek-func}
from a related point of view.

Therefore, given the decomposition of the moduli space in 
(\ref{moduli-decomp-bar}),  the expectation value of the 't Hooft loop operator
with magnetic charge $B=w$
takes the form
\begin{eqnarray}
 \hspace{-2mm}
  && \langle T(w)\rangle
 =
 \int da
\sum_{v}
% Z_\text{1-loop,pole}\left(ia-\frac v 2\right)
% Z_\text{1-loop,pole}\left(ia+\frac v 2\right)
%Z_{\text{1-loop}}(a,m;w; v)
Z_\text{mono}(i a, i m_f;w,v)
Z_\text{1-loop,eq}(i a, i m_f,v)
\nonumber\\
&&
%\hspace{10mm}
\times
\left|
Z_\text{cl}\left(ia
 -\frac{v}{2r},q
\right)
Z_\text{1-loop,pole}\left(
i a-\frac v {2r},im_f\right)
Z_\text{inst}\left(
i a
-\frac{v}{2r},{1\over r}+im_f,{1\over r},{1\over r},q
\right)
\right|^2
\,.
\label{eq:vev-decomposition}
\end{eqnarray}
%\TO{The first line is guessed using S-duality,
%but should be derived in an earlier section for $\theta\neq 0$.}
Except $Z_{\text{mono}}(i a,i m_f; w; v)$,
all the expressions in the integrand
were already calculated in the previous sections.
Our remaining task is to compute this factor.

\subsection{ADHM Construction of the Monopole Moduli Space}
\label{sec:ADHM}

% \TO{Compared with version 62, I made the replacement
% $
% %\bar z_1\rightarrow w_1, z_2\rightarrow w_2,
% \psi \rightarrow -\psi-\varphi, 
% %A\rightarrow -A,
% %\mathcal A\rightarrow -\mathcal A,
% \Phi\rightarrow -\Phi,
% B\rightarrow -B,
%  \vartheta
% \rightarrow
% \eta
% $ for the latitude, $\vartheta\rightarrow \nu$
% for the $U(1)_K$ weight. (Dec 28)
% }

To perform explicit calculations %based on the analysis
%in section \ref{sec:one-loop-screening}, 
we need an efficient way to 
describe
the monopole moduli space $\overline{\mathcal M}_\text{mono}$.
The  connection between monopoles and instantons
\cite{Kronheimer:MTh} reviewed in appendix \ref{sec:Kronheimer},
combined with the ADHM construction of instantons \cite{Atiyah:1978ri},
 provides a useful method to manipulate
the  monopole moduli space.

%Next we
Let us briefly review the ADHM construction of instantons in $\mathbb C^2$.
For simplicity we will take the gauge group $G$ to be $U(N)$.
The basic data in the construction are encoded in the complex
\begin{equation}
  0\rightarrow {\mathcal H} \stackrel{\alpha(z)}\rightarrow
{\mathcal H}\otimes U \oplus E_\infty \stackrel{\beta(z)}\rightarrow
{\mathcal H}\otimes \wedge^2 U \rightarrow 0\,,
\label{ADHM-complex}
\end{equation}
where ${\mathcal H}\simeq {\mathbb C}^k,
U\simeq {\mathbb C}^2,
E_\infty \simeq {\mathbb C}^N$.
On $U$ the $U(1)_K$ acts as $(z_1, z_2) \mapsto (e^{i\nu} z_1, e^{-i\nu}
z_2)$.
The $z$-dependent maps $\alpha(z)$ and $\beta(z)$ are given by
\begin{equation}
  \alpha(z)=
\left(  \begin{array}{c}
    z_2-B_2\\
-z_1+B_1\\
-J
  \end{array}
\right)\,,
~~~~
\beta(z)=(z_1-B_1,z_2-B_2,-I)\,,
\end{equation}
and their cohomology $E_z={\rm Ker}\,\beta(z)/{\rm Im}\,\alpha(z)$
is identified with 
the fiber of the gauge bundle (in the fundamental representation).
We are particularly interested in $E_{z=0}$ since it
encodes the singularity of the 't Hooft loop.
The $U(1)_K$ action on a vector space $V$ is specified by the character
$\chi(V)$, which is a Laurent polynomial of
$e^{i\nu}\in U(1)_K$.
The 't Hooft loop with charge $ w=i{\rm diag}(p_1,\ldots, p_N)$
corresponds to the case
\begin{equation}
  \chi(E_0)=\sum_{i=1}^N e^{i p_i\nu}.
\end{equation}
The $U(1)_K$ action on $(z_1,z_2)$ implies that
\begin{equation}
\chi(U)=e^{i\nu}+e^{-i\nu}\,.
\end{equation}
The characters of $\mathcal H$ and $E_\infty$ take the form
\begin{eqnarray}
  \chi(\mathcal H)=\Tr e^{i K\nu}\,,~~~~~~~~~~~~~~
\chi(E_\infty)=\Tr e^{i M\nu}\,,
\end{eqnarray}
where $K$ is a diagonal $k\times k$ matrix and
$M={\rm diag}(q_1,\ldots, q_N)$ is a diagonal $N\times N$ matrix related to
the coweight $ v=i(q_1,\ldots, q_N)$
 corresponding to the magnetic charge
at infinity.
Both $K$ and $M$ have integer entries that we choose
to be in the descending order.
The characters of various spaces are related as
\begin{equation}
  \chi(E_0)=\chi(E_\infty)+(\chi(U)-2)\chi({\mathcal H})\,,
\label{eq:char-relation}
\end{equation}
For given $ w=i(p_1,\ldots, p_N)$, the choice of
$K$ and $M$ is not necessarily 
unique, but we have the non-trivial condition that
the whole right hand side of
(\ref{eq:char-relation}) has only positive coefficients.

The moduli space $\mathcal M(w; v)$ is given as
 a hyperK\"ahler quotient of the space of $U(1)_K$-invariant ADHM data
$(B_1,B_2,I,J)$.
The action of $U(1)_K$ on $(B_1,B_2,I,J)$
can be read off from the complex (\ref{ADHM-complex})
and the action on $(z_1,z_2)$.
In the usual ADHM construction of
the instanton moduli space, we take a quotient 
by a certain action of the $U(k)$ group.
This action of $U(k)$ on the ADHM data
is induced from its natural action on $\mathcal H\simeq \mathbb C^k$.
The choice of $K$ breaks the  $U(k)$ symmetry
into the commutant subgroup $\prod_r U(k_r)$, where
$k=\sum_r k_r$ and $k_r$ is the number of
entries of the $r$-th largest integer in the diagonal of $K$.
% The space $\mathcal H\simeq \mathbb C^k$ decomposes
% as $\mathcal H=\oplus_r \mathcal H_r$ 
% with each $U(k_r)$ acting on
% $\mathcal H_r\simeq \mathbb C^{k_r}$.
% The ADHM data $B_1, B_2, I,J$
% decompose into elements of ${\rm Hom}$'s
Thus the moduli space is given as the hyperK\"ahler quotient
 \begin{eqnarray}
\mathcal M( w, v)=
\left   \{
 (B_1,B_2,I,J)
 \left|
\begin{array}{ccc}
   B_1+[K,B_1]&=&0\\
 -  B_2+[K,B_2]&=&0\\
 K I - I M&=&0\\
 M J - J K&=&0\\
 \end{array}
\right.
\right\}
{\Big /}{\Big /}{\Big /}
\prod_r U(k_r)\,.
 \end{eqnarray}
The hyperK\"ahler quotient denoted by ``$///$''  can be implemented
by imposing the ADHM equations
\begin{eqnarray}
  \mu_{\mathbb C}&\equiv& [B_1,B_2]+IJ=0\,,
\label{eq:complex-ADHM}
\\
\mu_{\mathbb R}&\equiv& [B_1^\dagger, B_1]+
[B_2^\dagger, B_2]+
II^\dagger-J^\dagger J=0
\end{eqnarray}
and then considering the solutions up to the action of $\prod_r U(k_r)$.
Or alternatively, if we are only interested in the complex structure,
we can drop the real equation $\mu_{\mathbb R}=0$ and divide
by the complexified group $\prod_r U(k_r)_{\mathbb C}$.
A resolution $\overline{\mathcal M}( w, v)$
%\TO{Less important: this may not be unique?}
of the moduli space
can be achieved by setting $\mu_{\mathbb R}$ to a non-zero
constant matrix instead of requiring it to vanish.

The $U(1)_{J+R}$-fixed points can be found by demanding that
for any $  e^{i\ve} \in  U(1)_{J+R}$ there
exists $e^{\phi}\in \prod_r U(k_r)$ such that%
\footnote{%
In our convention the $U(1)_{J+R}$ acts both on $I$ and $J$ as $e^{i\ve/2}$,
implying that it also acts on $E_\infty$ as $e^{i\ve/2}$.
% \TO{Less important: we could have chosen 1 for $I$ and $e^{i\ve}$ for $J$, but our 
% choice seems actually necessary in applying
% the Nekrasov-Shadchin prescription.
% It would be nice to understand this point better.}
}%
% \footnote{%
% \label{foot:lie-eigen}
% In our notation an element of a Lie algebra
% is an anti-hermitian matrix, 
% and its eigenvalues are expressed as  real variables
% multiplied by $i$: for example, $\phi={\rm diag}(i\phi_1,\ldots, i\phi_k)$.
% We sometimes abuse the notation as in $a={\rm diag}(ia,-ia)$ for
% $G=SU(2)$. \VP{Need to combine this footnote and \ref{sec:lie-alg-conv}}
% }
\begin{eqnarray}
 e^{i\ve/2} e^{\phi}B_se^{-\phi}
&=&B_s\,,~~~~~~~~~~~ s=1,2\,,
\nonumber\\
e^{i\ve/2} e^{\phi} I &=&I\,,
\label{eq:fixed-point-eq}
\\
e^{i\ve/2} J e^{-\phi}&=&J\,.
\nonumber
\end{eqnarray}
% \TO{The sign of $i\ve$ for $I$ and $J$ disagree with (3.7) of hep-th/0206161
% as well as (4.1) and (4.7) of hep-th/0502180.
% But I hope I'm not crazy, 
% their formulas are incompatible with the ADHM equation $[B_1, B_2]+IJ=0$.
% Our choice of signs agrees with p.12 of hep-th/0211108.
% }\VP{I agree, Shadchin has wrong signs there}

By construction,
the fixed points of $U(1)_{J+R}$ in $\overline{\mathcal M}( w; v)$
 automatically correspond to the fixed points of $U(1)_K\times U(1)_{J+R}$
in the instanton moduli space.
The fixed points in the instanton moduli space
were classified in \cite{MR1711344}, and they were found to sit on the boundary
components of the moduli space
corresponding to small instantons.
This in turn implies that the $U(1)_{J+R}$-fixed points on the monopole
moduli space sit on the bubbling locus.
We also know from the experience with instantons that
the fixed points of  $U(1)_K\times U(1)_{J+R}\times G\times G_{\text F}$
coincide with the fixed points
of $U(1)_K\times U(1)_{J+R}$.
% \TO{Less important: if there is a quick proof without referring to instantons
%  we might give it here,
% otherwise we'd leave it as it is.}

At each fixed point, the ratio
$Z_{\text{1-loop}}( w; v)/Z_{\text{1-loop}}( v; v)$ 
can be calculated
from
 the weights of the equivariant group action
on 
the tangent space and the Dirac zeromodes.
The ADHM construction provides  a concrete procedure
to derive such weights.

The tangent space can be described by considering the
linearization of the ADHM system.  Namely, let us consider
the complex
\begin{eqnarray}
&&  0\rightarrow
\left\{\delta\phi\in \text{Lie}\left(\prod_r U(k_r)_{\mathbb C}
\right)\right\}
\stackrel{h_1}{\rightarrow}
\left   \{
 (\delta B_1,\delta B_2,\delta I,\delta J)
 \left|
\begin{array}{ccc}
   \delta B_1+K \delta B_1-(\delta B_1) K&=&0\\
 - \delta B_2+K \delta B_2-(\delta B_2) K&=&0\\
 K \delta I - (\delta I) M&=&0\\
 M \delta J - (\delta J) K&=&0\\
 \end{array}
\right.
\right\}
\nonumber\\
&&\hspace{15mm}
\stackrel{h_2}{\rightarrow} \{
\delta \mu_{\mathbb C}
\in\text{End}\,\mathcal H
\otimes \wedge^2 U\,|\, [K,X]=0\}
\rightarrow 0\,,
\label{eq:ADHM-linearized-complex}
\end{eqnarray}
where the two maps $h_1$ and $h_2$ are the linearizations 
of the $\prod_r U(k_r)_{\mathbb C}$ transformation
and the complex ADHM equation $\mu_{\mathbb C}=0$:
\begin{eqnarray}
  h_1(\delta\phi)&=&
((\delta \phi) B_1-B_1\delta\phi,
(\delta \phi) B_2-B_2\delta\phi, (\delta\phi) I, -J\delta \phi
)\,,
\nonumber\\
h_2 (\delta B_1,\delta B_2,\delta I,\delta J)
&=&[\delta B_1, B_2]+ [B_1,\delta B_2]
+(\delta I) J+I\delta J\,.
\end{eqnarray}
The tangent space of the moduli space at the point $(B_1,B_2,I,J)$
is given by the cohomology $\ker h_2/{\rm im}\, h_1$.
The fixed-point equations for the
action of
$ U(1)_{J+R}\times G
\times\prod_r U(k_r)$
\begin{eqnarray}
  i\frac \ve 2 B_s+[\phi, B_s]&=&0\,,~~~~~~~~~~~~~~s=1,2\,,
\nonumber\\
i\frac \ve 2 I+\phi I-I \hat a&=&0\,,
\label{eq:fixed-point-eq-extended}
\\
i\frac \ve 2 J-J\phi+\hat a J&=&0\,,
\nonumber
\end{eqnarray}
determine $\phi$ as a function of $\ve$ and $\hat a$,
{\it i.e.},
they define a homomorphism
$U(1)_{J+R}\times G
\rightarrow\prod_r U(k_r)$
 at each fixed point.
Thus we have an action of $U(1)_{J+R}\times G$ on the complex
(\ref{eq:ADHM-linearized-complex}), 
and the character on the tangent
space is given as
\begin{eqnarray}
 - \Tr_{V_1}(g)
+ \Tr_{V_2}(g)
 - \Tr_{V_3}(g)\,,
\label{eq:character-tan}
\end{eqnarray}
where $V_1,V_2,V_3$ are the three vector spaces that appear
in (\ref{eq:ADHM-linearized-complex}) and $g=(e^{i\ve}, e^{i\hat a})
\in
U(1)_{J+R}\times G$.%
% \footnote{%
% The quantity (\ref{eq:character-tan}) is an analog of
% $\text{ind}(\bar D)_\text{inst}$ in (\ref{ind-Dbar-inst}).
% The argument in appendix \ref{sec:Nek-func} implies that, 
% the difference between the finite part of
% $\text{ind}(D_\text{Bogo})$ and twice  (\ref{eq:character-tan}) 
% does not affect the final result $Z_\text{mono}(w,v)$.
% }
% For the hypermultiplet zero-modes in the adjoint representation,
% we can use the relation between the 

There is another method, heuristic but efficient, which can be used to compute the
weights on the tangent space based on the character on the space of holomorphic functions.
It is best explained in the example we consider next.

\subsection{Example: $SU(2)$ $\mathcal N=2^*$ Theory}

%\subsubsection{$SU(2)$ $\mathcal N=2^*$ Theory}
\label{sec:SU2ex}

% For $SU(2)$,
% it seems that for a given magnetic charge $p$ 
% of the 't Hooft loop (this implies that $\chi(E_0)=X^{p/2}+X^{-p/2}$),
% there is a unique choice of $\chi(E_\infty)$ and $\chi({\mathcal H})$
% that have only terms with non-negative coefficients:
% \begin{equation}
%   \chi(E_\infty)=X^{\frac{p-2}2}+X^{-\frac{p-2}2}\,,~~
% \chi({\mathcal H})=X^{\frac{p-2}2}
% +X^{\frac{p-4}2}+\ldots
% +X^{-\frac{p-2}2}\,.
% \end{equation}
% Four-dimensional instantons descend to
% three-dimensional monopoles.
% Thus monopole bubbling occurs at a small instanton singularity, {\it i.e.}, when
% the complex (\ref{ADHM-complex}) splits.

For $G=SU(2)$, we can label the coweights with integers (corresponding to twice the spin).
Also we slightly modify the ADHM construction
above and allow $ w,  v$ and $K$ to have
half odd integers.
We define the integers $p\geq 0$ and $q$ by%
\footnote{%
Denoting reduced magnetic charge by $q$ should not confusion
with the instanton parameter $e^{2\pi i\tau}$ as the latter
does not appear in this subsection.
}
\begin{eqnarray}
   w=i
(p/2, -p/2)\,,~~~~~~~~~~
 v=i(q/2, -q/2)\,.
\end{eqnarray}
Since $ v$ is associated to $ w$,
$p-q$ is non-negative and even.
The constraint  (\ref{eq:char-relation}) then implies that
$k=p-1$
and also that
\begin{eqnarray}
\chi(\mathcal H)
&=&%\frac{(e^{i\frac{p}2 \vartheta}-e^{i\frac{q}2 \vartheta})
%(1-e^{-i \frac{p+q}2\vartheta})
\frac{
e^{i\frac{p}2 \nu}+e^{-i\frac{p}2 \nu}
-e^{i\frac{q}2 \nu}-e^{-i\frac{q}2 \nu}
}{(e^{i\frac 1 2\nu}-e^{-i\frac 1 2\nu})^2}\,.
\label{eq:chiH}
\end{eqnarray}
As a character, 
$\chi(\mathcal H)$ 
is a polynomial
with positive coefficients for
$-p\leq q\leq p$.
For ease of writing
we will assume $q\geq 0$,
and sum over $q<0$ in the end
remembering that the Weyl group acts as $q\rightarrow -q, 
\hat a\rightarrow -\hat a$.
We can then write 
%(\ref{eq:chiH}) as
\begin{eqnarray}
e^{i K\nu}
=\chi(\mathcal H)
&=&
e^{i(\frac p 2 -1)\nu}+\ldots
+\frac{p-q}2  e^{i\frac q 2 \nu}
+\ldots
+\frac{p-q}2  e^{-i\frac q 2 \nu}
+\ldots
+e^{-i(\frac p 2 -1)\nu}\,,
%   K={\rm diag}\left(\frac{p-2}2,
% \frac{p-4}2,
% \ldots,
% \frac{-(p-2)}2
% \right)\,.
\end{eqnarray}
where in the last expression
the coefficient of the exponential increases from $1$ to
$\frac{p-q}2$  monotonically, stays constant,
and then decreases monotonically to 1.

In order to illustrate the analysis, we start with the simplest non-trivial
case that involves monopole screening,
namely $ w =i(1,-1),  v =(0,0)$
corresponding to $p=2, q=0$.
We now explicitly work out the details of calculations
involving $\overline{\mathcal M}(2;0)$.
In this case the constraint
(\ref{eq:char-relation}) is solved by
\begin{eqnarray}
  \chi(E_0)=e^{i\nu}+e^{-i\nu}\,,
~~~~~~~~~~~
\chi(E_\infty)=2\,,
~~~~~~~~~~~
  \chi(\mathcal H)=1\,.
\end{eqnarray}
Let us write $B_1=(b_1), B_2=(b_2), I= (i_1,i_2), J=(j_1,j_2)^T$.
The non-trivial 
$U(1)_K$  action is given by $b_1\rightarrow e^{-i\nu} b_1,
b_2\rightarrow e^{i\nu} b_2$.
Thus a $U(1)_K$ invariant instanton
has to be centered at the origin, {\it i.e.}, $b_1=b_2=0$.
The remaining variables satisfy the ADHM equations
\begin{eqnarray}
  i_1j_1+i_2j_2&=&\xi_{\mathbb C}\,, \label{ADHM-ex-complex}
\\
|i_1|^2+|i_1|^2-|j_1|^2-|j_2|^2&=&\xi_{\mathbb R}\,,
\end{eqnarray}
and are subject to the $U(k)=U(1)$ equivalence relation
\begin{eqnarray}
  (i_1,i_2,j_1,j_2)\sim (e^{i\phi} i_1, e^{i\phi}i_2,
e^{-i\phi} j_1, e^{-i\phi}j_2)\,.
\label{eq:2to0equiv}
\end{eqnarray}
We have introduced the deformation parameters $\xi=(\xi_{\mathbb C}, \xi_{\mathbb R})$.
%In a commutative gauge theory $\xi=0$, and we give the explicit solution that follows from the ADHM procedure in appendix \ref{sec:explicit-soln}.

The moduli space $\mathcal M(2;0)$ can be smoothed by turning on $\xi$.
Using a hyperK\"ahler rotation we can set $\xi_{\mathbb C}=0$ and $\xi_\mathbb{R}>0$.
Then $(i_1,i_2)$ cannot vanish.
The equation (\ref{ADHM-ex-complex}) can be solved
by introducing a charge-$(-2)$ variable $\mu$ via $(j_1,j_2)=
\mu(i_2,-i_1)/\sqrt{|i_1|^2+|i_2|^2}$.
We see that $(i_1,i_2,\mu)$ are essentially the  variables
for $T^* \mathbb P^1$ that appear in the
gauged linear sigma model description \cite{Witten:1993yc}.

The fixed points of the $U(1)_{J+R}\times G$ action are found by demanding that the ADHM data are
invariant up to (\ref{eq:2to0equiv}):
\begin{eqnarray}
(i_1,i_2, j_1, j_2)=(
e^{i(\phi+\frac\ve 2-\hat a)}i_1,
e^{i(\phi+\frac \ve 2+\hat a)}i_2,
e^{i(-\phi+\frac \ve 2+\hat a)}j_1,
e^{i(-\phi+\frac \ve 2-\hat a)}j_2
)\,,
\end{eqnarray}
where $e^{i\ve}\in U(1)_{J+R}$ and ${\rm diag}(e^{i\hat a},e^{-i\hat a})\in G=SU(2)$.
We find two fixed points $p_1$ and $p_2$:
\begin{equation}
\begin{array}{llllllll}
p_1:& i_2\neq 0\,,& i_1=j_1=j_2=0\,, &\phi=-\hat a-\frac \ve 2\,,
\\
p_2: &i_1\neq 0\,, &i_2=j_1=j_2=0\,, &\phi=\hat a-\frac \ve 2\,.
\end{array}
\end{equation}
At each fixed point, we have the complex (\ref{eq:ADHM-linearized-complex})
with the vector spaces
$  V_1=\{\delta\phi\}\simeq \mathbb C,
V_2=\{(\delta i_1,\delta i_2,\delta j_1,\delta j_2)\}\simeq \mathbb C^4,
V_3\simeq \mathbb C$ representing the tangent space.
The weights of
$U(1)_{J+R}\times G$
are given by
\begin{eqnarray}
  \Tr_{V_1}(g)&=&1\,,\\
\Tr_{V_2}(g)&=&
\left\{
  \begin{array}{ll}
    e^{-2i\hat a}+1+e^{2i\hat a+i\ve}+e^{i\ve}&\text{ at $p_1$}\,,\\
1+e^{2i\hat a}+e^{i\ve}+e^{i\ve-2i\hat a}&\text{ at $p_2$}\,,
  \end{array}
\right.\\
\Tr_{V_3}(g)&=&e^{i\ve}\,.
\end{eqnarray}
Thus at $p_1$ the character on the tangent space is $e^{-2i\hat a}+e^{2i\hat a+i\ve}$,
corresponding to the weights $(e^{-2i\hat a},e^{2i\hat a+i\ve})$.
At $p_2$ the weights are $(e^{2i \hat a},e^{i\ve-2i\hat a})$.

At $p_1$, we get an extra contribution to the index 
$\text{ind}(D_\text{Bogo})$ in  (\ref{ind-Bogo}):
\begin{eqnarray}
&&
\hspace{5mm}
  \text{ind}(D_\text{Bogo})
\rightarrow
  \text{ind}(D_\text{Bogo})
+
 \text{ind}(D_\text{Bogo})_\text{mono}
\,,~~~~~
\\
&&
 \text{where }~~ \text{ind}(D_\text{Bogo})_\text{mono}
=-\frac{1+e^{-i\ve}}2(e^{-2i\hat a}+e^{2i\hat a+i\ve})\,.
\label{ind-mono-p1}
\end{eqnarray}
Here the factor $(1+e^{-i\ve})/2$ has the same origin
as in (\ref{ind-SD-reg}).

% We also need to tensor with the space of periodic functions on $S^1$:
% \begin{eqnarray}
% \sum_{n\in \mathbb Z}e^{i n\ve}   \text{ind}(D_\text{Bogo})_\text{mono}
% =
% 2\sum_{n\in \mathbb Z}e^{i n\ve}(e^{2i\hat a}+e^{-2i\hat a})\,.
% \end{eqnarray}
We also get the extra contribution
for the adjoint hypermultiplet.
To understand this, note the relations among the indices
of the Dirac, self-dual, 
and Dolbeault complexes in four dimensions 
\begin{eqnarray}
  \text{ind}(D_\text{SD})&=&\frac{1+e^{i\ve_1+i\ve_2}}2\text{ind}(\bar D)\,,
\label{DSD-Dolb}
\\
\text{ind}(D_\text{Dirac})&=& e^{\frac i 2 (\ve_1+\ve_2)}
\frac{e^{i\hat m}+e^{-i\hat m}}2
\text{ind}(\bar D)\,.
\end{eqnarray}
Since the indices for the Bogomolny and Dirac-Higgs complexes
are obtained from $ \text{ind}(D_\text{SD})$
and $\text{ind}(D_\text{Dirac})$ by averaging over $U(1)_K$ respectively,
they are related as
\begin{eqnarray}
\left(
e^{\frac i 2\ve} + e^{-\frac i 2\ve}
\right)
  \text{ind}(D_\text{DH})&=&
\left(
e^{i\hat m} + e^{-i\hat m}
\right)
  \text{ind}(D_\text{Bogo})\,.
\end{eqnarray}
The index that leads to the fluctuation determinants
is
\begin{eqnarray}
&&  \sum_{n\in\mathbb Z}e^{in\ve} \text{ind}(D_\text{Bogo})
-  \sum_{n\in\mathbb Z}e^{i(n+1/2)\ve} \text{ind}(D_\text{DH})
\nonumber\\
&=&
  \sum_{n\in\mathbb Z}e^{in\ve} 
\left(
1-\frac{e^{i\hat m}+e^{-i\hat m}}2
\right)
\text{ind}(D_\text{Bogo})
\label{ind-Bogo-DH}
\end{eqnarray}
% Due to the relation between the Dirac-Higgs complex
% and the Bogomolny complex,
% which follows from the relations among
% Dirac, self-dual, and Dolbeault complexes in four dimensions,
% the contribution is given by
% \begin{eqnarray}
% &&  \text{ind}(D_\text{DH})
% \rightarrow
%   \text{ind}(D_\text{DH})
% +  \text{ind}(D_\text{DH})_\text{mono}\,,
% \\
% &&
%  \text{ind}(D_\text{DH})_\text{mono}
% = (e^{i\hat m}+e^{-i\hat m})
% e^{-i\ve/2}(e^{-2i\hat a}+e^{2i\hat a+i\ve})\,.
% \end{eqnarray}
% As discussed in section \ref{sec:one-loop-eq},
% the hypermultiplet is anti-periodic after the field redefinition
% (\ref{field-redef}).  Therefore we tensor with the space of anti-periodic functions:
% \begin{eqnarray}
% -  \sum_{n\in \mathbb Z} e^{i(n+1/2)\ve}
% \text{ind}(D_\text{DH})_\text{mono}
% =
% -\sum_{n\in\mathbb Z}e^{in\ve}
% (e^{i\hat m}+e^{-i\hat m})
% (e^{2i\hat a}+e^{-2i\hat a})\,.
% \end{eqnarray}
According to the rule  
$\sum c_j e^{ w_j(\hat a,\hat m,\ve)}
\rightarrow \prod w_j(\hat a, \hat m, \ve)^{c_j}$,
this
%$\text{ind}(D_\text{Bogo})_\text{mono}$ in (\ref{ind-mono-p1}) 
leads to
the one-loop determinant 
\begin{eqnarray}
&& \prod_{n\in \mathbb Z}
\frac
{ 
[(n\ve+\hat m+2\hat a)
(n\ve+\hat m-2\hat a)
(n\ve-\hat m+2\hat a)
(n\ve-\hat m-2\hat a)
]^{1/2}
}
{
(n\ve+2\hat a)(n\ve-2\hat a)
}
\nonumber\\
&=&\frac
{
\sin\left(2\pi r  \hat a +\pi r \hat m\right)
\sin\left(2\pi r\hat a -\pi r \hat m\right)
}
{\sin^2(2\pi r \hat a)}
\end{eqnarray}
where we have used that $\ve=1/r$.
The second fixed point $p_2$ contributes the same amount.
Thus
\begin{eqnarray}
  Z_\text{mono}( \hat a, \hat m;2,0)
=
2\frac{\sin[\pi r(2\hat a+\hat m)]
\sin[\pi r(2\hat a-\hat m)]
}{
\sin^2[2\pi r \hat a]}\,.
\label{eq:2to0sines}
\end{eqnarray}

There is another
method based on contour integrals
as applied in \cite{Nekrasov:2004vw, Shadchin:2005mx}
to instantons.
Let us temporarily ignore the matter contribution.
In this approach,\footnote{%
It requires no explicit resolution of singularities,
and therefore can be applied to any group that admits
an ADHM construction.
}
we compute the character of the space holomorphic 
%sections of the Dirac zero-mode bundle
functions
on the moduli space $\mathcal M=\mathcal M(p;q)$,
identify it with the index of the Dolbeault operator
on the resolved moduli space
and read off the weights.
The holomorphic 
%sections
functions 
depend on $B_1,B_2, I, J$,
and we need to take into account the complex ADHM equation
(\ref{eq:complex-ADHM}) and the quotient
by the group $\prod_r U(k_r)$.
Schematically, the character is computed by averaging
over $h\in \prod_r U(k_r)$,
\newcommand{\ch}{\mathrm{ch}}
\begin{eqnarray}
  \ch(g)=\frac{1}{\rm Vol}\int dh \frac{\det_\text{equations}(1-e^{\rm weight}h)}
{\det_\text{variables}(1-e^{\rm weight}h )}\,,
\end{eqnarray}

where the determinants are taken in the spaces of equations
and variables and ${\rm Vol}$ is the volume of
$\prod_r U(k_r)$.
For $\mathcal M(2;0)$,
\begin{eqnarray}
  \ch(g)=\int_0^{2\pi}\frac{d\phi}{2\pi}
\frac{1-e^{i\ve}}{
(1-e^{i\frac 1 2\ve-i\hat  a+i\phi} )
(1-e^{i\frac 1 2\ve+i\hat a+i\phi} )
(1-e^{i\frac 1 2\ve+i\hat a-i\phi} )
(1-e^{i\frac 1 2\ve-i\hat a-i\phi} )
}\,.
\end{eqnarray}
To evaluate the integral by residues we need to
specify the precise contour.
Following \cite{Nekrasov:2004vw}
we assume that ${\rm Im}\,\ve>0$ and treat $\phi$ and $\hat a$ as
real variables. 
we find two poles in $z=e^{i\phi}$, and the character is given as
\begin{eqnarray}
  \ch(g)
=
\frac{1}{(1-e^{-2i\hat a})(1-e^{i\ve+2i\hat a})}
+
\frac{1}{(1-e^{2i\hat a})(1-e^{i\ve-2i\hat a})}
\,. \label{eq:p=2char}
\end{eqnarray}
Given the weights we found above, (\ref{eq:p=2char})
 is consistent with the identification of the character
with the index
\begin{eqnarray}
  {\rm ind}
(\bar\partial)
&\equiv&
\sum_{k=0}^{\dim \mathcal M} (-1)^k \Tr_{H^{0,k}_{\bar\partial}(\mathcal M)}
(g)
\nonumber\\
&=&
\mathop{\sum_{P:\text{ fixed}}}_{\text{points}}
\frac{1}{\prod_{j}(1-e^{w_j(P)})}\,,
\end{eqnarray}
where $j$ runs over the holomorphic tangent directions.

After this practice, let us now include the matter contribution.
It is convenient to consider the so-called $\chi_y$-genus%
:\footnote{%
The  $\chi_y$-genus 
also appeared in the instanton calculus
for $\mathcal N=2^*$ theory \cite{Hollowood:2003cv}.
}%

\begin{eqnarray}
  \chi_y(\mathcal M)
&=&\sum_{k,l\geq 0} y^k (-1)^l \Tr_{H^{k,l}_{\bar\partial}(\mathcal M
%,\wedge^k T^*\mathcal M
)}(g)
\nonumber\\
&=&
\mathop{\sum_{P:\text{ fixed}}}_{\text{points}}
\prod_{j}\frac{1- y e^{w_j(P)}}{
1-e^{w_j(P)}}\,,
\label{chi-y-genus}
\end{eqnarray}
with $y=e^{i\hat  m}$.
% \TO{Just a comment, no action necessary: I took the expression for $\chi_y$ in terms of weights
% from Hollowood-Iqbal-Vafa, and that's what we need.
% I translated that expression into the one in terms of cohomology,
% which at first glance differs from theirs.
% In their (6.6), $(-1)^p$ should be a typo for $(-1)^q$.
% If their $T^*$ denotes the holomorphic cotangent bundle
% and the cohomology is with respect to $\bar\partial$,
% our expression is consistent with theirs.
% }
Each weight $w_j(P)$ will be of the form
\begin{eqnarray}
  w_j=i n_j \hat a+ \frac i 2 l_j \ve\,,
\end{eqnarray}
where $n_j$ and $l_j$ are integers.
(\ref{DSD-Dolb}) implies that
the contribution to $  \text{ind}(D_\text{Bogo})$ at the fixed point $P$ is given by
$-\frac{1+e^{-i\ve}}2\sum_j e^{w_j}$.
Then the contribution to (\ref{ind-Bogo-DH}) is given by
\begin{eqnarray}
\sum_{n\in \mathbb Z} e^{in\ve}
\left(\frac{e^{i\hat m}+e^{-i\hat m}}2-1\right)
\sum_j e^{i n_j\hat  a+ \frac i 2 l_j \ve}\,.
\end{eqnarray}
Summing over the fixed points $P$, the contribution to the 
path integral  is
\begin{eqnarray}
\hspace{-8mm}
 && Z_\text{mono}(\hat a,\hat m;w,v)
=
\mathop{\sum_{P:\text{ fixed}}}_\text{points}
\prod_j \prod_{n\in\mathbb Z}
\frac
{
(n\ve+  \hat m+n_j \hat a+l_j\ve /2)^{1/2}
(n\ve- \hat m+n_j \hat a+l_j\ve /2)^{1/2}
}
{
(n\ve+n_j \hat a+l_j \ve/2)
}
\nonumber\\
&&
\hspace{22mm}
=
\mathop{\sum_{P:\text{ fixed}}}_\text{points}
\prod_j 
\frac
{
\sin^{1/2}[\pi (n_j r \hat a+r \hat m+l_j/2)]
\sin^{1/2}[\pi (n_j r \hat a-r \hat m+l_j/2)]
}
{
\sin[\pi (n_j r \hat a+l_j/2)]
}\,,
\end{eqnarray}
where we recall that $\ve=1/r$.
On the other hand, the $\chi_y$ genus in
(\ref{chi-y-genus})
can be written as
\begin{eqnarray}
\hspace{-5mm}
  \chi_y(\mathcal M)
=
e^{\frac i 2 (\dim_\BC \mathcal M)\hat m}
\mathop{\sum_{P:\text{ fixed}}}_\text{points}
\prod_j 
\frac
{
\sin^{1/2}[\frac 12(n_j \hat a+\hat m+l_j \ve/2)]
\sin^{1/2}[\frac 12 (n_j \hat a-\hat m+l_j \ve/2)]
}
{
\sin[\frac 12 (n_j \hat a+l_j \ve/2)]
}\,.
\end{eqnarray}
Thus we find that
\begin{eqnarray}
Z_\text{mono}(\hat a;w,v)
&=&
\left.
e^{-\frac i 2 (\dim_\BC \mathcal M)\hat m}
\chi_y( \mathcal M(w;v))
\right|_{(\ve,\hat m,\hat a)\rightarrow (2\pi,2 \pi   r\hat  m, 2\pi   r \hat a)}\,.
\label{eq:chi-ytoratio}
\end{eqnarray}
This is why the $\chi_y$-genus is useful for us.

We now calculate the $\chi_y$ genus using the ADHM construction
of the monopole moduli space.
Locally at the origin of the space of ADHM data,
the space of holomorphic sections is
the tensor product of the space of holomorphic functions
and the space of Dirac zeromodes.
(\ref{chi-y-genus}) corresponds to 
$\Tr [ \det(1-y g)]$,
where the trace is over the holomorphic functions
and the determinant is over the zeromodes.
Since the space of zeromodes is given by
the cohomology of the complex
(\ref{eq:ADHM-linearized-complex}),
the determinant over the zeromodes is
given by $\det_{V_2}(\cdot)/\det_{V_1}(\cdot)\det_{V_3}(\cdot)$.
Thus
\begin{eqnarray}
&& 
  \chi_y(\mathcal M)
%  \mathop{\sum_{p:\text{ fixed}}}_{\text{points}}
% \prod_{\alpha}\frac{1- y e^{w_\alpha(p)}}{
% 1-e^{w_\alpha(p)}}
\nonumber\\
&=&
\frac{1}{\rm Vol}\int dh \frac{\det_\text{equations}(1-e^{\rm weight}h)}
{\det_\text{variables}(1-e^{\rm weight}h )}
\frac{\det_{V_2}(1-y e^{\rm weight}h)}{\det_{V_1}(1-y e^{\rm weight}h)
\det_{V_3}(1-y e^{\rm weight}h)}\,.
\end{eqnarray}
In the case of $\mathcal M(2;0)$,
\begin{eqnarray}
&&  \chi_y(\mathcal M(2;0))
\nonumber\\
&=&
\int_0^{2\pi}\frac{d\phi}{2\pi}
\frac{(1-e^{i\ve})
(1-e^{i\hat m+i \frac 1 2\ve-i\hat a+i\phi})
(1-e^{i\hat m+i\frac 1 2\ve+i\hat a+i\phi})
}{
(1-e^{i\frac 1 2 \ve-i\hat a+i\phi} )
(1-e^{i\frac 1 2 \ve+i\hat a+i\phi} )
(1-e^{i\frac 1 2 \ve+i\hat a-i\phi} )
(1-e^{i\frac 1 2 \ve-i\hat a-i\phi} )
}
\nonumber
\\
&&
\hspace{10mm}
\times
\frac{
(1-e^{i\hat m+i\frac 1 2\ve+i\hat a-i\phi})
(1-e^{i\hat m+i\frac 1 2\ve-i\hat a-i\phi})
}{
(1-e^{i\hat m})(1-e^{i\hat m+i\ve})
}
\nonumber
\\
&=&\frac{(1-e^{i\hat m-2i\hat a})(1-e^{i\hat m+i\ve+2i\hat a})}
{(1-e^{-2i\hat a})(1-e^{i\ve+2i\hat a})}
+
\frac{(1-e^{i\hat m+2i\hat a})(1-e^{i\hat m+i\ve-2i\hat a})}
{(1-e^{2i\hat a})(1-e^{i\ve-2i\hat a})}
\,. \label{eq:p=2chi-y}
\end{eqnarray}
We note that (\ref{eq:2to0sines}) is indeed obtained from
(\ref{eq:p=2chi-y})
using the relation (\ref{eq:chi-ytoratio}).

The magnetic charge $p$ can be screened by monopoles and get reduced to 
$q$, also an even integer.
We set $l:=p-q$.
The moduli space $\mathcal M(p;q)$
can be described using the ADHM construction as follows.
The action of $U(1)_K
%\prod_r U(k_r)
$ is specified by
the matrix
\begin{eqnarray}
  K=\left(
    \begin{array}{ccccccccccc}
      \frac{p-2}2 \mathbb I_{1\times 1} & 
\\
& \frac{p-4}2 \mathbb I_{2\times 2}
\\
&&
\ddots
\\
&&&\frac{p-l}2 \mathbb I_{\frac{l}2\times\frac l 2}
\\
&&&&\ddots
\\
&&&&& -\frac{p-l}2 \mathbb I_{\frac{l}2\times\frac l 2}
\\
&&&&&&\ddots
\\
&&&&&&& -\frac{p-4}2 \mathbb I_{2\times 2}
\\
&&&&&&&&-\frac{p-2}2\mathbb I_{1\times 1}
    \end{array}
\right)
\,,
\end{eqnarray}
and the action of $G$ by
\begin{eqnarray}
M=
\left(
    \begin{array}{ccccccc}
\frac{q}2
%\frac{p-l}2
&0 
\\
0&
-\frac{q}2
%-\frac{p-l}2
    \end{array}
\right)\,.
\end{eqnarray}
The conditions of $U(1)_K$-invariance
\begin{eqnarray}
  B_1+[K,B_1]&=&0\,,\\
-  B_2+[K,B_2]&=&0\,,\\
K I - I M&=&0\,,\\
M J - J K&=&0
\end{eqnarray}
require that the ADHM matrices take the following form:
\begin{eqnarray}
  B_1=
\left(
    \begin{array}{ccccccc}
0&
\\
B_{21}&0
\\
0&B_{32}&0
\\
&\ddots&\ddots&\ddots
\\
&&0&
B_{p-1,p-2}&0
    \end{array}
\right)\,,
~~~~~~
~~~~~
  B_2=
\left(
    \begin{array}{ccccccc}
0&\tilde B_{12} &0&&
\\
&0 &\tilde B_{23}&\ddots&
\\
&&0&\ddots&0
\\
&&&\ddots&\tilde B_{p-2,p-1}&\\
&&&&
0
\end{array}
\right)\,,
\\
I=\left(
  \begin{array}{ccccccc}
0&0
\\
\vdots&\vdots\\
0&0
\\
I_{l/2,1}&0\\
\vdots&\vdots
\\
0&
I_{p- l/ 2,2}
\\
0&0
\\
\vdots&\vdots
\\
0&0
  \end{array}
\right)\,,
~~~~~~~
J=\left(
  \begin{array}{cccccccccccccc}
0&\cdots&J_{1,l/2}   &\cdots&0&\cdots&0
\\
0&\cdots&0&\cdots&J_{2,p-l/2}    &\cdots&0
  \end{array}
\right)\,.
\end{eqnarray}
We consider the space of solutions to the complex ADHM equation
\begin{eqnarray}
  [B_1,B_2]+IJ=0
\end{eqnarray}
% becomes
% \begin{eqnarray}
% -\tilde B_{12} B_{21}&=&0\,,
% \\
% B_{21}\tilde B_{12}-  \tilde B_{23} B_{32}&=&0\,,
% \\
% &\vdots&\\
% B_{\frac l 2,\frac l 2-1}\tilde B_{\frac l 2-1,\frac l 2}
% -
% \tilde B_{\frac l 2,\frac l 2+1} B_{\frac l 2+1,\frac l 2}
% +
% \end{eqnarray}
and then take the quotient by the complexification of the group $
\prod_{r=1}^{p-1} U(k_r)$
with 
\begin{eqnarray}
(k_1,k_2,\ldots,k_{p-1})
:=(1,2,\ldots,l/2,\ldots,l/2,\ldots,1)\,.
\end{eqnarray}
Counting shows that the resulting space $\mathcal M(p; q)$ has complex dimension $l$.
This  is the singular moduli space of monopoles on $\mathbb R^3$
in the presence of a singular monopole of charge $p$ at the origin
with the boundary condition that the fields 
look like the charge $p-l$ monopole at infinity.
The group $U(1)_{J+R}$ generated by $Q^2$
and the maximal torus $U(1)_\infty\subset SU(2)$ of the group of
global gauge transformations act on
$\mathcal M(p; q)$
according to
\begin{eqnarray}
  B_s &\rightarrow& e^{i\frac12\ve}B_s\,,~~~~s=1,2\,,
\\
I&\rightarrow &e^{i\frac12\ve}I 
\left(
  \begin{array}{cc}
    e^{-i\hat a}&
\\
&e^{i\hat a}
  \end{array}
\right)\,,
\\
J&\rightarrow &
e^{i\frac12\ve}\left(
  \begin{array}{cc}
    e^{i\hat a}&
\\
&e^{-i\hat a}
  \end{array}
\right)
J\,.
\end{eqnarray}
% We now compute the character of $U(1)_{J+R}\times U(1)_\infty$
% on the space of functions on $\overline{\mathcal M}(p; q)$.
% The character is expected to coincide
% with the index ${\rm ind}(\bar\partial)$
% with the equivariant parameters $(e^{i\ve},e^{i a})$.

% The prescription of Nekrasov and Shadchin formally gives
% the character as the group averaging
% of $\prod(1-\text{weight(equation)})/
% \prod(1-\text{weight(variable)})$:
We thus obtain a contour integral expression for the
$\chi_y$-genus of $\mathcal M(p;q)$:
\begin{eqnarray}
&&
\hspace{-12mm}
\chi_y(\mathcal M)
%Ch_{\overline{\mathcal M}(p; q)}(\ve, a)
%\nonumber\\
% &&
=
  \frac{1}{\prod_{r=1}^{p-1}k_r!}
\oint 
\prod_{r=1}^{p-1}
\prod_{i=1}^{k_r}\frac{dz_{r,i}}{2\pi i z_{r,i}}
\prod_{i<j}\frac{-z_{r,i}}{z_{r,j}}
\left(1-z_{r,j}/z_{r,i}\right)^2
\nonumber\\
&&
~~~~\times
\prod _{r=1}^{p-1} \prod _{i,j}
\frac{
    \left(1-e^{ i \ve } z_{r,i}/z_{r,j}\right)
}{
\left(1-e^{i\hat m} z_{r,i}/z_{r,j}\right)
\left(1-e^{i\hat m } e^{i\ve}z_{r,i}/z_{r,j}\right)
}
\nonumber\\
&&~~~~~~~
\times
 \frac{
\displaystyle \prod_{i=1}^{k_{\frac{l-2}{2}}} 
(1-e^{i\hat m}
e^{i\hat  a+i \frac12   \ve }
/z_{\frac{l-2}{2},i})
 \left(1-
e^{im}
e^{i \frac12 \ve-i\hat  a} z_{\frac{l-2}{2},i}
\right) 
% \prod_{i=1}^{k_{p-\frac{l}{2}-1}} 
% \left(1-
% e^{im}e^{i a+i
%     \ve } 
% z_{
% p
% -1-l/2,i}
% \right)
%  \left(1-e^{im}
% e^{i   \ve -i a}/z_{p-1-l/2
% ,i}\right)
}{
\displaystyle \prod_{i=1}^{k_{\frac{l-2}{2}}} 
(1-e^{i\hat  a+i \frac12  \ve }
/z_{\frac{l-2}{2},i})
 \left(1-e^{i \frac12 \ve-i\hat  a} z_{\frac{l-2}{2},i}
\right) 
% \prod_{i=1}^{k_{p-\frac{l}{2}-1}} 
% \left(1-e^{i a+i
%     \ve } 
% z_{
% p
% -1-l/2,i}
% \right)
%  \left(1-
% e^{i   \ve -i a}/z_{p-1-l/2
% ,i}\right)
}
\nonumber\\
&&
\times
 \frac{
\displaystyle 
% \prod_{i=1}^{k_{\frac{l-2}{2}}} 
% (1-e^{im}
% e^{i a+i 
%     \ve }
% /z_{\frac{l-2}{2},i})
%  \left(1-
% e^{im}
% e^{i 
%  \ve-i a} z_{\frac{l-2}{2},i}
% \right) 
\prod_{i=1}^{k_{p-\frac{l}{2}-1}} 
\left(1-
e^{im}e^{i\hat  a+i
\frac12    \ve } 
z_{p-1-l/2,i}
\right)
 \left(1-e^{i\hat m}
e^{i \frac12 \ve -i\hat  a}/z_{p-1-l/2
,i}\right)
}{
\displaystyle
%  \prod_{i=1}^{k_{\frac{l-2}{2}}} 
% (1-e^{i a+i 
%     \ve }
% /z_{\frac{l-2}{2},i})
%  \left(1-e^{i 
%  \ve-i a} z_{\frac{l-2}{2},i}
% \right) 
\prod_{i=1}^{k_{p-\frac{l}{2}-1}} 
\left(1-e^{i\hat  a+i\frac12    \ve } 
z_{
p
-1-l/2,i}
\right)
 \left(1-
e^{i\frac12  \ve -i\hat  a}/z_{p-1-l/2
,i}\right)
}
\nonumber\\
&&
\times
\frac{
\displaystyle
\prod _{r=1}^{p-2} \prod_{i=1}^{k_r} \prod _{j=1}^{k_{r-1}}
 \left(1-
e^{i\hat m}
e^{i\frac12 \ve}
    z_{r,i}/z_{r-1,j}\right)
 \prod_{r=1}^{p-2} \prod_{i=1}^{k_r}
    \prod_{j=1}^{k_{r+1}} \left(1-
e^{i\hat m}
e^{i\frac12 \ve }
z_{r,i}/z_{r+1,j}
\right)
}{
\displaystyle
\prod _{r=1}^{p-2} \prod_{i=1}^{k_r} \prod _{j=1}^{k_{r-1}}
 \left(1-e^{i\frac12 \ve }
    z_{r,i}/z_{r-1,j}\right)
 \prod_{r=1}^{p-2} \prod_{i=1}^{k_r}
    \prod_{j=1}^{k_{r+1}} \left(1-e^{i \frac12\ve }
z_{r,i}/z_{r+1,j}
\right)
}\,.
\end{eqnarray}
The first line on the right hand side represents the
Haar measure on $\prod_r U(k_r)$, which would be clearer
if the integral is written in terms of $\phi_{r, i}$ such that $z_{r,i}=e^{i\phi_{r,i}}$.
% To this contour integral we need to supply a prescription
% for which poles to pick up.
We again choose to use the prescription
% of Nekrasov and Shadchin, 
where we integrate over each $z_{r,i}$ along the unit circle $|z_{r,i}|=1$, 
assuming that $\hat a\in \mathbb R$ and ${\rm Im}\,\ve>0$.
The integral can be evaluated by residues, and the computation can be automated as a Mathematica code.
Applying the rule (\ref{eq:chi-ytoratio}), we 
find
experimentally%
\footnote{%
We have checked this for $(p,q)=(2,0), (3,1),(4,2),(5,3), (4,0),(6,2)$, and $(6,0)$.
} that
\begin{eqnarray}
  Z_\text{mono}(\hat a, \hat m;p,q)=
\frac{p!}{(\frac{p-q}2)!(\frac{p+q}2)!}
\times
\left\{
  \begin{array}{ll}
\displaystyle
  \frac{\cos^{\frac{p-q}2}[\pi r( 2 \hat a+\hat m)]
\cos^{\frac{p-q}2}[\pi r(2\hat  a-\hat m)]
}{\cos^{p-q}[2\pi r\hat  a]}
&\text{ for $p$ odd,}
\\
\displaystyle
\frac{\sin^{\frac{p-q}2}[\pi r( 2 \hat a+\hat m)]
\sin^{\frac{p-q}2}[\pi r(2\hat  a-\hat m)]
}{\sin^{p-q}[2\pi r\hat  a]}
&\text{ for $p$ even.}
  \end{array}
\right.
\label{mono-SU2}
\end{eqnarray}
Combined with (\ref{Eq-SU2}),
the dependence of
$Z_\text{mono}(i a, i m;p,q)Z_\text{1-loop,eq}(ia,im;q)$
on $q$ is in fact only in the binomial coefficient.

% \TO{Important: we assumed that at each fixed point the number of exponentials
% with $+a$ coincides with the number of exponentials with $-a$.
% }

% \TO{Old:
% More generally, we find using Mathematica%
% \footnote{%
% This was done for $(p,q)=(2,0), (4,2), (4,0), (6,2), (6,0),(8,2),(8,0), (10,2)$.
% The case $(p,q)=(10,2)$ involves 589 terms and takes 600 Mb of memory and $10^{14}$ CPU cycles to compute the binomial coefficient.
% }
%  that \TO{ up to signs}
% \begin{eqnarray}
% Ch_{\mathcal M(p;q)}|_{\ve\rightarrow 0, a\rightarrow \pi i a_E}=
% \frac{p!}{(\frac{p-l}2)!(\frac{p+l}2)!}
% \frac{1}{\sinh^l(2\pi r a_E)}\,.
% \end{eqnarray}
% Assuming that all the fixed points contribute the same amount
% to this character, this implies that the number of fixed points
% is given by the binomial coefficient
% $\frac{p!}{(\frac{p-l}2)!(\frac{p+l}2)!}$.
% By taking into account the adjoint matter contribution, we have
% \begin{eqnarray}
%   Z_{\text{1-loop}}(p;p-l)/  Z_{\text{1-loop}}(p-l;p-l)=
% \frac{p!}{(\frac{p-l}2)!(\frac{p+l}2)!}
% \frac{\sin^{\frac{p-q}2}[\pi r( 2 a+m)]\sin^{\frac{p-q}2}[\pi r(2 a-m)]
% }{\sin^l[2\pi r a]}\,.
% \end{eqnarray}
% }
% There should be a better derivation of the result
% using a method in Shadchin's thesis,
% based on the fact that the Dirac zero-modes can be obtained
% in terms of the ADHM data.  Such a method would be
% necessary to deal with fundamental matter.}

We now put everything together. 
Including the terms with $q\leq 0$, we get 
\begin{eqnarray}
  \langle T_p\rangle&=&
\sum_{q=p,p-2,\ldots, -p}
\frac{p!}{(\frac{p-q}2)!(\frac{p+q}2)!}
\int da
\,
\left|
Z_\text{1-loop,pole }\left(ia-\frac{q}{4r}\right)
Z_\text{cl
%assical
}\left(ia-\frac{q}{4r}\right)
\right.
\nonumber\\
&&
\times
\left.Z_\text{inst}\left(ia-\frac{q}{4r}\right)
\right|^2\times
\left\{
  \begin{array}{ll}
\displaystyle
\frac{\cosh^{\frac{p}2}[\pi r( 2 a+m)]\cosh^{\frac{p}2}[\pi r(2 a-m)]
}{\cosh^{p}[2\pi r a]}
&\text{ for $p$ odd,}
\\
\displaystyle
\frac{\sinh^{\frac{p}2}[\pi r( 2 a+m)]\sinh^{\frac{p}2}[\pi r(2 a-m)]
}{\sinh^{p}[2\pi r a]}
&\text{ for $p$ even.}
  \end{array}
\right.
\label{eq:N=2*result}
\end{eqnarray}
This is the complete gauge theory result for
't Hooft loops in $SU(2)$ $\mathcal N=2^*$ theory.

This analysis, with the philosophy described, can be extended to other gauge theories.

 \section{Gauge Theory Computation vs Toda CFT}
\label{sec:compare}

In this section we compare the results of our gauge theory analysis for the expectation
value of 't Hooft loop operators in $\cN=2$ gauge theories on $S^4$ with    formulae in 
\cite{Alday:2009fs,Drukker:2009id,Gomis:2010kv}, which were  obtained from computations in two dimensional Liouville/Toda CFT. As we shall see, for the theories
 for which we explicitly carry out the comparison, we find beautiful agreement.

In \cite{Alday:2009fs, Drukker:2009id} a dictionary was put forward  relating the exact expectation value of gauge theory loop operators in $\cN=2$ gauge theories on $S^4$ and  Liouville/Toda correlation functions in the presence of  Liouville/Toda loop operators (topological defects). This enriches the AGT correspondence \cite{AGT},  which identifies  the gauge theory partition function   
with a correlation function  in Liouville/Toda (see also \cite{Wyllard:2009hg}), to encompass more general observables. The   identification in \cite{Alday:2009fs, Drukker:2009id} has yielded explicit predictions for the exact expectation value of 't Hooft loop operators in $\cN=2$ gauge theories on $S^4$.

We compare the Liouville/Toda results for 't Hooft operators in $\cN=2^*$ with the corresponding gauge theory computations for both the one-loop determinants as well as for the non-perturbative contributions due to monopole screening.

 \subsection{'t Hooft Loop Determinants from Toda CFT}
\label{sec:compareone-loop}

We now  explicitly compare the results obtained for 't Hooft operators in the $N=2^*$ theory -- corresponding to an $\cN=2$  $SU(N)$ vectormultiplet with a massive hypermultiplet in the adjoint representation -- with   loop operator  computations in Toda CFT  on the once-punctured torus. For a 't Hooft loop labeled by a magnetic weight $B=h_1$ -- corresponding to the fundamental representation of $SU(N)$  -- the Toda CFT calculation yields \cite{Gomis:2010kv}
 \begin{equation}
\label{vevTN2starA}
\int da\, C(ia,im) \overline{Z_\text{cl}(ia,q)} \overline{Z_\text{inst}(ia,1+im,q)}\sum_{k=1}^N T_k(ia, im)
Z_\text{cl}(ia-h_k,q)   Z_\text{inst}(ia-h_k,1+im,q)\,,
\end{equation}
where 
\begin{equation*}
T_k(ia, im)=\frac{1}{N}\prod_{1\leq j\leq N}^{j\neq k}
  \frac{\Gamma(i (a_j-a_k))\Gamma(2+ i(a_j-a_k))}
{\Gamma(1+ i(a_j-a_k)  
-   i  m)\Gamma(1 + i(a_j-a_k)  + i m)}\,,
\label{monodromy}
\end{equation*}
$m$ is the mass of the adjoint hypermultiplet and  $h_i$ are the $N$ weights of the fundamental representation of $SU(N)$.%
\footnote{%
Explicitly, $h_i=(\delta_{ij}-1/N)_{j=1}^N$.
} The result in \rf{vevTN2starA} is expressed as much as possible in terms of gauge theory quantities introduced in previous sections. The factor $Z_\text{cl}(ia,q)$ is the classical contribution to Nekrasov's equivariant instanton partition function \rf{eq:class-contr}
 \begin{equation}
 Z_{\text{cl}}(ia, q)=\exp\left[\pi i   \tau   a\cdot  a \right]=\exp\left[\pi i   \tau \sum_{l=1}^N a_l^2 \right]\,,
 \end{equation}
while  $Z_\text{inst}(ia,1+i m,q)$ is  the instanton contribution \rf{nekrinst}.\footnote{To lighten notation we have set $r=1$, have omitted  the $\ve_1,\ve_2$ dependence of the  instanton partition function $Z_{\text{inst}}(\hat a , \hat m, \ve_1, \ve_2, q)\rightarrow Z_{\text{inst}}(\hat a , \hat m, q)$ and also used that $\hat m =1+ im$ \rf{massmap}.} Finally $C(ia,im)$ is the Toda CFT three-point function\footnote{This is the three-point function of two non-degenerate and one semi-degenerate primary operators in Toda CFT when the background charge $b=1$.}
 relevant for the once-punctured
 torus\ description of $\cN=2^*$ (\`a la \cite{Gaiotto:2009we})
 \beq
C(ia,im)={\prod_{\alpha>0} \Upsilon_{b=1}(-i \alpha\cdot a)\Upsilon_{b=1}(i\alpha\cdot a)\over
\prod_{i,j=1}^N \Upsilon_{b=1}(1+i (h_i-h_j)\cdot a +i m)}\,,
\eeq
with $\alpha$ the roots of the $SU(N)$ Lie agebra.
Since $\Upsilon_{b=1}(x)=G(x)G(2-x)/2\pi$, with  $G(x)$ being the Barnes $G$-function \rf{Barnesdef}
and because  $\alpha=h_i-h_j$ for $j>i$ if $\alpha>0$ we obtain\footnote{In this section, in order to avoid cluttering formulas, we
drop inessential overall numerical factors.}
\beq
C(ia,im)={\prod_{\alpha}G( i\alpha\cdot a)G(2+ i\alpha\cdot a)\over  \prod_{\alpha>0}\prod_{\pm,\pm} G(1\pm i\alpha\cdot a\pm i m)}\,.
\eeq
We note that $C(ia,im)$ is precisely given by the square of the one-loop factor in Nekrasov's partition function of $\cN=2^*$ in $\bR^4$ (see \rf{loopcombination}
and \rf{n=2test})
\beq
C(ia,im)= \left| Z_\text{1-loop,pole}(ia,im)\right|^2\,,
\eeq
with 
\begin{eqnarray}
Z_{\text{1-loop,pole}}(ia,im)
&=&
Z_{\text{1-loop,pole}}(-ia,-im)=\overline{Z_{\text{1-loop,pole}}(ia,im)}
\nonumber\\
&=&
\left[{\prod_{\alpha}G( i\alpha\cdot a)G(2+ i\alpha\cdot a)\over  \prod_{\alpha>0}\prod_{\pm,\pm} G(1\pm i\alpha\cdot a\pm i m)}
\right]^{1/2}
\,.
%Z_{\text{1-loop}}(ia)=\prod_{\alpha>0}{\left[\prod_{\pm,\pm} \Gamma_2(1\pm i\vev{a,\alpha}\pm i\hat m)\right]^{1/2}\over \Gamma_2(1+ i\vev{a,\alpha})\Gamma_2(1- i\vev{a,\alpha})}\,.
\label{loopfactor}
\end{eqnarray}
 Thus we can write the Toda loop correlator as
 \begin{eqnarray}
\label{vevTN2starB}
&&\int da    \left| Z_\text{1-loop,pole}(ia,im)\right|^2  \overline{Z_\text{cl}(ia,q)} \overline{Z_\text{inst}(ia,{1+im},q)}
\nonumber\\
&&
\hspace{45mm}
\times\sum_{k=1}^N T_k(ia,  m)
Z_\text{cl}(ia-h_k,q)   Z_\text{inst}(ia-h_k,{1+ im},q)\,.
\end{eqnarray}
We note that the result is given by the sum of $N$ terms, associated to the $N$ weights of the fundamental representation of $SU(N)$. Each of the $N$ weights yields an identical contribution, and therefore we can focus on the contribution of the highest weight term, labeled by $h_1$. It is important to remark at this point that genuine new contributions appear for loop operators labeled   by a representation with highest weight $B$ for which not all weights are in the Weyl orbit of $B$ (non-minuscule representations). These contributions correspond precisely to the non-perturbative contributions due to monopole screening encountered in our gauge theory analysis! Section \ref{sec:screenn} demonstrates   for 't Hooft loops with higher magnetic weight $B$ that Liouville theory precisely reproduces the non-perturbative screening contributions discussed in section \ref{sec:ADHM}.

Focusing on the  highest weight vector contribution, we trivially rewrite the answer as
 \begin{equation}
 \begin{aligned}
\label{vevTN2starC}
\int da\,& Z_\text{1-loop,pole}(-ia,-im){Z_\text{cl}(-ia,\bar{q})}Z_\text{inst}(-ia,  1-im,\bar{q})\times T_1(ia,im)\\
 \times & Z_\text{1-loop,pole}(ia,im)  
Z_\text{cl}(ia-h_1,q)   Z_\text{inst}(ia-h_1,{1+i m},q)\,.
\end{aligned}
\end{equation}
Without encountering any residues, we now shift the contour of integration $ia\rightarrow ia +h_1/2$   to express the answer in a more symmetric form
 \begin{equation}
 \begin{aligned}
\label{vevTN2starD}
\int da\, &|Z_\text{cl}(ia-h_1/2,q)   Z_\text{inst}(ia-h_1/2,{ 1+ im},q)|^2
\\
&  Z_\text{1-loop,pole}(-ia-h_1/2,-im) Z_\text{1-loop,pole}(ia+h_1/2,im)    ~T_1(ia+h_1/2,  im)
   \,.
\end{aligned}
\end{equation}

Our next goal is to rewrite the second line in \rf{vevTN2starD}   as a complete square of a function with the same shifted argument $ia-h_1/2$
as in the first line times a remainder, which we   denote by $E(ia,im)$
\begin{equation}
 \begin{aligned}
\label{vevTN2starnewnot}
\int da\, & |Z_\text{cl}(ia-h_1/2,q)  Z_\text{1-loop,pole}(ia-h_1/2,im) Z_\text{inst}(ia-h_1/2,{ 1+ im},q)|^2 \times E(ia,im)  
   \,.
\end{aligned}
\end{equation}

 To anticipate where this path will  leads us when comparing with our gauge theory analysis, the complete square contributions   reproduce    the classical, one-loop and instanton contributions that arise from the north and south poles of $S^4$, while the remainder captures the contribution from the equator!

In order to determine $E(ia,im)$ in \rf{vevTN2starnewnot} we need to calculate 
\beq
E(ia,im)={Z_\text{1-loop,pole}(ia+h_1/2,im)\over Z_\text{1-loop,pole}(ia-h_1/2,im)}\,T_1(ia+h_1/2,im)\,.
\label{ratioshifts}
\eeq
The ratio of one-loop factors can be determined by recalling that
 $a_j= a\cdot h_j$, so that  the shifts $ia\pm h_1/2$ in the arguments in \rf{ratioshifts} are   given by (since  $ h_i\cdot h_j=\delta_{ij}-1/N$)
\beq
 \begin{aligned}
ia_j&\rightarrow ia_j\mp1/N\qquad  j\neq 1 \\
ia_1&\rightarrow ia_1\pm1/2\mp1/N\,.
 \end{aligned}
\eeq
Therefore, only  $a_{1j}\equiv a_1-a_j$ shifts, by $ia_{1j}\rightarrow ia_{1j}\pm 1/2$.
Since $\alpha=h_i-h_j$ for $j>i$ if $\alpha>0$, we decompose  the product over positive roots appearing in  \rf{loopfactor}
 \beq
 \prod_{\alpha>0}\cdot=\prod_{j=2}^N\cdot \prod_{2\leq i<j\leq N}\,,
 \eeq
 comprising the splitting of positive roots into $h_1-h_j$ and the rest.
Therefore only the factors $\prod_{j=2}^N\cdot$ shift, the rest cancel between the numerator and denominator in \rf{ratioshifts}. We find
\beq
 \begin{aligned}
{Z_\text{1-loop,pole}(ia+h_1/2,im)\over Z_\text{1-loop,pole}(ia-h_1/2,im)}=\prod_{j=2}^N&\left[\prod_{\pm}{G({1\over2}+ia_{1j}\pm i  m)G({3\over 2}-ia_{1j}\pm i  m)
\over
G({3\over2}+ia_{1j}\pm i  m)G({1\over 2}-ia_{1j}\pm i  m)}\right]^{1/2}\\
&\left[{G({1\over2}+ia_{1j})G({5\over2}+ia_{1j})G(-{1\over 2}-ia_{1j})G({3\over2}-ia_{1j})\over
G(-{1\over2}+ia_{1j})G({3\over2}+ia_{1j})G({1\over 2}-ia_{1j})G({5\over2}-ia_{1j})}\right]^{1/2}\,,
\end{aligned}
\eeq
which since
$G(z+1)/ G(z)=\Gamma(z)$ equals
\beq
\prod_{j=2}^N\left[\prod_{\pm}{\Gamma({1\over 2}-ia_{1j}\pm i  m)\over \Gamma({1\over2}+ia_{1j}\pm i m)}\right]^{1/2}\left[{\Gamma(-{1\over2}+ia_{1j})\Gamma({3\over2}+ia_{1j})\over
\Gamma(-{1\over2}-ia_{1j})\Gamma({3\over2}-ia_{1j})}\right]^{1/2}
\,.
\eeq
Using the explicit form of the monodromy operators $T_k(ia,m)$ in \rf{monodromy} we arrive at
\beq
 \begin{aligned}
%{Z_\text{1-loop}(ia+h_1/2)\over Z_\text{1-loop}(ia-h_1/2)}\,T_1(ia+h_1/2,\hat m)=
E(ia,im)=
\prod_{j=2}^N\left[{\Gamma(-{1\over2}+ia_{1j})\Gamma({3\over2}+ia_{1j})\Gamma(-{1\over2}-ia_{1j})\Gamma({3\over2}-ia_{1j})\over \prod_{\pm} \Gamma({1\over2}+ia_{1j}\pm i  m) \Gamma({1\over 2}-ia_{1j}\pm i  m)}\right]^{1/2}\,,
 \end{aligned}
\eeq
which by Euler's reflection formula $\Gamma(z)\Gamma(1-z)={\pi \over \sin(\pi z)}$ yields
\beq
 \begin{aligned}
E(ia,im)= \prod_{j=2}^N \left[ {\sin(\pi( {1\over 2}+i a_{1j}-im))\sin(\pi( {1\over 2}-i a_{1j}-im))\over \sin(\pi( {1\over 2}+i a_{1j}))\sin(\pi( {1\over 2}+i a_{1j}))}\right]^{1/2}\,.
 \end{aligned}
\eeq
The result can be written in a more covariant form to arrive at the final answer
\beq
\begin{aligned}
E(ia,im)=&\prod_{\alpha>0}{\sin^{{|\alpha\cdot B|\over 2}}\left(\pi\left[{\alpha\cdot B\over 2}+i\alpha\cdot a -i m\right]\right)\sin^{{|\alpha\cdot B|\over 2}}\left(\pi\left[{\alpha\cdot B\over 2}-i\alpha\cdot a-i m\right]\right)\over \sin^{|\alpha\cdot B|}\left(\pi\left[{\alpha\cdot B\over 2}+i\alpha\cdot a\right]\right)}\\
=&{\prod_{w\in \rm{adj}}\sin^{{|w\cdot B|\over 2}}\left(\pi\left[{w\cdot B\over 2}+iw\cdot a -i m\right)\right]\over \prod_{\alpha>0} \sin^{|\alpha\cdot B|}\left(\pi 
\left[{\alpha\cdot B\over 2}+i\alpha\cdot a\right]\right)}\,.
\label{equatorfinal}
\end{aligned}
 \eeq
This is precisely the gauge theory formula for the equatorial one-loop determinant  \rf{1-loop-equator}.

This shows that the  Toda prediction for the expectation of the 't Hooft loop operator labeled by the fundamental representation in the $\cN=2^*$ theory with  $SU(N)$ gauge group precisely agrees with our gauge theory computation. We identify in the  Toda correlator   the factor $ |Z_\text{cl}(ia-h_1/2,q)  Z_\text{1-loop,pole}(ia-h_1/2,im) Z_\text{inst}(ia-h_1/2,{ 1+ im},q)|^2$ in \rf{vevTN2starnewnot} with the gauge theory contributions arising from the north and south poles of $S^4$ (see \rf{eq:thooft-result-extra}), while comparison of  \rf{equatorfinal} with   \rf{1-loop-equator} demonstrates that indeed $E(ia,m)$ precisely captures the gauge theory contribution from the equator, so that
\beq
E(ia,im)=Z_\text{1-loop,eq}(i a, im, h_1)\,.
\eeq

The Toda calculation of \cite{Gomis:2010kv} can be extended to describe 't Hooft operators with higher magnetic weight in $\cN=2^*$. We have checked that the Toda calculation for $B=2h_1$ also exactly reproduces the gauge theory prediction.

\subsection{Monopole Screening from Liouville Theory}
\label{sec:screenn}

We now  specialize to the $A_1$ Toda theory, {\it i.e.},
Liouville theory.
We compare the results in \cite{Drukker:2009id,Alday:2009fs},
which are the special case of the general Toda calculations above,
with
the non-perturbative contributions from monopole screening
in $SU(2)$ $\mathcal N=2^*$ theory computed in section \ref{sec:SU2ex}

In Liouville theory, the 't Hooft loop expectation value is given 
in terms of shifted conformal blocks.
To simplify formulas we set $r=1$ without loss of generality, 
  also set $b$ to 1 and adapt the normalization of
 \cite{Alday:2009fs}
\begin{eqnarray}
  Z_\text{L}(\hat a, im, \tau)\equiv
% Z_\text{cl}(\hat a ,q) Z_\text{inst}(\hat a ,{ 1+ im},q)=
e^{-2\pi i  \tau\,\hat a^2}
\mathcal F\left(
1+ \hat a,
1+ im,\tau
\right)\,,
\end{eqnarray}
where $\mathcal F(\alpha, \alpha_e,\tau)$
is the conformal block of the 1-punctured torus with modulus $\tau$
in the standard normalization \cite{Belavin:1984vu},
with internal and external Liouville momenta $\alpha$ and $\alpha_e$.%
\footnote{%
The shift by $1$ in $\alpha_e=1+im$
was clarified in \cite{Okuda:2010ke}.
}
Up to a normalization constant, it was shown in 
\cite{Drukker:2009id,Alday:2009fs}
that the loop operator expectation value is given by%
\footnote{%
The complex conjugate of $\bar\tau$ appears
with a minus sign because $\bar \tau$ enters into $\overline{Z_\text{L}}$
through $\overline{e^{2\pi i \tau}}=e^{-2\pi i \bar\tau}$.
In this subsection we avoid using the symbol $q$ to denote $e^{2\pi i \tau}$,
in order to avoid confusion with screened magnetic charge $q$.
}
\begin{eqnarray}
&& \langle (\mathcal L_{1,0})^p\rangle
\nonumber\\
&=&
\int_{\hat a\in i\mathbb R}
 d\hat a \,
C(1+\hat a,1-\hat a,1+i m)
\overline{ Z_\text{L}(\hat a,im, \tau)}
[(\mathcal L_{1,0})^p\cdot Z_\text{L}](\hat a, im,\tau)
\nonumber\\
&=&
\int_{\hat a\in i\mathbb R}
 d\hat a \,
C(1+\hat a,1-\hat a,1+i m)
 Z_\text{L}(-\hat a, -im,- \bar\tau)
[(\mathcal L_{1,0})^p\cdot Z_\text{L}](\hat a, im, \tau)
\,.
\label{eq:N=2*lioiuville}
\end{eqnarray}

 The Liouville loop operator $\mathcal L_{1,0}$ acts as
a difference operator.
For any meromorphic function $f(\hat a)$, let us define the operators
$\hat h_\pm$ as multiplication by the functions $h_\pm(\hat a)$:
%that appear in
%(\ref{L10}):
\beq
\begin{aligned}
%\begin{eqnarray}
  (\hat h_+\cdot f)(\hat a)&\equiv
\frac{\Gamma(-2\hat a)\Gamma(2-2 \hat a)}
{\Gamma(-2\hat a+1+im)\Gamma(-2\hat a+1-im)} f(\hat a)
\equiv 
h_+(\hat a) f(\hat a)
\,,
\\
  (\hat h_-\cdot f)(\hat a)&\equiv
\frac{\Gamma(2 \hat a)\Gamma(2+2 \hat a)}
{\Gamma(2\hat a+1+im)\Gamma(2\hat a+1-im)}f(\hat a)
\equiv
h_-(\hat a) f(\hat a)
\,.
%\end{eqnarray}
\end{aligned}
\eeq
We also define the shift operator $\Delta$:
\begin{eqnarray}
  (\Delta\cdot f)(\hat a):=f(\hat a-1/4)\,.
\end{eqnarray}
With these definitions the Liouville loop operator
is defined by
\beq
\mathcal L_{1,0}=\hat h_+ \Delta^2 
+\hat h_- \Delta^{-2}\,,
\eeq
and the higher powers of  $\mathcal L_{1,0}$ take the form
\begin{eqnarray}
  (\mathcal L_{1,0})^p=\sum_{q=p,p-2,\ldots, -p}
\hat h_{p,q}\, \Delta^{2q}\,,
\end{eqnarray}
where $\hat h_{p,q}$ is multiplication by a function $h_{p,q}(\hat a)$.
This function can be determined by the recursion relation
\begin{eqnarray}
  h_{p+1,q}(\hat a)= h_+(\hat a) [\Delta^2\cdot h_{p,q-1}](\hat a)
+h_-(\hat a) [\Delta^{-2}\cdot  h_{p,q+1}](\hat a)\,.
\end{eqnarray}
The solution is given by
\beq
\begin{aligned}
  h_{p,q}(\hat a)
=&
\left(
\prod_{r=1}^{|q|}\frac{\Gamma(-2 {\rm sgn}(q) \hat a+r-1)
\Gamma(-2{\rm sgn}(q)\hat a+r+1)}
{\Gamma(-2{\rm sgn}(q)\hat a+r+im)\Gamma(-2{\rm sgn}(q)\hat a+r-im)}
\right) 
\\
&
\hspace{10mm}
\times
\frac{p!}{\left(\frac{p+q}2\right)!
\left(\frac{p-q}2\right)!
}
\frac{
\sin^{\frac{p-|q|}2}(2\pi \hat a+\pi i m)
\sin^{\frac{p-|q|}2}(2\pi \hat a-\pi i m)
}
{\sin^{p-|q|}(2\pi \hat a)}
\,.
\end{aligned}
\label{hpq}
\eeq

Up to $\hat a$-independent factors,
the Liouville three-point function is related to the gauge theory one-loop
determinant
%quantities are related to the gauge theory quantities
as 
 \begin{eqnarray}
%   Z_\text{L}(\hat a,im, q)&=& Z_\text{cl}(\hat a ,q) Z_\text{inst}(\hat a ,{ 1+ im},q)\,,
% \\
C(1+\hat a,1-\hat a, 1+im)&=&
\left|Z_\text{1-loop,pole}(\hat a, im)\right|^2
\nonumber
\\
&=&
Z_\text{1-loop,pole}(-\hat a, -im)
Z_\text{1-loop,pole}(\hat a, im)
\end{eqnarray}
for $\hat a\in i\mathbb R$, 
where $Z_\text{1-loop,pole}(\hat a, \hat m)$ is given in
(\ref{SU2-1loop-pole}).
Thus the Liouville correlator (\ref{eq:N=2*lioiuville}) becomes
\beq
\begin{aligned}
    \langle (\mathcal L_{1,0})^p\rangle
=&
\int_{\hat a\in i\mathbb R} d\hat a \,
Z_\text{1-loop,pole}(-\hat a, -im)
Z_\text{L}(-\hat a, -im, - \bar \tau)
\\
&
\hspace{20mm}
\times
Z_\text{1-loop,pole}(\hat a, im)
\sum_q\left[\hat h_{p,q} \Delta^{2q}
\cdot Z_\text{L}\right]
(\hat a, im, \tau )\,.
\end{aligned}
\eeq
Assuming that we can shift the contour without picking
up residues,%
\footnote{%
We have checked the validity of the contour deformation numerically
by comparing with the S-dual Wilson loop expectation values.}
we can write this as
 \beq
 \begin{aligned}
     \langle (\mathcal L_{1,0})^p\rangle
 &=
 \sum_q 
 \int_{\hat a\in i\mathbb R} d\hat a \,
[\Delta^{-q}\cdot h_{p,q}](\hat a)
\frac
{[\Delta^{-q}\cdot  Z_\text{1-loop,pole}](\hat a, im) }
{[\Delta^{q}\cdot Z_\text{1-loop,pole}](\hat a, im) }
\\
&
\hspace{10mm}
\times 
[\Delta^{q}\cdot Z_\text{1-loop,pole}](-\hat a, -i m)
[\Delta^{q}\cdot Z_\text{L}](-\hat a, -i m,- \bar \tau)
\\
&
\hspace{20mm}
 \times
 [\Delta^{q}\cdot Z_\text{1-loop,pole}](\hat a, im)
[\Delta^{q}\cdot  Z_\text{L}](\hat a, im, \tau )
 \,.
 \end{aligned}
 \eeq
Using (\ref{SU2-1loop-pole}) and (\ref{hpq}),
we can calculate the combination in the first line
and obtain a simple result:
\begin{eqnarray}
&& 
[\Delta^{-q}\cdot h_{p,q}](\hat a)
\frac
{[\Delta^{-q}\cdot  Z_\text{1-loop,pole}](\hat a) }
{[\Delta^{q}\cdot Z_\text{1-loop,pole}](\hat a) }
\nonumber
\\
&&
\hspace{10mm}
=
\frac{p!}{\left(\frac{p+q}2\right)!
\left(\frac{p-q}2\right)!
}
\times
\left\{
  \begin{array}{ll}
\displaystyle
\frac{
\sin^{\frac{p}2}(2\pi \hat a+\pi i m)
\sin^{\frac{p}2}(2\pi \hat a-\pi i m)
}
{\sin^{p}(2\pi \hat a)}
&\text{ for $p$ odd,}
\\    
\displaystyle
\frac{
\cos^{\frac{p}2}(2\pi \hat a+\pi i m)
\cos^{\frac{p}2}(2\pi \hat a-\pi i m)
}
{\cos^{p}(2\pi \hat a)}
&\text{ for $p$ even,}
  \end{array}
\right.
\end{eqnarray}
Comparing this with (\ref{Eq-SU2}) and (\ref{mono-SU2}),
this is precisely $Z_\text{mono}(ia,im;p,q)Z_\text{1-loop,eq}(ia,im,q)$.

Using the relation \cite{AGT}
\beq
   Z_\text{L}(\hat a,im, \tau)
= Z_\text{cl}(\hat a , e^{2\pi i\tau}) Z_\text{inst}(\hat a ,{ 1+ im},e^{2\pi i\tau})\,,
\eeq
we thus obtain
\begin{eqnarray}
\langle( \mathcal L_{1,0})^p\rangle
&=&
\sum_{q=p,p-2,\ldots,-p} 
\frac{p!}{
\left(\frac{p+q}2\right)!
\left(\frac{p-q}2\right)!
}
\int d a\,
Z_\text{1-loop,eq}(ia,im,p)
\label{eq:Liouville-result}\\
&&
\hspace{10mm}
\times
\left|
Z_\text{cl}\left(ia -\frac q 4, e^{2\pi i\tau}\right)
Z_\text{1-loop,pole}\left(ia -\frac q 4, im\right)
Z_\text{inst}\left(ia -\frac q 4, im, e^{2\pi i\tau}\right)
\right|^2
\,.
\nonumber
\end{eqnarray}
After reintroducing the dimensionful parameter $r$,
the gauge theory result
(\ref{eq:N=2*result}) for $\langle T_p\rangle$
and
the Liouville theory expression
(\ref{eq:Liouville-result})
for 
$\langle (\mathcal L_{1,0})^p\rangle$
precisely agree, including the monopole screening contributions!

We note that the charge $p$ 't Hooft loop $T_p$
corresponds to the Liouville operator  $(\mathcal L_{1,0})^p$.
Thus our charge $p$ 't Hooft loop $T_p$ equals the power
$(T_1)^p$ of  the 't Hooft loop that is S-dual
to the spin $1/2$ Wilson loop,
and differs from the S-dual of the spin $p/2$ Wilson loop.
The origin of the power 
is in the natural resolution of the Bogomolny
moduli space.
As explained in \cite{Kapustin:2006pk},
the moduli space of   solutions describing
an array of $p$ minimal 't Hooft loops $T_{p=1}$
develops a singularity when two of the loop operators
collide.
In the limit that all of them are on top of each other,
the magnetic charge of the 't Hooft loop is $p$.
Said
another way,
the singularity of the moduli space can be resolved
by replacing the charge $p$ 't Hooft loop with
a collection of slightly displaced minimal 't Hooft loops.%
\footnote{%
As noted in \cite{Pestun:2007rz},
the localization supercharge $Q$ is indeed compatible
with parallel loop operators each located at a fixed latitude.
}

 \section{Conclusion
}
\label{sec:conclu}

We performed an exact localization
calculation for the expectation value of 
 supersymmetric  't Hooft loop opertors  in 
 $\mathcal N=2$ supersymmetric gauge theories on
 $S^4$.
 These results combined with the exact computation of  Wilson loop expectation values \cite{Pestun:2007rz}
 constitute a suite of exact calculations for  the 
 simplest loop operators in these gauge theories and allow
 for a quantitative study of S-duality for this rich class of gauge theory observables.

A 't Hooft loop was defined by specifying a boundary condition
of the fields in the path integral.  We integrated over the non-singular and
singular solutions to the saddle point equations in the localization computation.
We integrated over the
fluctuations of the fields
around the singular monopole background (\ref{dyonic})
that represents an infinitely heavy monopole with a circular worldline.

 In the leading classical approximation the expectation value 
 was obtained by evaluating the on-shell action in the non-singular background (\ref{dyonic}),
 and the only perturbative quantum corrections%
 \footnote{%
 The one-loop determinants are the unique perturbative corrections
 with respect to the localization action $Q\cdot V$.
 All the perturbative corrections with respect to the physical action \cite{Gomis:2009ir}
 are reproduced by integrating over the zero-mode $a$. }
 in the localization path integral are the one-loop determinants computed using
 the Atiyah-Singer index theorem, arising from the north pole, south pole and equator.
 The 't Hooft loop expectation value receives two types of
 non-perturbative corrections.
 The first is from instantons and anti-instantons  localized at the north and south poles
 as in \cite{Pestun:2007rz}, arising because our localization saddle point
 equations become $F^+=0$ and $F^-=0$ there. 
 One new feature in our calculation is that the Nekrasov instanton partition functions
 at the poles have their argument shifted due to the 't Hooft loop
 background.
 The second type of non-perturbative correction
 occurs as new saddle point field configurations,
 where smooth monopoles in the bulk of $S^4$
 screen the charge of the singular monopole inserted along the loop.
  These arise from non-abelian solutions to the Bogomolny equations
 $D\Phi=*F$, which describe  the saddle point equations in the equator.
 The field configurations were identified as the
 fixed points of an equivariant group action on the moduli space
 of solutions of the Bogomolny equations.

 In this paper we have focused on the computation of 't Hooft operators for
which the magnetic charge and the electric charge vectors are parallel, where the electric
charge is acquired by the Witten effect, due to the non-vanishing topological angle $\theta$. The 
techniques introduced here, however,  can be used to compute general dyonic Wilson-'t Hooft operators. 
The   new ingredient for a dyonic operator is the insertion of a Wilson loop for the unbroken gauge group
preserved by the singular monopole background.

We compared our gauge theory calculations
with some of the predictions in \cite{Drukker:2009id, Alday:2009fs, 
Gomis:2010kv} obtained from computations with topological defects in  Liouville and Toda field theories, and 
 found a perfect match for all
comparisons we have performed.
The physical observables in
Liouville/Toda  theory are known to be invariant
under the modular transformations 
(or more generally under the Moore-Seiberg groupoid)
that are identified with the S-duality transformations
in gauge theory.
Thus our results prove S-duality invariance of the $\mathcal N=2$
gauge theories,
 in the sector of physical observables
involving Wilson and 't Hooft loop operators.
In turn, the progress we made on gauge theory loop operators
 provides motivation to study in more depth the two-dimensional observables.
 In particular the computational techniques for
 {\it topological webs},
 -- 
 the defects involving trivalent vertices
 -- 
 are to be developed in order to make a useful comparison
 with more complicated loop operators in higher-rank gauge theories.

In our study an important role was played by the equivariant index
for the moduli space of solutions of the Bogomolny equations in the presence of a singular
monopole background, created by the 't Hooft operator.
The analysis was similar to that of the instanton moduli space
that led to the Nekrasov partition function,
and we defined the quantity $Z_\text{mono}$ which is 
an analogous physical quantity in the monopole case.
It is possible to generalize and formalize
 the definition of $Z_\text{mono}$ by setting up 
a localization scheme on $S^1\times \mathbb R^3$
\cite{Ito-Okuda-Taki}.

The localization techniques we developed for 't Hooft operators
should also admit generalizations to other supersymmetric disorder operators,
such as monopole and vortex loop operators in three dimensions
and surface operators in four dimensions.
For example, it would be interesting to
formulate a path integral framework that realizes the mathematical
calculations \cite{Braverman:2004vv,Braverman:2004cr,2009InMat.178..299N,2008arXiv0812.4656F} for instantons
in the presence of singularities representing  surface operators.
Also, the localization framework for $\mathcal N=2$ theories on $S^4$
should apply to surface operators preserving 
two-dimensional $\mathcal N=(2,2)$ supersymmetry.
Localization calculations for such observables should help understand
disorder operators 
 in the broad duality web involving
quantum field theories in diverse dimensions.

\vskip+2cm

\section*{Acknowledgements}
We would like to thank Nadav Drukker, Davide Gaiotto, Bruno Le Floch, Yuji Tachikawa, J\"org Teschner and Masato Taki for
helpful discussions.
We would also like to thank the KITP for providing us with a productive environment where part of this work was performed. 
The research at KITP was supported in part by DARPA under Grant No.
HR0011-09-1-0015 and by the National Science Foundation under Grant
No. PHY05-51164. 
T.O. and V.P. also thank 
the Aspen Center for Physics for providing a stimulating
environment.
 J.G. is grateful to  the Theory Group at CERN
for warm hospitality and support. He would also like to thank  LPTENS and
LPTHE in Paris for warm hospitality and the FRIF for support.
Research at the Perimeter Institute is supported in part by the Government of Canada through NSERC and by the Province of Ontario through MRI. 
J.G. also acknowledges further support from an NSERC Discovery Grant and from an ERA grant by the Province of Ontario. 
The research of T.O. was supported in part
by Grant-in-Aid for Scientific Research (B) No. 20340048 from the Japan Society for the Promotion of Science.
The research of V.P. has been partially
supported by a Junior Fellowship
from the Harvard Society of Fellows, and grants NSh-3035.2008.2 and
RFBR07-02-00645.

  \vfill\eject

 %%%%%%%%%%%%%%%%
  \appendix
  
  \section{Supersymmetry and Killing Spinors}
\label{kill}

 The spinors in this paper transform in  a   representation  of $Spin(10)$,  whose generators are constructed from the Clifford algebra $Cl(10)$
\begin{equation}
\{\gamma^M,\gamma^N\}=2\eta^{MN}\qquad~~~~\hbox{where} ~~M=1,\ldots,9,0\,.
\end{equation} 
We take the Euclidean metric $\eta^{MN}=\delta^{MN}$.
 In the chiral representation  
  \begin{equation}
 \gamma^M=\left(
\begin{array}{cc}
 0& \tilde\Gamma^M \\
\Gamma^M&0\end{array}
\right)\,,
 \end{equation}
where $\tilde\Gamma^M\equiv(\Gamma^1,\ldots,\Gamma^9,
-\Gamma^0)$,  and $\Gamma^{M}, \tilde \Gamma^{M}$ are $16\times 16$ matrices which satisfy
\begin{equation}
\tilde\Gamma^M\Gamma^N+\tilde\Gamma^N\Gamma^M=2\delta^{MN}\,,\qquad \Gamma^M\tilde\Gamma^N+\Gamma^N\tilde\Gamma^M=2\delta^{MN}\,.
\end{equation}
The matrices $\tilde\Gamma^M$ and $\Gamma^M$ act respectively on the
negative and positive chirality spinors of $Spin(10)$ since
\begin{equation}
\gamma^{(10)}\equiv-i\gamma^1\ldots\gamma^9\gamma^0=\left(
\begin{array}{cc}
-i \tilde\Gamma^1\Gamma^2\ldots \tilde\Gamma^9\Gamma^0& 0 \\
0&-i \Gamma^1\tilde\Gamma^2\ldots \Gamma^9\tilde\Gamma^0\end{array}
\right)=
\left(
\begin{array}{cc}
1& 0 \\
0&-1\end{array}
\right)
\,.
\end{equation}

 In Euclidean signature, which we use in this paper,   ten
 dimensional spinors are complex. We choose a basis   in which
 $\Gamma^{1},\ldots, \Gamma^9$ are real and $\Gamma^0$
 imaginary. To describe $\Gamma^{M}$ explicitly it is convenient
 to break $\SO(10)$ to $\SO(8) \times \SO(2)$ and use the octonionic
 construction of the Clifford algebra   $Cl(10)$. 
For the explicit expressions which are needed  for explicit
construction of the supersymmetry equations in components we use matrices
as defined in appendix A of  \cite{Pestun:2007rz} with a certain
permutation of spacetime indices. If $\underline\Gamma^{M}$ are the matrices
in \cite{Pestun:2007rz}, then the present $\Gamma^{M}$ are given by
\begin{equation}
\begin{aligned}
  \Gamma^M = \underline \Gamma^{M+1} \quad \text{for} \quad  M = 1,2,3,5,6,7 \\
  \Gamma^{4} = \underline \Gamma^{1} \quad  \Gamma^{8} =
  \underline \Gamma^{5} \quad \Gamma^{9} = \underline \Gamma^{9}
  \quad \Gamma^{0} = i \underline \Gamma^{0}.
\end{aligned}
\end{equation}
The factor of $i$ appears in the relation to $\Gamma^{0}$ because our
present conventions   use the Euclidean metric $\eta^{MN} = \delta^{MN}$, while 
\cite{Pestun:2007rz} used the  Lorentz metric with $\eta^{00} = -1$.

The supersymmetry parameter $\epsilon$ and gaugino $\Psi$ in the $\cN=2$ vectormultiplet are positive chirality spinors of $Spin(10)$,
while hyperino $\chi$ in the $\cN=2$ hypermultiplet is a negative
chirality spinor; they are subject to the projections
\beq
\Gamma^{5678}\epsilon=-\epsilon\,,\qquad\Gamma^{5678}\Psi=-\Psi\,, \qquad\Gamma^{5678}\chi=\chi\,,
\eeq
 where $\Gamma^{5678}\epsilon=  \tilde\Gamma^{5}\Gamma^{6}\tilde\Gamma^{7}\Gamma^8\epsilon$.

The conformal Killing spinor equation  \rf{killspin} in the $B_3\times S^1$ metric \rf{ballmetric} is
 \begin{eqnarray}
  \nabla_\mu \epsilon &=&\tilde{\Gamma}_\mu \tilde\epsilon
    \label{killapp1}\\
  \tilde{\Gamma}^\mu \nabla_\mu \tilde\epsilon &=& -{1\over 4r^2}{1\over \left(1-{|\vec x|^2\over 4r^2}\right)} \epsilon\,. 
  \label{killapp2}
\end{eqnarray}
In the vielbein basis $e^{\hat i}=e^i=dx^i$ and $e^{\hat 4}=r \left(1-{|\vec x|^2\over 4r^2}\right) d\tau$, 
the non-zero components of the spin connection are
\begin{equation}
 w^{\hat 4i}= -w^{i\hat 4}=-{x^i\over 2r}d\tau\qquad i=1,\ldots,3\,.
\end{equation}
Equation \rf{killapp2} implies that $\tilde\epsilon=\epsilon_c(\tau)$ while the first three equations in \rf{killapp1}   imply  that $\epsilon= \epsilon_{s}(\tau)+x^i\tilde\Gamma_i \epsilon_c(\tau)$. The solution to the equation
\begin{equation}
 \nabla_\tau \epsilon =\tilde{\Gamma}_\tau \tilde\epsilon
\end{equation}
is \rf{killsol}
\begin{equation}
 \epsilon=\cos(\tau/2)\left(\hat\varepsilon_s+x^i\tilde\Gamma_i\,\hat\varepsilon_c\right)+\sin(\tau/2)\,\tilde\Gamma^4\left(2r\, \hat\varepsilon_c+{x^i\over 2r}\Gamma_i\,\hat\varepsilon_s\right)\,,
\label{eq:epsilon_sc}
 \end{equation}
 with $\hat\varepsilon_s$ and $\hat\varepsilon_c$   two constant ten dimensional Weyl spinors of opposite chirality.

\section{Lie Algebra Conventions}
\label{sec:lie-alg-conv}

Let $G$ be a compact Lie group and $\g$ the Lie algebra of
$G$. 
As a vector space $\g$ is isomorphic to $ \BR^{\,\dim G}$. 
In our conventions, for a gauge theory with gauge group $G$, the fields
$A_{\mu}, F_{\mu \nu}$ and $\Phi_A$ ($A=0,9$)
of  the vectormultiplet
take values in $\g$. In particular, we write the
covariant derivative as $D\equiv D_{A} = d + A$ and the curvature as $F_{\mu \nu}
= [D_{\mu}, D_{\nu}] $. 
If $G$ is $\U(N)$ or $SU(N)$, 
the basis $\{T_{\alpha}\}$ of the Lie algebra 
$\mathfrak{g}$
% = \mathfrak{u}(N)$ 
can be
represented by $N \times N$ antihermitian matrices. 
Given the basis,
%$\{T_{\alpha}\}$ of $\g$, 
the real coordinates $a^{\alpha}$ of
an element $a \in \g$ are defined by the expansion $a = a^{\alpha}
T_{\alpha}$.
Let $\g_{\BC} =\g \otimes \BC$ be the complexification of $\g$. An
element $a = x + i y$ of $\g_{\BC}$, where $x, y \in \g$, 
can be written as $a  = a^{\alpha} T_{\alpha}$
with $a^{\alpha}$ being complex numbers. 
We say that  an element $a$ of
$\g_{\BC}$ 
is real if the coordinates $a^{\alpha}$ are real.
Complex conjugation acts by conjugating the coefficients:
$a=a^\alpha T_\alpha\rightarrow \overline{a^\alpha} 
T_\alpha$.

If $G$ is a compact  
Lie group, then the Lie algebra $\g$ of $G$
can be equipped with a positive definite bilinear form 
$(\bullet\,,\bullet): \g
\times \g \to \BR$ invariant under the adjoint  action of $G$.
Such a bilinear form $ \g \times \g \to \BR$ 
is  defined uniquely up to a scaling, and extends holomorphically
to $\g_\BC\times \g_\BC \rightarrow \BC$.
For $\g = \mathfrak{u}(N)$
and $\g =\mathfrak{su}(N)$,
we choose $(\bullet\, ,\bullet)$, also donoted by $\bullet \cdot \bullet$, to be given by 
minus the trace in the fundamental
representation: $ (a,b) = - \Tr a b$.

The basis elements $T^{\alpha}$ in the Cartan algebra $\mathfrak{t}$ of $\mathfrak{u}(N)$
can be represented by the diagonal $N \times N$ matrices
\begin{equation}
  T_{\alpha} = i \diag(0,\dots, 0, 1, 0, \dots, 0)\,,
\end{equation}
where $1$ is at the position $\alpha$. Since in this basis the bilinear form
is the identity matrix, $- \Tr T_{\alpha} T_{\beta} = \delta_{\alpha \beta}$,  
we do not distinguish between contravariant and covariant Lie algebra indices.
For an element $a = a^\alpha T_\alpha $  of $\mathfrak{t}$ we refer to $a$ using the following
equivalent notations 
\begin{equation}
   (a_1, \dots, a_N)  \leftrightarrow a  \leftrightarrow  a_\alpha T_\alpha 
  =
  \begin{pmatrix}
   i a_1 & 0 & \dots \\
    0    & i a_2 & \dots \\
    \vdots & \vdots & \ddots 
  \end{pmatrix}\,,
\label{mat-notation}
\end{equation}
where $a_{\alpha}$ are real.
When dealing with complexification $\mathfrak{t}_{\BC}$ we allow $a_{\alpha}$ to
be complex. 
The notation (\ref{mat-notation}) is also used for $\g=\mathfrak{su}(N)$.
For example, the Nekrasov instanton partition function $Z_{\text{inst}}$
takes a complex element $\hat a$ of $\mathfrak{t}_{\BC}$,
{\it i.e.} the Coulomb parameter, as one of its
  arguments. 
We use equivalently the following forms
  referring to $Z_{\text{inst}}$ evaluated at $\hat a$
\begin{equation}
  Z_{\text{inst}}(\hat a;\ve_1,\ve_2) = Z_{\text{inst}}( (\hat a_1,
  \dots, \hat a_N), \ve_1, \ve_2) = Z_{\text{inst}}( \hat a_1,
  \dots, \hat a_N; \ve_1, \ve_2)\,.
\end{equation}
It should be clear from the context what $\hat a$ refers to in the main text.

\section{Coordinates and Weyl Transformations on $S^4$}
\label{coords}

The $SO(5)$ isometry of the round metric on $S^4$ is made manifest  by the induced metric on the following hypersurface in $\bR^5$
\begin{equation}
X_1^2+\ldots+X_5^2=r^2\,.
\end{equation}
 In this paper, a certain $U(1)_J\subset SO(5)$ isometry of $S^4$ generated by the generator $J$ plays a key role. It acts on the embedding coordinates as
 \beq
 \begin{aligned}
 X_1+iX_2&\rightarrow e^{i\ve}(X_1+iX_2)\\
X_3+iX_4&\rightarrow e^{i\ve}(X_3+iX_4)\,,
\end{aligned}
\eeq
 and its fixed points   $X_5=\pm r$ define the north and south pole of $S^4$.
The following coordinates are of use in the paper:

\vskip+5pt
\noindent {\it Latitude Coordinates}:
The metric is given by
\begin{equation}
ds^2=r^2(d\vartheta^2+\sin^2\vartheta d\Omega_3)\,
\end{equation}
where $d\Omega_n$ is the metric on the unit $S^n$ and $\vartheta$ is the latitude angle on $S^4$, with $\vartheta=0,\pi/2$ and $\pi$ corresponding to the north pole,   equator and south pole respectively. The embedding coordinates are 
\beq
\begin{aligned}
X_a&=r\sin\vartheta\,  n_a\qquad a=1,2,3,4\\
X_5&=r\cos\vartheta\,,
\end{aligned}
\eeq
where $n_a$ is a unit vector in $\bR^4$ parametrizing $S^3$.

The $U(1)_J$ action induced by $J$ is realized by the Hopf fibration. Consider $S^3: |w_1|^2
+ |w_2|^2 = 1$, $(w_1,w_2) \in \BC^2$ and the $\U(1)$ action  $(w_1,
w_2) \mapsto (e^{i \ve}w_1, e^{i \ve} w_2)$. Introduce   angular
coordinates on $\BC^2$:  $w_1=\rho \cos\frac\eta 2  e^{i \psi}$ and 
$w_2=\rho \sin\frac\eta 2  e^{i\psi+i\varphi}$ so that the $U(1)_J$ acts by shifts $\psi \rightarrow \psi+\ve$, and consider the map
$\bar \BC^2 \otimes \BC^2 \to \BR^3$
\begin{equation}
  \vec{x} = \bar w \vec{\sigma}  w = \rho^2 ( \sin \eta \cos  \varphi,
  \sin \eta \sin \varphi, \cos \eta),
\end{equation}
so that $(\rho^2, \eta, \varphi)$ are the spherical  coordinates on   
$\BR^3$. 
Rewriting the flat metric on $\BC^2$ in the $(\rho, \eta, \varphi,
\psi)$ coordinates we get
\begin{equation}
  ds^2 = d \rho^2 + \rho^2\left( \frac {1} {4} d \eta^2 
+ \frac 1 2 (1 -  \cos \eta) d \varphi^2 + (1- \cos \eta) d \varphi d \psi + d \psi^2
\right)\,.
\end{equation}
 The unit $S^3$  is at $\rho =1$ with metric 
\begin{equation}
  d\Omega_3=
 \frac {1} {4} d \eta^2 + \frac 1 2 (1 -
  \cos \eta) d \varphi^2 +  (1- \cos \eta) d \varphi d \psi + d
  \psi^2 =
\frac14 d\Omega_2+(d\psi+\omega)^2\,,
\label{eq:Hopf}
\end{equation}
where 
\begin{equation}
  d\Omega_2 = d \eta^2 + \sin^2 \eta\, d \varphi^2
\end{equation}
and 
\begin{equation}
  \omega = \frac 1 2 (1 - \cos \eta) d\varphi.
\end{equation}
In these coordinates, the $U(1)_J$ vector field is $v = \frac{1}{r}\frac{\p}{\p \psi}$, 
and the dual 1-form
used in section \ref{sec:vanishing-theorem} is 
\begin{equation}
  \tilde v = {dx^{\mu} h_{\mu \nu}v^{\nu} \over v^{\mu} v_{\mu}} = 
r(d \psi + \omega)\,.
\end{equation}
The one-form $\omega$ satisfies $d\omega=\frac12{\rm vol}(S^2)$.

\vskip+5pt
\noindent {\it $S^2\times S^1$ Foliation Coordinates}:
The metric is given by
\begin{equation}
ds^2=r^2(d\xi^2+\sin^2\xi d\Omega_2+\cos^2\xi d\tau^2)\,
\end{equation}
where $\tau$ is the coordinate on $S^1$ and $0\leq\xi\leq \pi/2$. The embedding coordinates are given by
\beq
\begin{aligned}
X_1+iX_2&=r\cos\xi e^{i\tau}\\
X_3+iX_4&=r\sin\xi\sin\alpha e^{i\phi}\\
X_5&=r\sin\xi\cos\alpha\,,
\end{aligned}
\eeq
where $d\Omega_2=d\alpha^2+\sin^2\alpha d\phi^2$.  The $U(1)_J$ symmetry generator $J$ acts by shifts
\beq
\begin{aligned}
\tau&\rightarrow\tau+\ve\\
\phi&\rightarrow\phi+\ve\,.
\end{aligned}
\eeq
In these coordinates the north and south pole are at $(\xi=\pi/2,\alpha=0)$ and $(\xi=\pi/2,\alpha=\pi)$ respectively.

\vskip+5pt
\noindent {\it $B_3\times S^1$ Foliation Coordinates}:
The metric is given by
\begin{equation}
ds^2={\sum_{i=1}^3dx_i^2\over \left(1+{|\vec x|^2\over 4r^2}\right)^2} + 
r^2 {\left(1-{|\vec x|^2\over 4r^2}\right)^2\over \left(1+{|\vec x|^2\over 4r^2}\right)^2}d\tau^2\,,
\end{equation}
and $|\vec x|^2\leq 4r^2$ defines the three-ball $B_3$. The embedding coordinates are given by
\beq
\begin{aligned}
X_1+iX_2&=r{\left(1-{|\vec x|^2\over 4r^2}\right)\over \left(1+{|\vec x|^2\over 4r^2}\right)}e^{i\tau}\\
X_I&={x_{I-2}\over \left(1+{|\vec x|^2\over 4r^2}\right)}\qquad I=3,4,5\,. 
\label{bs1cooord}
\end{aligned}
\eeq
The $U(1)_J$ symmetry generator $J$ acts by 
\beq
\begin{aligned}
x_1+ix_2&\rightarrow e^{i\ve}(x_1+ix_2)\\
\tau&\rightarrow\tau+\ve\,.
\end{aligned}
\eeq
In these coordinates the north and south pole are at $\vec x=(0,0,2r)$ and 
$\vec x=(0,0,-2r)$ respectively.

\section{$Q$-Invariance of the 't Hooft Loop Background}
\label{sec: soluns}

 The background created by a circular 't Hooft loop with magnetic weight $B$ located at $\vec x=0$ in the $B_3\times S^1$ metric \rf{ballmetric} takes the same form as that of a static  't Hooft line in flat spacetime \rf{scalarpole}  (for $\theta=0$)
\begin{equation}
\label{eq:monopole}
  \begin{aligned}
  F_{jk} &= -\frac B 2 \ep_{ijk} \frac{x_i}{|\vec x|^3}  \\
  \Phi_9 &= \frac{B}{2 |\vec x|}\,. 
  \end{aligned}
\end{equation}
Since $B\in \mathfrak{t}$ takes values in the Cartan subalgebra of the gauge group $G$, the singularity is abelian in nature.

We can verify that the the deformed monopole equations (\ref{eq:sp1}, \ref{eq:sp2}, \ref{eq:sp3})
are solved by the 't Hooft loop background (\ref{eq:monopole}).
For example, let's consider the first spatial equation (\ref{eq:sp1}). In the background (\ref{eq:monopole})
$F_{1\hat4},F_{3\hat 4}, K_1, D_{\hat 4}\Phi_9$ vanish. 
We group the remaining   terms to make the structure of cancellation
obvious using  that  (\ref{eq:monopole}) satisfies
\begin{equation}
  D_i \Phi_9 =\partial_i \Phi_9= -\frac{B x_i}{2 |\vec x|^3} =\frac 1 2 \epsilon_{ijk}F_{jk}\qquad i,j,k=1,2,3\,,
  \label{BPS}
\end{equation}
where the first equality is due to the abelian nature of the background \rf{eq:monopole}.
Evaluation yields for (\ref{eq:sp1})
\begin{eqnarray}
&& 4r^2(-D_1 \Phi_9 + F_{23}) + (-x_1^2 + x_2^2 + x_3^2) (D_1 \Phi_9 + F_{23})
+(-2x_1 x_2) (D_2 \Phi_9 - F_{13}) 
\nonumber\\
&&
\hspace{85mm}+(-2 x_1 x_3) (D_3 \Phi_9 + F_{12})+
- 2 x_1  \Phi_9  
\nonumber\\
&=&0+ (-x_1^2 + x_2^2 +x_3^2) \frac{-Bx_1}{|\vec x|^3}
+(-2 x_1 x_2) \frac{-Bx_2}{|\vec x|^3}
+(-2 x_1 x_3) \frac{-Bx_3}{|\vec x|^3} 
- \frac{ Bx_1}{|\vec x|} 
\nonumber\\
&=&
  \frac{Bx_1 x^2}{|\vec x|^3} - \frac{ Bx_1}{|\vec x|}
=0\,.
\end{eqnarray}
The cancellation in the second deformed monopole equation (\ref{eq:sp2})
 is exactly the same with
replacement of indices $1 \to 2$. 
In the third equation (\ref{eq:sp3}), the relative signs are different, but again all
terms cancel similarly. In analyzing the last equation (\ref{eq:sp3}) we also find that 
$\Phi_0$ can be turned on as long as  
\beq
K_3= -{\Phi_0/r\over 1+ {|\vec x|^2\over 4r^2} }\,.
\label{constraint}
\eeq
This observation plays an important role in finding the most general solution to the saddle point equations, as discussed in section \ref{sec:vanishing-theorem}.

Similarly, it is very easy to show that the    invariance
equations \rf{eq:invariance}
 are satisfied by the background \rf{eq:monopole}. For these only $D_{i} \Phi_9$ and $F_{jk}$   contribute and
 cancel elementarily due to formula \rf{BPS}. These equations also exhibit that $\Phi_0$ has a zeromode, given by 
\begin{equation}
\label{eq:Phi_0-profile}
  \Phi_0 = \frac{a}{1 + {|\vec x|^2\over 4r^2}}\,,
\end{equation}
and therefore, due to \rf{constraint}
\beq
K_3= -{a/r\over \left(1+ {|\vec x|^2\over 4r^2}\right)^2}\,,
\label{constrainta}
\eeq
where $a\in \mathfrak{t}$ is constant. In comparison with \cite{Pestun:2007rz} the profile of $\Phi_0$ is not constant in $B_3\times S^1$. However, since the metric on $B_3\times S^1$ and $S^4$ are related by a Weyl transformation with $\Omega= \left(1+{|\vec x|^2\over 4r^2}\right)$, it follows from \rf{transform} that the Weyl transformation makes $\Phi_0$ constant in $S^4$, as found in  \cite{Pestun:2007rz} .

It is straightforward to show that the background created by the 't Hooft loop when $\theta\neq 0$  \rf{dyonic}
\beq
\begin{aligned}
  F_{jk} &= -\frac B 2 \ep_{ijk} \frac{x_i}{|\vec x|^3}\,, \qquad F_{i\hat 4}=-ig^2\theta {B\over 16\pi^2} \frac{x_i}{|\vec x|^3}\,,\\
  \Phi_9 &= \frac{B}{2 |\vec x|}\,,\qquad\qquad\quad \Phi_0 = -g^2\theta {B\over 16\pi^2} {1\over  |\vec x|}\,,
  \label{appmon}
\end{aligned}
\eeq
solves the localization equations $Q\cdot \Psi=0$
As we have already demonstrated that the terms involving $\Phi_0$ and $F_{jk}$ cancel in the  invariance
equations \rf{eq:invariance}  and deformed monopole equations \rf{eq:sp1}\rf{eq:sp2}\rf{eq:sp3}, we just have to exhibit cancellation of the terms involving 
$\Phi_0$ and $F_{i4}$. Since      the 't Hooft loop background
 is $\tau$ independent and abelian (i.e. $[\Phi_0,\Phi_9]=0$),
  we are just left to verify from the  invariance
equations \rf{eq:invariance} that
\beq
\begin{aligned}
{1\over 2r} F_{i4}+\left[D_i, {i\over 2}\left(1+{|\vec x|^2\over 4r^2}\right)\Phi_0\right]&=0\qquad i=1,2,3\\
x_1F_{42}-x_2F_{41}&=0\,.
\end{aligned}
\eeq
Using that $F_{i4}=r\left(1-{|\vec x|^2\over 4r^2}\right)F_{i\hat 4}$ and   $D_i\left(\left(1+{|\vec x|^2\over 4r^2}\right)\Phi_0\right)=-{x_i\over |\vec x|^2}\left(1-{|\vec x|^2\over 4r^2}\right)\Phi_0={i\over r}F_{i4}$, we conclude that \rf{appmon} solves the equations  \rf{eq:invariance}.

We now verify that the   deformed monopole equations \rf{eq:sp1}\rf{eq:sp2}\rf{eq:sp3}
are solved by $\Phi_0$  and $F_{i\hat 4}$ of the 't Hooft loop background (\ref{appmon}).
From  \rf{eq:sp1}\rf{eq:sp2} we get
\beq
-x_3 F_{i\hat 4}+x_i F_{3\hat 4}=0\qquad i=1,2\,,
\eeq 
which is trivially satisfied by (\ref{appmon}).
From \rf{eq:sp3} the relevant equation is
\beq
    i  {\Phi_0}  -{x_1} F_{1\hat4}
 -  {x_2} F_{2\hat 4} 
 -  {x_3}  F_{3\hat4} =0
\eeq
which is indeed solved by the 't Hooft loop background (\ref{appmon}).

This concludes the explicit check that the direct sum of the   
monopole background configuration (\ref{appmon})
and the $\Phi_0$ zeromode profile (\ref{eq:Phi_0-profile}) with the associated auxiliary field   $K_3$ \rf{constrainta} solve  the localization
equations $Q\cdot \Psi = 0$.

\section{Hypermultiplets in General Representations}
\label{sec:matter-general}

In this appendix we will derive the formula (\ref{eq:index-univ})
of the one-loop index 
for hypermultiplets in an arbitrary representation.

We will do this by generalizing, and also applying in a suitable way, 
the formula (\ref{eq:index-adj}) 
that is valid for the adjoint representation.
Let us begin with $\mathcal N=2^*$ theory in flat space
which we regard as a dimensional reduction of the super Yang-Mills
in ten dimensions.
The group $SO(4)$ that rotates the 5678 directions factorizes
into the product of the R-symmetry group $SU(2)_R$ and the flavour
symmetry group $SU(2)_\text{F}$.

In order to derive (\ref{eq:index-univ}) for
complex and real representations, let us take the gauge group 
to be $U(2)$.
Applying the adjoint formula (\ref{eq:index-adj})  to this case,
we find the index%
\footnote{%
We neglect the terms with zero weights.
}
\begin{equation}
  -  \frac { 
e^{\frac 1 2 (i \ve_1 + i \ve_2) }
}
 {(1 -e^{i\ve_1})(1-e^{i\ve_2})}
\frac{e^{i \hat m} + e^{-i \hat m}}{2}
\left(
e^{i(\hat a_1-\hat a_2)}+ e^{-i(\hat a_1-\hat a_2)}
\right)\,,
\label{eq:index-U2-hyper}
\end{equation}
where ${\rm diag}(e^{i \hat a_1}, e^{i\hat a_2})$ and 
${\rm diag}(e^{i \hat m},e^{-i \hat m})$ parametrize
the maximal tori of $U(2)$ and $SU(2)_\text{F}$ that
we denote by $U(1)_1\times U(1)_2$ and $U(1)_\text{F}$ respectively.
Under $U(1)_1\times U(1)_2\times U(1)_\text{F}$,
off-diagonal fields in the hypermultiplet transform in representations
with charges $(+1,-1,\pm 1)$ and their complex conjugate.
The trick is to consider a new $\mathcal N=2$ theory that
is obtained by setting all the off-diagonal components of the $U(2)$ adjoint
fields in the vector multiplet to zero,
regarding $G'=U(1)_1$ as a new gauge group.
We also project the hypermultiplet fields onto those with charges
$(+1,-1,+ 1)$ and their conjugate,
and regard $U(1)'_\text{F}\equiv [U(1)_2\times U(1)_\text{F}]_\text{diag}$
as a new flavour group.
The hypermultiplet index for the new theory
is obtained from (\ref{eq:index-U2-hyper}) by keeping the relevant terms:
\begin{equation}
  -  \frac { 
e^{\frac 1 2 (i \ve_1 + i \ve_2) }
}
 {2(1 -e^{i\ve_1})(1-e^{i\ve_2})}
\left(
e^{i \hat a'-i \hat m'}
+
e^{-i \hat a'+i \hat m'}
\right)\,,
\label{eq:index-U1}
\end{equation}
where  $\hat a'\equiv \hat a_1$, and
the Coulomb parameter $\hat a_2$ and the original mass parameter $\hat m$
have combined into a new mass parameter $\hat m'\equiv \hat a_2-\hat m$.
Thus we have derived the formula (\ref{eq:index-univ}) for the
spacial case of gauge group $U(1)$ and a single charged hypermultiplet.
Noting that a general complex irreducible representation 
of an arbitrary gauge group $G$ can be thought of as embedding $G$ into
$U(\dim R)$ whose maximal torus 
is $U(1)^{\dim R}$, this $U(1)$ result implies the formula
(\ref{eq:index-univ}) for any complex $R$.

Similarly any strictly real irreducible representation
defines an embedding of $G$ into $SO(\dim R)$
with maximal torus $SO(2)^{[\dim R/2]}$.
Noting that the vector representation
of $SO(2)\simeq U(1)$ gives the minimal
real irreducible representation,
the $U(1)$ formula (\ref{eq:index-U1}) also generalizes to
(\ref{eq:index-univ}) for any real representation $R$.

To treat the case where $R$ is a pseudo-real representation,
let us begin with $\mathcal N=4$ theory with
gauge group $SU(3)$ and perform a projection as follows.
We pick a subgroup $G'=SU(2)$ of $SU(3)$ as a new gauge group, and 
denote its commutant by $U(1)'$.
We parametrize the maximal torus of $G'\times U(1)'$ by
${\diag}(e^{i(\hat a+\hat b)}, e^{i(-\hat a+\hat b)}, e^{-2i\hat b})$.
Let us keep only the vectormultiplet fields for $G'$.
Under the embedding 
\beq
SU(3)\times SU(2)_\text{F}
\supset
SU(2)\times U(1)'\times U(1)_\text{F}
\eeq
where $U(1)_\text{F}$ is the maximal torus of $SU(2)_\text{F}$,
 the hypermultiplet splits as 
\beq
(\text{adj},{\bf 2})
\rightarrow \ldots\oplus
 {\bf 2}_{+1,+1}
\oplus {\bf 2}_{+1,-1}
\oplus {\bf 2}_{-1,+1}
\oplus {\bf 2}_{-1,-1}
\oplus\ldots\,.
\eeq
We project the hypermultiplets onto $ {\bf 2}_{-1,+1}$
and its conjugate ${\bf 2}_{+1,-1}$.
Picking the diagonal $U(1)'_\text{F}\equiv
[U(1)'\times U(1)_\text{F}]_\text{diag}$,
we get half-hypermultiplets in the pseudo-real representation
${\bf 2}$ of gauge group $SU(2)$ with flavour symmetry
$SO(2)\simeq U(1)'$.
This is the minimal case involving a pseudo-real representation.
We can obtain the hypermultiplet index in the present case 
by keeping relevant terms in the adjoint formula (\ref{eq:index-adj}):
\begin{eqnarray}
&&-  \frac { 
e^{\frac 1 2 (i \ve_1 + i \ve_2) }
}
 {(1 -e^{i\ve_1})(1-e^{i\ve_2})}
\frac{ e^{i\hat m}+ e^{-i\hat m}}2
\left(
e^{2i\hat a}+e^{-2i\hat a}
+e^{i\hat a+i\hat b}+e^{-i\hat a-i\hat b}
+e^{-i\hat a+i\hat b}+e^{i\hat a-i\hat b}
\right)
\nonumber
\\
&\rightarrow &
-  \frac { 
e^{\frac 1 2 (i \ve_1 + i \ve_2) }
}
 {2(1 -e^{i\ve_1})(1-e^{i\ve_2})}
(e^{i\hat a'+i\hat m'}+e^{-i\hat a'+i\hat m'}+e^{i\hat a'-i\hat m'}
+e^{ -i\hat a'-i\hat m'})
\,,
\label{eq:index-SU2fund}
\end{eqnarray}
where the arrow indicates the projection and we have defined
$\hat m'\equiv \hat m -\hat b$.
The expression (\ref{eq:index-SU2fund}) 
is a special case of (\ref{eq:index-univ}).
For any gauge group $G$, a pseudo-real
representation 
defines a homomorphism from $G$ to the group $Sp(\dim R)$
whose Cartan subalgebra is isomorphic to that of 
$Sp(2)^{\frac 1 2 \dim R}=SU(2)^{\frac 1 2 \dim R}$.
The Cartan subalgebra of flavour $SO(2N_\text{F})$
is isomorphic to that of $SO(2)^{N_\text{F}}\simeq U(1)^{N_\text{F}}$.
Thus this minimal case (\ref{eq:index-SU2fund}) implies the formula
(\ref{eq:index-univ}) for any pseudo-real representation $R$.

\section{Singular Monopoles and Instantons}
\label{sec:Kronheimer}

Solutions of the Bogomolny equations are related
to $U(1)$ invariant instantons
\cite{Kronheimer:MTh}.
Let us consider a gauge field $\mathcal A$
in $\mathbb R^4\simeq \mathbb C^2$.
We regard the four-dimensional space $\mathbb C^2$
as a $U(1)$ fibration over $\mathbb R^3$
using the map
\begin{eqnarray}
  (z_1,z_2)\mapsto ( z_1, \bar z_2)\vec\sigma 
\left(
  \begin{array}{c}
    \bar z_1\\
z_2
  \end{array}
\right)=:\vec x
\label{z-x}
\end{eqnarray}
from $\mathbb C^2$ to $\mathbb R^3$.
The right-hand side is invariant under $(z_1,z_2)\rightarrow
(e^{-i\nu} z_1, e^{i\nu}z_2)$.
We will denote this symmetry group  by $U(1)_K$.
If $\psi$ is a coordinate of the $U(1)_K$ orbits,
the four-dimensional metric is given by
\begin{eqnarray}
  ds^2_{\mathbb C^2}=
\frac{1}{4x}
(
d \vec x ^2+
4 x^2(d\psi+\omega)^2)\,,
\label{eq:4Dmetric}
\end{eqnarray}
where 
\begin{eqnarray}
x=|\vec x|
\end{eqnarray}
and $\omega$ is a 1-form on $\mathbb R^3$ such that
\begin{eqnarray}
2 d\omega={\rm vol}(S^2)
\end{eqnarray}
is the volume form on the unit two-sphere.\footnote{%
For example, if we take angular parametrization 
$z_1=x^{1/2} \cos\frac\eta 2  e^{-i\psi}$ and 
$z_2=x^{1/2}\sin\frac\eta 2  e^{i\psi+i\varphi}$, 
then $\omega = \frac 1 2 (1- \cos \eta) d \varphi$.
}
In accord with this fibration structure,
we decompose the four-dimensional gauge field as
\begin{eqnarray}
\mathcal A=A+ 2x(d\psi+\omega) \Phi\,,
\label{ea:4DA3DAphi}
\end{eqnarray}
where $A=A_i dx^i$ and $\Phi$ are the connection and a
scalar on $\mathbb R^3$.
If we assume that $A$ is independent of $\psi$,
or equivalently invariant under the $U(1)_K$ action,
the four-dimensional curvature 
$ \mathcal F=d\mathcal A+\mathcal A\wedge \mathcal A$
decomposes as
\begin{eqnarray}
  \mathcal F&=&d\mathcal A+\mathcal A\wedge\mathcal A
\nonumber\\
&=& F- 2x(d\psi+\omega)\wedge  D\Phi+\frac \Phi x
(x^2 {\rm vol}(S^2)+2x dx\wedge (d\psi+\omega))\,,
\label{eq:4DF}
\end{eqnarray}
where $F=dA+A\wedge A$ and $D=d+[A,\cdot\, ]$
are the three-dimensional curvature and covariant derivative.
Its dual with respect to the four-dimensional metric
(\ref{eq:4Dmetric})
 is given by%
\footnote{%
To compute the Hodge star, 
we need to know the orientation of $\mathbb C^2$
in terms of our coordinates.
The standard orientation of $\mathbb C^2$ 
corresponds to
the {\it sign} of the volume form
${\rm vol}(\mathbb C^2)
\propto -  dx d\eta d\varphi d\psi
$ in the
angular parametrization. Indeed, at 
 $(x,\eta,\varphi,\psi)=(1,0,0,0)$ we have
$d{\rm Re}z_1\supset dx, d{\rm Im}z_1\supset 
-d\psi,
d{\rm Re} z_2\supset d\eta,
d{\rm Im} z_2\supset d\varphi
$,
so ${\rm vol}(\mathbb C^2) \propto -dx d\eta d\varphi d\psi  $.
The three-dimensional volume form is ${\rm vol}(\mathbb R^3)=x^2\sin\eta
dx d\eta d\varphi$.
}

\begin{eqnarray}
 *_4 \mathcal F&=&
-(*_3 F)\wedge 2x(d\psi+\omega)
- *_3 D\Phi
-\frac \Phi x
\left(x^2 {\rm vol}(S^2)+2xdx\wedge (d\psi+\omega)\right)\,.
\label{eq:4DstarF}
\end{eqnarray}
Comparing (\ref{eq:4DF}) and (\ref{eq:4DstarF}) we see that
 the anti-self-duality equations $\mathcal F^+=0$ in four dimensions
is equivalent to
the Bogomolny equations
\begin{eqnarray}
  F=*_3 D\Phi
\end{eqnarray}
in three dimensions.
Thus $U(1)_K$-invariant instantons
are in a one-to-one correspondence with
solutions of the Bogomolny equations.

To be more precise, we need to specify the boundary conditions
we impose in three and four dimensions.
In three dimensions, we require that
the Higgs field $\Phi$ vanishes at infinity.
As we see from (\ref{ea:4DA3DAphi}) this is indeed necessary if
$\mathcal A$ at infinity becomes pure gauge $g^{-1}dg$
with $g: S^3\rightarrow G$ depending only
on the angular directions of $\mathbb C^2$.

To understand the appropriate boundary condition
at the origin, let us consider
the trivial background $\mathcal A=0$
on $\mathbb C^2$.
Let $ w$ be a coweight of the gauge group $G$.
We recall that a coweight is an element of the Lie algebra,
and discretely quantized in such a way that the exponential
$e^{ B \psi}$ is invariant under $\psi\rightarrow \psi+2\pi$.
A singular gauge transformation by $e^{B \psi}$ induces
a non-trivial field 
\begin{eqnarray}
  \mathcal A= 
%g_{B}^{-1}dg_{B}
e^{-B\psi} d e^{B\psi}
=- B\, \omega+2x (d\psi+\omega)\frac{B
}{2x}\,,
\end{eqnarray}
{\it i.e.},
\begin{eqnarray}
  A=-B
\,\omega\,,~~~F=-\frac{B}2 {\rm vol}(S^2)\,,~~~~
\Phi=\frac{B}{2x}\,.
\end{eqnarray}
This is precisely the 't Hooft operator background in the transverse
directions to the loop.
If we start with a general gauge field, after the singular gauge
transformation by $e^{ B\psi}$, 
the group $U(1)_K$ acts as an isometry that shifts $\psi$
as well as a linear transformation on the fibers of the gauge bundle.
In general, a smooth gauge field on $\mathbb C^2$ in variant under the $U(1)_K$
group action
becomes
a field configuration in three dimensions
that obeys the boundary condition appropriate for the 't Hooft loop.
The linear transformation on the fiber at the origin
encodes the magnetic charge of the 't Hooft operator.
In fact, one can reverse the logic and
use this connection with instantons to {\it define} the precise boundary
conditions for singular solutions of the Bogomolny equations, which
is otherwise difficult to specify.
See for example \cite{MR1624279}, where this definition
of boundary conditions was concretely used to compute the
dimension of the moduli space by suitably applying the
index theorem.

\section{Instanton Partition Functions for  $U(N)$}
\label{sec:Nek-func}

For $G=U(N)$, the localization calculation represents the 
instanton partition function $Z_\text{inst}$
as a sum over the set of the $\U(1)_{\epsilon_1}\times \U(1)_{\epsilon_2} \times \U(1)^N$-fixed  points on
the moduli space of non-commutative instantons on $\mathbb C^2$.
For each fixed point, we need to compute the equivariant Euler 
character of the self-dual complex
\begin{equation}
\label{eq:sd-complex2}
D^\text{vm}
 :  \Omega^{0} \otimes {\ad(\g)} \stackrel{D}{\to} \Omega^{1}\otimes \ad(\g) \stackrel{D_{+}}{\to}
   \Omega^{2 +} \otimes {\ad(\g)}\,.
\end{equation}

Note that we can decompose 
the complexified spaces of differential forms
as
 $\Omega^0_\BC\simeq \Omega^{0,0}$, $\Omega^1_\BC\simeq \Omega^{1,0}\oplus
\Omega^{0,1}$, 
$\Omega^{2+}_\BC\simeq \Omega^{2,0}\oplus\Omega^{0,0} \kappa \oplus \Omega^{0,2}$,
where $\kappa$ is the K\"ahler form.
Using Hodge duality,  we also  have the relations $\Omega^{2,2}\simeq \Omega^{0,0}$
and $\Omega^{2,1}\simeq \Omega^{1,0}$.
It follows that the complexification of the self-dual complex (\ref{eq:sd-complex2})
is isomorphic to the Dolbeault complex
\begin{equation}
\label{eq:Dolb-complex}
\bar D
: \Omega^{0,0} \otimes {\ad(\g)} \stackrel{\bar D}{\to}
  \Omega^{0,1}\otimes \ad(\g) \stackrel{\bar D}{\to}
   \Omega^{0,2} \otimes {\ad(\g)}
\end{equation}
twisted by $\Omega^{0,0}\oplus \Omega^{2,0}$.
The index of the
 self-dual complex (\ref{eq:sd-complex2}) differs from the index of the
 Dolbeault complex (\ref{eq:Dolb-complex}) by a factor
accounting for complexification,
and another 
computing the
 weights of the toric $\U(1)_{\epsilon_1}\times \U(1)_{\epsilon_2}$ action on the fiber of $\Omega^{0,0} \oplus
 \Omega^{2,0}$ at the origin:
 \begin{equation}
\label{eq:chern-ch-self}
   \ind (
D^\text{vm}
  ) = 
\frac{1 + t_1^{-1} t_2 ^{-1}}2
\text{ind}(\bar D)\,.
 \end{equation}

Mathematically it is sometimes more convenient to consider the
 torsion free sheaves,
which are known to be in a one-to-one correspondence with
non-commutative instantons. 
Deformations of the 
 torsion free sheaves are captured by the Dolbeault complex
(\ref{eq:Dolb-complex}).
Each fixed point is labeled
by an $N$-tuple of Young diagrams $\vec Y = (Y_1, \dots, Y_N)$. Each
partition $Y_{\alpha}$ defines an ideal sheaf of rank one $\CalE_{Y}$ in the standard
way \cite{MR1711344}. Let $V_{Y}$ be the space of holomorphic
sections of $\CalE_{Y}$. For $Y = (\lambda_1 \geq \lambda_2 \dots \geq
\lambda_{\lambda_1'})$,
where $\lambda_i$ and $\lambda'_i$ are the number of squares in the $i$-th column
and row respectively,
 the basis of $V_{Y}$ is given by monomials
$z_1^{i-1} z_2^{j-1}$ for all $(i, j)$ such that $j > \lambda_{i}$. (The
counting of squares in each Young diagram starts from $(i,j) = (1,1)$).   In
other words the basis in the $V_{Y}$ is enumerated by the squares
outside of the Young diagram $Y$. Each basis element $z_1^{i-1}
z_2^{j-1}$ generates an eigenspace of the torus $T=\U(1)_{\epsilon_1}\times \U(1)_{\epsilon_2}$ with eigenvalue
$t_1^{1-i} t_2^{1-j}$, where $(t_1, t_2) = (e^{i \ve_1}, e^{i \ve_2})$.
Therefore the character of $V_{Y}$ as a $\U(1)_{\epsilon_1}\times \U(1)_{\epsilon_2}$-module
is 
\begin{equation}
  \ch(V_{Y}) = \sum_{(i,j) \not \in Y} t_1^{1 - i} t_2 ^{1 -j} = 
\frac{ 1}{(1 - t_1^{-1})(1- t_2^{-1})} - \chi(Y)\,,
\end{equation}
where 
\begin{equation}
  \chi(Y) = \sum_{(i,j) \in Y} t_1^{1 - i} t_2 ^{1 - j}\,.
\end{equation}
For a transposed Young diagram $Y^*$, we have
\begin{equation}
  \chi(Y^*) = \sum_{(i,j) \in Y} t_1^{i-1} t_2 ^{j-1}\,.
\end{equation}

For each fixed point $\vec Y$ we need to compute the equivariant index
 of the twisted Dolbeault complex (\ref{eq:Dolb-complex})
in the background of the connection defined by $\vec Y$. Since $\ad
(\g) = \bar {\bf N} \otimes {\bf N}$ where $\bf N$ is the fundamental representation, 
the adjoint-valued cohomology space of $\bar D$ is the tensor
product of the $V^{*}_{Y}$ and $V_{Y}$ modules over the ring of
holomorphic functions. Hence 
\begin{multline}
\label{eq:character}
\text{ind}(\bar D)
 = \ch ( V^{*}_{Y} \otimes_{\CalO} V_{Y}) = \ch(V^{*}_Y) \ch
  (V_{Y}) / \ch (\CalO^{*})
 \\ = \sum_{i, j = 1}^{N} s_{i}^{-1} s_j
\left ( \frac{ 1}{(1 - t_1)(1- t_2)} - \chi(Y^{*}_{i}) \right) 
\left ( \frac{ 1}{(1 - t_1^{-1})(1- t_2^{-1})} - \chi(Y_j)
\right)(1-t_1)(1-t_2) 
\end{multline}
where $s_i = e^{i \hat a_i}$. 
We can extract the common infinite part independent of $\vec Y$
\begin{equation}
\label{eq:chern-ch-dolb}
\text{ind}(\bar D)_{\text{1-loop}}
 = \sum_{i, j = 1}^{N} s_i^{-1}
   s_j \frac {1}{(1 - t_1^{-1})(1- t_2^{-1})}\,,
\end{equation}
and denote the remainder in (\ref{eq:character})
by $\text{ind}(\bar D)_{\text{inst}}$:
\begin{equation}
 \text{ind}(\bar D) 
=: \text{ind}(\bar D)_{\text{1-loop}} 
+ \text{ind}(\bar D)_{\text{inst}}\,.
\label{ind-Dbar-inst}
\end{equation}

To convert the index
(Chern character)
$\text{ind}(D^\text{vm})$
to the fluctuation determinant (Euler character),
 we need to expand in
  powers of $(t_1,t_2)$ and take the product of weights
according to the rule
\begin{eqnarray}
\label{rule-app}
\sum_\alpha c_\alpha e^{w_\alpha(\ve_1,\ve_2,\hat a)}
\rightarrow \prod_\alpha w_\alpha(\ve_1,\ve_2,\hat a)^{c_\alpha}\,.  
\end{eqnarray}
Notice that $\text{ind}(\bar D)$
and $t_1^{-1}t_2^{-1} \text{ind}(\bar D)$
in (\ref{eq:chern-ch-self})
are exchanged by
$(\ve _1,\ve _2, \hat a) \rightarrow 
(-\ve _1,- \ve _2, -\hat a)$.
For the common one-loop factor $Z_\text{1-loop}$ at the north pole, 
it is
important to use 
\begin{eqnarray}
  \text{ind}(D^\text{vm})_\text{1-loop}
=
\frac{1 + t_1^{-1} t_2 ^{-1}}2
\text{ind}(\bar D)_\text{1-loop}\,,
\end{eqnarray}
rather than $ \text{ind}(\bar D)_\text{1-loop}$,
before expanding in positive powers of $t_1, t_2$
as we did in section \ref{sec:one-loop}.%
\footnote{%
Recall that one applies the positive and negative expansions
to the north and south poles, respectively.
Because of this, in the absence of a 't Hooft loop,
the {\it product} of north and south pole
contributions to the one-loop factor
obtained from $\text{ind}(\bar D)$
is the same as the one from  $\text{ind}(D^\text{vm})$.
}
For the finite instanton part $Z_\text{inst}$ computed by the rule
(\ref{rule-app}), however, 
the result obtained from
\begin{eqnarray}
  \text{ind}(D^\text{vm})_\text{inst}
=
\frac{1 + t_1^{-1} t_2 ^{-1}}2
\text{ind}(\bar D)_\text{inst}
\end{eqnarray}
is identical to the result from $\text{ind}(\bar D)_\text{inst}$
because the signs that appear from
$(\ve _1,\ve _2, \hat a) \rightarrow 
(-\ve _1,- \ve _2, -\hat a)$ cancel out in the product.
Thus the instanton partition function can be computed either from
the self-dual complex or the Dolbeault complex.

\vfill\eject

\bibliography{refs}
\end{document}